\documentclass[letterpaper,english,prl,twocolumn,amsmath,amssymb,superscriptaddress]{revtex4-2}
\usepackage[T1]{fontenc}
\usepackage{babel}
\usepackage{prettyref}
\usepackage{bm}
\usepackage{dsfont}
\usepackage{amsmath}
\usepackage{amssymb}
\usepackage{graphicx}
\usepackage[unicode=true,
 bookmarks=true,bookmarksnumbered=false,bookmarksopen=false,
 breaklinks=false,pdfborder={0 0 1},backref=false,colorlinks=false]
 {hyperref}
\hypersetup{pdftitle={The Distribution of Unstable Fixed Points in Chaotic Neural Networks},
 pdfauthor={Jakob Stubenrauch, Christian Keup, Anno Kurth, Moritz Helias, Alexander van Meegen},
 colorlinks=true,linkcolor=black,citecolor=black,urlcolor=black,filecolor=black}

\makeatletter

\usepackage{tikz}
\usetikzlibrary{calc}

\usepackage[latin1]{inputenc}

\newrefformat{eq}{\hyperref[#1]{Eq.~(\ref{#1})}}
\newrefformat{cap}{\hyperref[#1]{Fig.~\ref{#1}}}
\newrefformat{fig}{\hyperref[#1]{Fig.~\ref{#1}}}
\newrefformat{tab}{\hyperref[#1]{Table ~\ref{#1}}}
\newrefformat{sec}{\hyperref[#1]{Section~\ref{#1}}}
\newrefformat{sub}{\hyperref[#1]{Section~\ref{#1}}}
\newrefformat{cha}{\hyperref[#1]{Chapter~\ref{#1}}}

\usepackage{MnSymbol}
\DeclareMathAlphabet\mathbb{U}{msb}{m}{n}

\makeatother

\begin{document}
\global\long\def\T{\mathrm{T}}%
\global\long\def\L{\mathrm{L}}%
\global\long\def\l{\langle}%
\global\long\def\r{\rangle}%
\global\long\def\rr{\rangle\rangle}%
\global\long\def\tr{\mathrm{tr}}%
\global\long\def\lr#1{\left\langle #1\right\rangle }%
\global\long\def\vect#1{\bm{#1}}%
\global\long\def\matr#1{\mathbf{#1}}%
\global\long\def\uint#1{\int\mathrm{d}#1\,}%
\global\long\def\dint#1#2#3{\int_{#2}^{#3}\mathrm{d}#1\,}%
\global\long\def\pint#1{\int\mathcal{D}#1\,}%
\global\long\def\C{C}%
\global\long\def\Ct{\tilde{C}}%
\global\long\def\k{\ell}%
\global\long\def\evalat#1{\left.#1\right|}%
\global\long\def\xt{\tilde{x}}%
\global\long\def\erf{\mathrm{erf}}%
\global\long\def\clip{\mathrm{clip}}%
\global\long\def\bx{\boldsymbol{x}}%
\global\long\def\by{\boldsymbol{y}}%
\global\long\def\bz{\boldsymbol{z}}%
\global\long\def\bJ{\mathbf{J}}%
\global\long\def\bEta{\boldsymbol{\eta}}%
\global\long\def\bM{\mathbf{M}}%
\global\long\def\bX{\mathbf{X}}%
\global\long\def\bSigma{\boldsymbol{\Sigma}}%
\global\long\def\cN{\mathcal{N}}%
\global\long\def\cO{\mathcal{O}}%
\global\long\def\bphi{\phi(\boldsymbol{x})}%
\global\long\def\id{\mathds{1}}%
\global\long\def\bLambda{\boldsymbol{\Lambda}}%
\global\long\def\diag{\mathrm{diag}}%
\global\long\def\cD{\mathcal{D}}%
\global\long\def\reals{\mathbb{R}}%
\global\long\def\asymeq{\overset{.}{=}}%
\global\long\def\bs{\boldsymbol{s}}%

\title{The Distribution of Unstable Fixed Points in Chaotic Neural Networks}
\author{Jakob Stubenrauch}
\email{Corresponding author. {jakob.stubenrauch@rwth-aachen.de}}

\affiliation{Institute of Neuroscience and Medicine (INM-6) and Institute for Advanced
Simulation (IAS-6) and JARA-Institute Brain Structure-Function Relationships
(INM-10), J\"ulich Research Centre, J\"ulich, Germany}
\affiliation{Department of Physics, Faculty 1, RWTH Aachen University, Aachen,
Germany}
\author{Christian Keup}
\affiliation{Institute of Neuroscience and Medicine (INM-6) and Institute for Advanced
Simulation (IAS-6) and JARA-Institute Brain Structure-Function Relationships
(INM-10), J\"ulich Research Centre, J\"ulich, Germany}
\affiliation{Department of Physics, Faculty 1, RWTH Aachen University, Aachen,
Germany}
\author{Anno C.~Kurth}
\affiliation{Institute of Neuroscience and Medicine (INM-6) and Institute for Advanced
Simulation (IAS-6) and JARA-Institute Brain Structure-Function Relationships
(INM-10), J\"ulich Research Centre, J\"ulich, Germany}
\affiliation{Department of Physics, Faculty 1, RWTH Aachen University, Aachen,
Germany}
\author{Moritz Helias}
\affiliation{Institute of Neuroscience and Medicine (INM-6) and Institute for Advanced
Simulation (IAS-6) and JARA-Institute Brain Structure-Function Relationships
(INM-10), J\"ulich Research Centre, J\"ulich, Germany}
\affiliation{Department of Physics, Faculty 1, RWTH Aachen University, Aachen,
Germany}
\author{Alexander van Meegen}
\affiliation{Institute of Neuroscience and Medicine (INM-6) and Institute for Advanced
Simulation (IAS-6) and JARA-Institute Brain Structure-Function Relationships
(INM-10), J\"ulich Research Centre, J\"ulich, Germany}
\affiliation{Institute of Zoology, University of Cologne, 50674 Cologne, Germany}
\date{\today}
\begin{abstract}

We analytically determine the number and distribution of fixed points
in a canonical model of a chaotic neural network. This distribution
reveals that fixed points and dynamics are confined to separate shells
in phase space. Furthermore, the distribution enables us to determine
the eigenvalue spectra of the Jacobian at the fixed points. Despite
the radial separation of fixed points and dynamics, we find that nearby
fixed points act as partially attracting landmarks for the dynamics.
\end{abstract}
\maketitle
Chaotic dynamics are well understood in low dimensional systems but
are notoriously challenging in high dimensions \citep{Strogatz2014}.
In low dimensions, the first step in the analysis of a dynamical system
is to determine its fixed points in phase space, for example the two
unstable fixed points at the centers of the Lorenz attractor \citep{Lorenz63_130}.
For high-dimensional nonlinear systems, merely finding all fixed points
rapidly becomes prohibitive \citep{Press07}. Accordingly, the phase
space of high-dimensional chaotic systems is still largely terra incognita
(see \citep{Ros23_257401} for a recent exception) despite their ubiquitous
appearance across disciplines.

Here, we investigate the phase space of a particular high-dimensional
nonlinear system: a neural network. Neural networks inherently operate
outside equilibrium due to the asymmetric coupling \citep{Rabinovich06_1213,Sompolinsky88_2}
and exhibit chaotic dynamics \citep{Sompolinsky88_259,Vreeswijk96_1724}.
Concretely, we consider the canonical model of a chaotic neural network
proposed by \citep{Sompolinsky88_259}: $N$ nonlinearly connected
units $x_{i}(t)$ receiving a constant external input $\eta_{i}$
and obeying the dynamics
\begin{align}
\dot{x}_{i} & =-x_{i}+\sum_{j=1}^{N}J_{ij}\phi(x_{j})+\eta_{i}\label{eq:SCS_model}
\end{align}
with nonlinear transfer function $\phi(x)=\tanh(x)$, independent
and identically distributed (i.i.d.) coupling weights $J_{ij}\sim\cN(0,g^{2}/N)$,
and i.i.d.~external inputs $\eta_{i}\sim\cN(0,D)$. Due to the directional
nature of synapses, the coupling weights are asymmetric.

A major advantage of the recurrent network model \eqref{eq:SCS_model}
is that the analytical approach of dynamical mean-field theory \citep{Sompolinsky88_259,Crisanti18_062120}
(pedagogically reviewed in \citep{Helias20_970}) led to a deep understanding
of its dynamics at large $N$. Both without \citep{Sompolinsky88_259,Crisanti18_062120}
and with \citep{Schuecker18_041029} external input, the statistics
of the activity is well captured by a zero-mean Gaussian process with
self-consistent autocorrelation function and the system is chaotic
above a critical value of $g=g_{c}$ (without external input $g_{c}=1$).
Dynamical mean-field theory has also been successfully applied to
various extensions of the model \citep{molgedey92_3717,Stern14_062710,Kadmon15_041030,Aljadeff15_088101,Mastrogiuseppe17_e1005498,vanMeegen18_258302,Landau18_e1006309,Mastrogiuseppe18_609,Kusmierz_20_028101,Keup21_021064,vanMeegen21_158302,Wardak22}.
Furthermore, cross-correlations \citep{Clark23_118401} and the full
Lyapunov spectrum have been investigated recently \citep{Engelken23_043044}.
In summary, the high-dimensional chaotic dynamics of the model are
exceptionally well understood.

In contrast to the dynamics, the phase space and the fixed point structure
of the model received considerably less attention. The pioneering
work \citep{wainrib13_118101} showed that for $g=1+\varepsilon$,
$0<\varepsilon\ll1$ and in the absence of external input, the phase
space contains a number of fixed points that grows exponentially with
the system size $N$. Their finding that the maximum Lyapunov exponent
and the rate controlling the exponential increase of fixed points
to leading order in $\varepsilon$ coincide led the authors to hypothesize
a deep link between the abundance of fixed points and the chaoticity
of the dynamics. However, to investigate the relation between the
fixed points and the dynamics, a mandatory first step is to establish
the location of the fixed points.

In this Letter, we determine the spatial distribution of the fixed
points. On the technical level, this requires to compute the expected
zero-crossings of a Gaussian process with location dependent, i.e.,
non-homogeneous, statistics. Afterwards, we compare the geometries
of the fixed points and the dynamics to show that both are confined
to separate shells in phase space. Next, we leverage the distribution
of fixed points to investigate the stability of the local dynamics
at the fixed points, from which we deduce that the dynamics closely
passes the fixed points. Finally, we argue that the fixed points can
be used as landmarks to describe the dynamics symbolically. 

\emph{Spatial distribution of fixed points.}---Throughout the Letter
we assume that the network is in the chaotic regime $g>g_{c}$ and
that the number of units $N$ is sufficiently large to allow us focus
on the leading order behavior, which we express by the abbreviated
notation $a\overset{.}{=}e^{Nb}$ to denote $\lim_{N\to\infty}\frac{1}{N}\ln a=b$.

We use vector notation to write \prettyref{eq:SCS_model} as $\dot{\bx}=\by(\bx)$
with velocity $\by(\bx)=-\bx+\bJ\bphi+\bEta$. Since $\bJ$ and $\bEta$
are Gaussian, the velocity $\by(\bx)$ and the Jacobian $\by^{\prime}(\bx)=-\id+\bJ\,\diag[\phi^{\prime}(\bx)]$
are Gaussian processes (note that both $\by(\bx)$ and $\by^{\prime}(\bx)$
are non-homogeneous). Due to the randomness of $\by(\bx)$, the location
of the fixed points $\by(\bx)=0$ is described by a distribution $\rho(\bx)$.
This distribution counts how many fixed points are on average within
an infinitesimal volume in phase space. We determine $\rho(\bx)$
from the Kac-Rice formula \citep{Kac43_314,Rice45,Azais09}
\begin{align}
\rho(\bx) & =\left\langle \delta[\by(\bx)]\,\left|\det\by^{\prime}(\bx)\right|\right\rangle _{\bJ,\bEta}.\label{eq:KR_forFPD}
\end{align}
The expected number of fixed points $N_{\mathrm{fp}}$ follows from
the normalization constant $N_{\mathrm{fp}}=\int_{\reals^{N}}d\bx\,\rho(\bx)$.
The Jacobian determinant $\left|\det\by^{\prime}(\bx)\right|$ in
\prettyref{eq:KR_forFPD} ensures that every fixed point contributes
to the distribution with equal weight. \prettyref{eq:KR_forFPD} is
equivalent to a random matrix problem: Using Bayes' law to condition
on $\by(\bx)=0$, $\rho(\bx)$ can be rewritten into \citep[A.1]{suppmat}
\begin{align}
\rho(\bx) & =p_{\L}(\bx)\left\langle \left|\det\left[\bM(\bx)+\bX\,\bSigma(\bx)\right]\right|\right\rangle _{X_{ij}\sim\cN(0,N^{-1})}\label{eq:FPD_exact}
\end{align}
where the first factor $p_{\L}(\bx)=\mathcal{N}(\bm{x}\,|\,0,\kappa(\bx)+D)$
with $\kappa(\bx)=\frac{g^{2}}{N}\sum_{i=1}^{N}\phi(x_{i})^{2}$ is
the probability of the velocity to be zero and the second factor is
the expected determinant of a random matrix with mean $\bM(\bx)$
and covariance $\bSigma(\bx)^{\T}\bSigma(\bx)$ controlling for the
fluctuations of the velocity process. Here, $\kappa(\bx)+D$ is the
variance of the Gaussian process $\by(\bx)$ and $\bM(\bx)$, $\bSigma(\bx)^{\T}\bSigma(\bx)$
are the mean and variance, respectively, of the Gaussian process $\by^{\prime}(\bm{x})$
conditioned on $\by(\bx)=0$. Extending the technique introduced in
\citep{Sommers88}, and excluding singularities, the determinant is
given by $\left\langle \left|\det\left[\bM(\bx)+\bX\,\bSigma(\bx)\right]\right|\right\rangle \asymeq\exp\big(N\zeta(\bx)\big)$
with \citep[A.2]{suppmat}
\begin{align}
\zeta(\bx) & =-\frac{1}{2}z_{*}+\frac{1}{2N}\sum_{i=1}^{N}\ln[1+z_{*}g^{2}\phi^{\prime}(x_{i})^{2}]\label{eq:logdet_norank1}
\end{align}
where $z_{\ast}$ is the solution of
\begin{align}
1 & =\frac{1}{N}\sum_{i=1}^{N}\frac{g^{2}\phi^{\prime}(x_{i})^{2}}{1+z_{\ast}\,g^{2}\phi^{\prime}(x_{i})^{2}}.\label{eq:z_star_eq}
\end{align}
To summarize, the $N$-dimensional distribution of the fixed points
is $\rho(\bx)\asymeq\exp\big(-NS(\bx)\big)$ with
\begin{align}
S(\bx) & =\frac{q(\bx)}{2[\kappa(\bx)+D]}+\frac{1}{2}\ln\left\{ 2\pi[\kappa(\bx)+D]\right\} -\zeta(\bx)\label{eq:FPD_exponent_final}
\end{align}
where $q(\bx)=\frac{1}{N}\sum_{i=1}^{N}x_{i}^{2}$ and $\zeta(\bx)$
is determined by \prettyref{eq:logdet_norank1} and \prettyref{eq:z_star_eq}.

The fixed point distribution \prettyref{eq:FPD_exponent_final}
depends on $\bx$ only through network averages. Consequently, it
is permutation-symmetric, which implies an approximate independence
of coordinates $\int\rho(\bx)\,dx_{k+1}...dx_{N}\approx\prod_{i=1}^{k}\mu(x_{i})$
for $k\ll N$ \citep[B.3]{suppmat}. Furthermore, we can express
it as a functional $\rho(\bx)=\rho[\mu_{\bx}]$ of the \emph{empirical
measure} 
\begin{align}
\mu_{\bx}(y) & =\frac{1}{N}\sum_{i=1}^{N}\delta(y-x_{i}),
\end{align}
i.e., the empirical distribution of vector components of $\bx$. From
the\emph{ }expected empirical measure at the fixed points $\mu_{\ast}(y)=\left\langle \mu_{\bx}(y)\right\rangle _{\bx\sim\rho(\bx)}$
all network-averaged expectation values $\left\langle \frac{1}{N}\sum_{i=1}^{N}f(x_{i})\right\rangle _{\bx\sim\rho(\bx)}=\int dy\,\mu_{\ast}(y)f(y)$
can be computed. The expected empirical measure is given, for large
$N$, by the saddle point that maximizes $\rho[\mu_{\bx}]$ in function
space and admits the form \citep[B.1]{suppmat}
\begin{equation}
\mu_{\ast}(y)\propto\sqrt{1+\alpha\phi^{\prime}(y)^{2}}e^{-\frac{y^{2}+\gamma\phi(y)^{2}}{2\beta}},\label{eq:em_compact}
\end{equation}
for which the parameters $\alpha$, $\beta$, and $\gamma$ are determined
by $1=g^{2}\big\langle(\phi^{\prime}(y)^{-2}+\alpha)^{-1}\big\rangle_{\mu_{\ast}}$,
$\beta=g^{2}\big\langle\phi(y)^{2}\big\rangle_{\mu_{\ast}}+D$, and
$\gamma=g^{2}\big(1-\beta^{-1}\big\langle y^{2}\big\rangle_{\mu_{\ast}}\big)$
where expectation values have to be taken self-consistently w.r.t.~$\mu_{\ast}$.

We compare the empirical measure \prettyref{eq:em_compact} to the
distribution of vector components of numerically determined fixed
points. 
\begin{figure}
\includegraphics{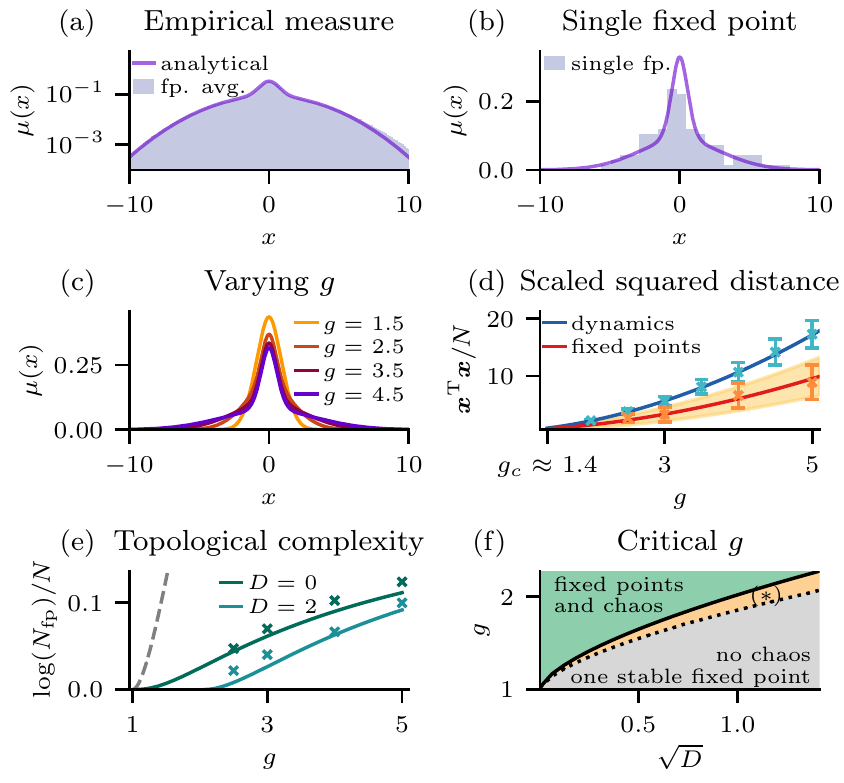}

\caption{\label{fig:emp_meas_and_dist_to_orig}Characterization of fixed points.
(a) Distribution of vector components of fixed points (empirical measure):
theoretical result (solid line) based on \prettyref{eq:em_compact}
and\textbf{ }histogram (bars) averaged across $4\times10^{4}$ fixed
points for a single realization of the coupling weights. (b) Same
as (a) for a randomly chosen single fixed point. (c) Empirical measure
for different values of $g$. (d) Scaled squared norm of fixed points
(theory: red solid line; theoretical finite-size standard deviation:
yellow shading; numerical results: orange error bars) and dynamics
(theory: blue solid line; numerical integration averaged over $15$
realizations per $g$: turquoise error bars). (e) Number of fixed
points (topological complexity; solid lines: theory, \prettyref{eq:topcomp};
crosses: numerical results, \citep[E]{suppmat}; gray dashed line:
theory for $g\rightarrow1^{+}$, $D=0$ by \citep{wainrib13_118101}).
(f) Transition to positive topological complexity (black dotted line)
and transition to chaos (black solid line) based on \citep{Schuecker18_041029}.
In the regime $(\ast)$ the phase space exhibits an exponential number
of fixed points but the dynamics are not chaotic.\textbf{ }Parameters:
$D=0.1$ for (a)--(d), $g=4$ for (a) and (b), $N=100$ for numerical
results and the shading in (d).}
\end{figure}
For the numerical results, we fix the realization of the random parameters
and employ a Levenberg-Marquart rootfinder starting from independent
normally distributed initial conditions until saturation, i.e., until
almost no new fixed points are found (see \citep[E]{suppmat}). We
see in \prettyref{fig:emp_meas_and_dist_to_orig}(a) that the theory
\prettyref{eq:em_compact} is in excellent agreement with the empirical
measure averaged over all fixed points found numerically in a single
realization of $\bJ$ (see \citep[Fig. 2(a)]{suppmat} for further
examples). Moreover, as shown in \prettyref{fig:emp_meas_and_dist_to_orig}(b),
even single fixed points closely resemble the expected empirical measure.

Indeed the probability distribution functional of the empirical measures
takes the form $P[\mu]\asymeq\exp(-NH[\mu])$ with an analytically
determined\emph{ }rate functional $H[\mu]$ \citep[B.2]{suppmat},
i.e., it obeys a large deviation principle \citep{Touchette09,Dembo10}.
The minimum of $H[\mu]$ is attained at the expected empirical measure
$\mu_{\ast}$. Since $P[\mu]$ quantifies both the variability within
a realization of the parameters as well as across realizations \citep[B.2]{suppmat},
akin to the law of total variance, deviations of $\mu$ from $\mu_{\ast}$
are rare for large $N$ even at the level of individual fixed points.
Mismatches between $\mu_{\bx}$ and $\mu_{\ast}$ for a fixed point
$\bx$ are thus finite size effects (see \citep[Fig. 2(b)]{suppmat}
for further examples).

\emph{Geometry of fixed point distribution.}---The excess kurtosis
of $\mu_{\ast}$ reflects the compromise between the two contributions
in the fixed point density \prettyref{eq:FPD_exact}: high probability
of a vanishing velocity, captured by $p_{\L}(\bx)$, and a steep expected
slope to increase the density of zero crossings, captured by the determinant.
The former leads to the broad Gaussian base and the latter to the
sharp peak.  Geometrically the excess kurtosis implies that the fixed
points are posed in the vicinity of spans of subsets of axes in phase
space.

The expected value of the \emph{scaled squared distance} $u_{\bx}=\frac{1}{N}\bx^{\T}\bx$,
which quantifies the distance to the origin, is $u_{\ast}=\int dy\,y^{2}\mu_{\ast}(y)$.
The distribution of the distance $P(u)$ inherits the exponential
form of $P[\mu]$ because $u$ is determined by the empirical measure;
formally, this is a consequence of the\emph{ }contraction principle
\citep{Touchette09}. Thus, $P(u)\asymeq\exp(-NI(u))$ where the rate
function is
\begin{equation}
I(u)=\inf_{\mu:\,\int y^{2}\mu(y)\,dy=u}H[\mu].\label{eq:rate_function_ssd}
\end{equation}
The rate function is again $\cO(1)$; hence, for $N\gg1$, the fluctuations
of the distance vanish and the fixed points are distributed on a thin
spherical shell with radius $\sqrt{Nu_{\ast}}$. In \prettyref{fig:emp_meas_and_dist_to_orig}(d),
we show the average distance and fluctuations based on \prettyref{eq:rate_function_ssd}
(see \citep[C.1]{suppmat}) for $N=100$.

To put the fixed point's distance to the origin into context with
the dynamics we leverage the result from dynamic mean-field theory
that the network-averaged variance $q[\bx(t)]=\frac{1}{N}\sum_{i=1}^{N}x_{i}(t)^{2}$
is self-averaging for stationary statistics with fluctuations vanishing
in the large $N$ limit \citep{Helias20_970,vanMeegen21_158302}.
Hence, also the trajectory is embedded in a thin shell around the
origin, which is of radius $\sqrt{Nq}$.

The confinement to a thin spherical shell is a generic feature of
high-dimensional, weakly correlated random variables \citep{Vershynin18}
but the radius depends on the underlying high-dimensional distribution.
Thus, we compare the radii of the two shells in \prettyref{fig:emp_meas_and_dist_to_orig}(d).
For all $g>g_{c}$, the fixed points shell is inside of the trajectories
shell. Furthermore, for $N\rightarrow\infty$, the overlap between
the shells vanishes and thus the trajectory is clearly separated from
the fixed points in phase space.

\emph{Number of fixed points.}---A core result of \citep{wainrib13_118101}
is that without noise, $D=0$, the system has a transition from a
single stable fixed point to an exponential number of unstable fixed
points $N_{\text{fp}}\asymeq\exp[cN]$ at $g_{c}=1$. The respective
rate $c$, the \emph{topological complexity}, is
\begin{equation}
c=\frac{\gamma-\alpha}{2g^{2}}-D\frac{\gamma}{2\beta g^{2}}+\ln\mathcal{Z}-\frac{1}{2}\ln(2\pi\beta)\label{eq:topcomp}
\end{equation}
where $\mathcal{Z}$ is the normalization of \prettyref{eq:em_compact}.
Asymptotically, at $D=0$, $c\to-\frac{1}{\pi}+\ln\big[1+\mathrm{erf}(\sqrt{1/\pi})e^{1/\pi}\big]$
for $g\to\infty$ and $c=\big(\frac{2}{3}\big)^{4}\epsilon^{3}+O(\epsilon^{4})$
for $g-1=\epsilon\to0$ \citep[C.3]{suppmat}. In \prettyref{fig:emp_meas_and_dist_to_orig}(e),
we see that the critical gain parameter $g_{c}$ grows with $D>0$;
the corresponding transition line is shown in \prettyref{fig:emp_meas_and_dist_to_orig}(f).
For $D\ll1$, the transition to an exponential number of fixed points
coincides with the transition to chaos. For larger noise strengths
$D$, however, a regime exists where the system has an exponential
number of fixed points yet the dynamics are not chaotic (see \citep{Krishnamurthy22}
for a similar observation). Both our theory and numerical results
are in agreement with the critical point $g_{c}=1$ for $D=0$ found
by \citep{wainrib13_118101} but the quantitative value of the topological
complexity differs clearly from the result by \citep{wainrib13_118101}
and is well-captured by our theory \prettyref{fig:emp_meas_and_dist_to_orig}(e).

\begin{figure}
\includegraphics{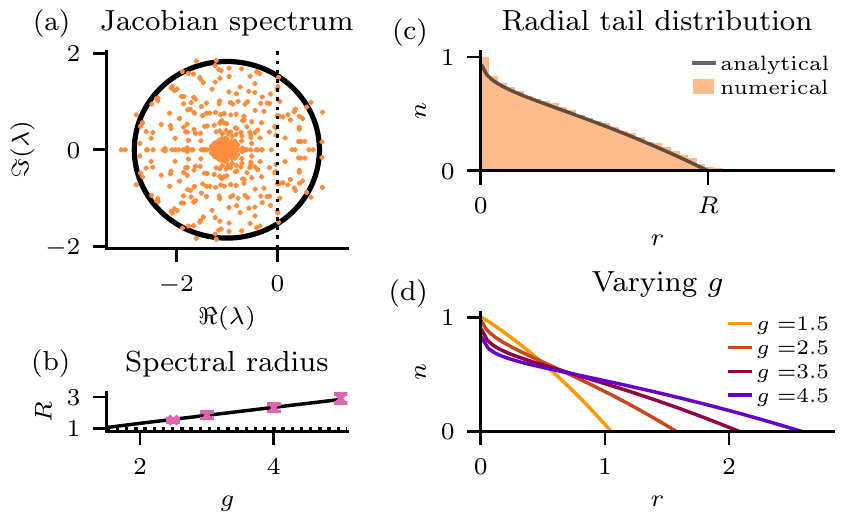}

\caption{\label{fig:Jacobian-spectrum}Jacobian spectrum at fixed points. (a)
Eigenvalues (orange dots) of the Jacobian at five randomly chosen
fixed points of one realization of $\protect\bJ$ and\textbf{ $\protect\bEta$}.
Every eigenvalue with positive real part corresponds to an unstable
mode of the fixed point. The support of the spectrum (black circle)
is determined by the spectral radius, \prettyref{eq:spectral_radius}.
(b) Spectral radius; theory (black line) and numerical results (pink
errorbars). (c) Radial tail distribution of eigenvalues; theory (black
line) based on \prettyref{eq:fraction_further_r} and histogram (orange
bars)\textbf{ }based on the eigenvalues shown in (a). (d) Theory from
(c) for varying $g$. Parameters: $N=100$, $g=3$ in (a) and (c),
$D=0.1$.}
\end{figure}

\emph{Stability of fixed points.}---We now consider the dynamics
in the vicinity of fixed points $\bx^{\ast}$. Local stability is
determined by the eigenvalues of the Jacobian at the fixed point $\by^{\prime}(\bx^{\ast})$
(see \prettyref{fig:Jacobian-spectrum}(a)): Each eigenvalue with
positive (negative) real part corresponds to an unstable (stable)
eigendirection of the fixed point. The Jacobian can be written as
$\by^{\prime}(\bx)=-\id+\bX\,\diag[g\phi^{\prime}(\bx)]$ with $X_{ij}\sim\cN(0,N^{-1})$;
the corresponding eigenvalue spectrum can be computed with the method
developed in \citep{Ahmadian15_012820} because $\diag[g\phi^{\prime}(\bx)]$
is invertible. For large $N$, the eigenvalue distribution of $\by^{\prime}(\bx)$
is centered around $-1+0i$ and confined within a circle of radius
$R(\bx)=g\sqrt{N^{-1}\phi^{\prime}(\bx)^{\T}\phi^{\prime}(\bx)}$.
At a fixed point, the contraction principle attests a large deviation
principle for the spectral radius, with the expected value given by
\begin{equation}
R_{\ast}=g\sqrt{\int dy\,\phi^{\prime}(y)^{2}\mu_{\ast}(y)}.\label{eq:spectral_radius}
\end{equation}
This radius is always $>1$ in the chaotic phase (\prettyref{fig:Jacobian-spectrum}),
indicating that for large $N$ all fixed points are unstable.

Within the support the distribution of eigenvalues is isotropic around
the center. We express the distribution by the\emph{ }fraction of
eigenvalues further than $r$ from the center $n_{\bx}(r)$, i.e.,
the radial tail distribution. It obeys, again, a large deviation principle
dominated by the solution $n_{\ast}(r)$ of
\begin{equation}
1=\int dy\,\mu_{\ast}(y)\frac{g^{2}\phi^{\prime}(y)^{2}}{r^{2}+n_{\ast}(r)\,g^{2}\phi^{\prime}(y)^{2}}.\label{eq:fraction_further_r}
\end{equation}
We present the solution in \prettyref{fig:Jacobian-spectrum}(c).
The unstable modes of fixed points are underrepresented relative to
a uniform spectrum. The overrepresentation of eigenvalues on the real
line {[}\prettyref{fig:Jacobian-spectrum}(a){]} and the smearing
of the spectral radius {[}\prettyref{fig:Jacobian-spectrum}(c){]}
are known finite size effects \citep{Edelman94_247,Rider14_1621}.

\begin{figure}
\includegraphics{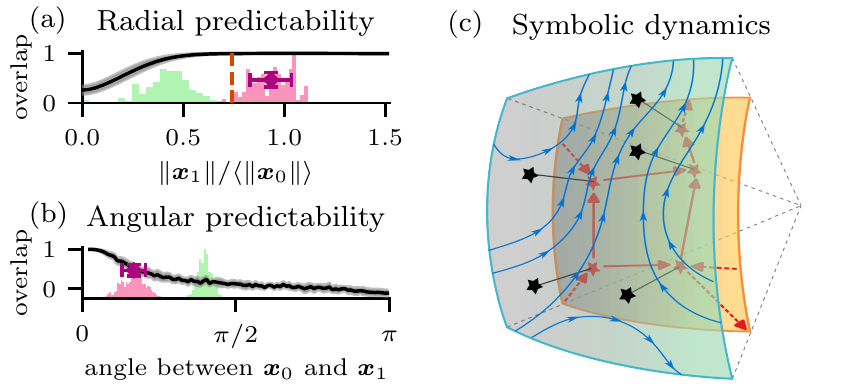}

\caption{\label{fig:Correlations}Impact of fixed points. (a) Pearson correlation
between $\protect\by(\protect\bx_{0})$, where $\protect\bx_{0}$
is on the attractor, and its linear predictor $\protect\by(\protect\bx_{1})+\protect\by^{\prime}(\protect\bx_{1})[\protect\bx_{0}-\protect\bx_{1}]$
for radially perturbed points $\protect\bx_{1}$ (black line), and
nearest fixed points (purple errorbar). Density of norm of the fixed
points that are closest to the dynamics (pink) and to random points
(green) and mean norm of all fixed points (dashed line). Parameters:
$g=5$, $D=0$, and $N=300$ (perturbations), and $N=100$ (fixed
points), shadings denote standard deviation. (b) Like (a) but for
angular perturbed points $\protect\bx_{1}=a\protect\bx_{0}+\sqrt{1-a^{2}}\|\protect\bx_{0}\|\protect\bz$,
where $z_{i}\protect\overset{\mathrm{i.i.d.}}{\sim}\protect\cN(0,1/N)$.
(c) Sketch of trajectories (blue lines) meandering around the projections
(black stars) onto the dynamics shell of some fixed points (red stars),
and the resulting symbolic dynamics (red arrows).}
\end{figure}

\emph{Impact of fixed points.}---In \citep{wainrib13_118101}, it
is conjectured that the dynamics meanders around the different fixed
points, first following their stable directions and then being repelled
along their unstable directions. The radial separation of fixed points
and dynamics seemingly contradicts this hypothesis.

The conjecture assumes that linearizing the velocity at the nearest
fixed point $\bx^{\ast}$ provides a satisfactory prediction of the
actual velocity $\by[\bx(t)]$. Given the linear predictor $\by_{1}\equiv\by(\bx^{\ast})+\by^{\prime}(\bx^{\ast})\left[\bx(t)-\bx^{\ast}\right]$
we quantify the accuracy of the prediction by the time-averaged Pearson
correlation between $\by[\bx(t)]$ and $\by_{1}$. Points that are
radially shrunk or stretched still predict the dynamics well {[}\prettyref{fig:Correlations}(a){]}
whereas points that are rotated by a fixed angle into a random tangential
direction quickly decline in predictive power {[}\prettyref{fig:Correlations}(b){]}.
On average, the correlation with the nearest fixed points' linear
predictor is approximately $0.5$ (purple error bar) which corroborates
the intuitive picture by \citep{wainrib13_118101}. The remaining
gap to perfect predictability must be predominantly attributed to
the angular, instead of the radial, separation (pink histograms)
although the angular separation is small compared to the angular separation
between fixed points and random control points that are statistically
equivalent to the dynamics (green histograms).

The small angular distance at which the dynamics passes the fixed
points results from a majority of attractive directions: In \citep[D.1]{suppmat}
we show that almost every sample from a sphere around a fixed point
moves towards it. In contrast, the expected constant velocity of equivalent
points $\boldsymbol{p}$, $p_{i}\overset{\text{i.i.d.}}{\sim}\mu_{\ast}$,
is non-zero which renders $\boldsymbol{p}$ highly repulsive. In
this sense, fixed points can be seen as landmarks of the dynamics
\citep{Rabinovich08_48}: The dynamics float above the shell of fixed
points, visiting the fixed points in a chain that symbolically describes
the dynamics {[}\prettyref{fig:Correlations}(c){]}.

\emph{Discussion.}---In this Letter we characterize the phase space
structure of a chaotic neural network using the distribution of fixed
points. We identify a decoupling of the chaos transition and the emergence
of unstable fixed points. We furthermore show a spatial separation
between fixed points and dynamics. Last, we establish the dynamic
role of the fixed points as attractive landmarks for the trajectory.

In high-dimensional linear dynamical systems, May's pioneering stability
analysis \citep{May72_413} enabled considerable insights into the
dynamics of ecosystems \citep{Allesina15_63}. In the nonlinear case,
the number of fixed points can be determined if the velocity is generated
by a homogeneous Gaussian potential \citep{Fyodorov04_240601}; in
this case, it is even possible to determine the number of minima of
the potential \citep{Bray07_150201,Fyodorov07_1081,Fyodorov12_167203}
with applications in deep learning \citep{Dauphin14_NeurIPS,Choromanska15_ICAIS}.
The non-potential case has been addressed in \citep{wainrib13_118101}
for the random network \eqref{eq:SCS_model} at $g=1+\varepsilon$
with $\varepsilon\ll1$, in \citep{Fyodorov16_6827} for a velocity
field based on a homogeneous Gaussian field (for which it is possible
to extend the analysis to the fraction of stable directions of fixed
points \citep{BenArous21_e2023719118}), and in \citep{Ros23_257401,Ros23_305003}
for a Lotka-Volterra model. Other non-homogeneous cases have been
studied in \citep{Fyodorov16a_124003} with dynamics constrained to
a sphere and in \citep{Fedeli21_022201} with a metastable model where
the distance of the fixed points to the origin determines which initial
conditions decay or escape. For a recent review on stationary points
of random fields see \citep{Ros23_HighDimensional}. Here, we go beyond
the previous results and determine distribution of fixed points, which
includes their number, of the random neural network \eqref{eq:SCS_model}
for arbitrary $g>g_{c}$. To this end, we extend methods from random
matrix theory \citep{Sommers88,Ahmadian15_012820} to compute the
determinant of non-hermitian random matrices with a correlation structure
including low-rank terms. The analysis is restricted to the average
number of fixed points, which provides an upper bound to the typical
number of fixed points \citep{Ros23_257401,Ros23_305003}. Our numerical
results suggest that the bound is rather tight; for the density of
fixed points we expect no difference between average and typical behavior,
in line with the excellent match of empirical and theoretical density.

The results presented here pave the way towards a mechanistic understanding
of the velocity field underlying high dimensional chaotic networks.
There are several directions for further research: First, it would
be interesting to extend the analysis to more structured networks,
e.g., in terms of low rank perturbations \citep{Mastrogiuseppe18_609},
levels of symmetry \citep{Sommers88,Marti18_062314,Berlemont22_bioRxiv},
or population structure \citep{Aljadeff15_088101,Kadmon15_041030,vanMeegen21_158302}.
Second, the frustration created by the quenched rotation between the
axes system, singled out by the element-wise application of the nonlinearity,
and the eigensystem of the connectivity creates the complexity of
the phase space. What is the geometric relation between the axes
system and the dynamics on the chaotic attractor? Last, deep insights
into trained neural networks are possible by analyzing their phase
space \citep{Sussillo13_626,Vyas20_249}. Here, we analyzed the phase
space of a random reservoir which already allows universal computation
if the readout is optimized \citep{Maass02_2531}---more generally,
learning with chaotic networks \citep{Poole16_3360,Keup21_021064,Farrell22_564}
is a direction of research that might be able to leverage the exponential
number of fixed points and the associated capability for sequence
processing.

\begin{acknowledgments}
We are grateful to G\"unther Palm and Alexandre Ren\'e for discussions
about chaotic dynamics. CK would like to acknowledge a helpful discussion
with Vittorio Erba about entropy. This work was partly supported by
the European Union Horizon 2020 Grant No.~945539 (Human Brain Project
SGA3), funded by the Deutsche Forschungsgemeinschaft (DFG, German
Research Foundation) - 368482240/GRK2416, the Helmholtz Association
Initiative and Networking Fund under project number SO-092 (Advanced
Computing Architectures, ACA), the German Federal Ministry for Education
and Research (BMBF Grant 01IS19077A), and the Excellence Initiative
of the German federal and state governments (ERS PF-JARA-SDS005).
Open access publication funded by the Deutsche Forschungsgemeinschaft
(DFG, German Research Foundation) -- 491111487
\end{acknowledgments}

\end{document}


\global\long\def\T{\mathrm{T}}%
\global\long\def\L{\mathrm{L}}%
\global\long\def\l{\langle}%
\global\long\def\r{\rangle}%
\global\long\def\tr{\mathrm{tr}\,}%
\global\long\def\lr#1{\left\langle #1\right\rangle }%
\global\long\def\vect#1{\bm{#1}}%
\global\long\def\matr#1{\mathbf{#1}}%
\global\long\def\uint#1{\int\mathrm{d}#1\,}%
\global\long\def\dint#1#2#3{\int_{#2}^{#3}\mathrm{d}#1\,}%
\global\long\def\pint#1{\int\mathcal{D}#1\,}%
\global\long\def\C{C}%
\global\long\def\Ct{\tilde{C}}%
\global\long\def\k{\ell}%
\global\long\def\evalat#1{\left.#1\right|}%
\global\long\def\xt{\tilde{x}}%
\global\long\def\erf{\mathrm{erf}}%
\global\long\def\clip{\mathrm{clip}}%
\global\long\def\bx{\boldsymbol{x}}%
\global\long\def\by{\boldsymbol{y}}%
\global\long\def\bz{\boldsymbol{z}}%
\global\long\def\bJ{\mathbf{J}}%
\global\long\def\bEta{\boldsymbol{\eta}}%
\global\long\def\bM{\mathbf{M}}%
\global\long\def\bem{\mathbf{m}}%
\global\long\def\bX{\mathbf{X}}%
\global\long\def\bSigma{\boldsymbol{\Sigma}}%
\global\long\def\bsigma{\boldsymbol{\sigma}}%
\global\long\def\bZ{\mathbf{Z}}%
\global\long\def\cN{\mathcal{N}}%
\global\long\def\cO{\mathcal{O}}%
\global\long\def\bphi{\phi(\boldsymbol{x})}%
\global\long\def\id{\mathds{1}}%
\global\long\def\bLambda{\boldsymbol{\Lambda}}%
\global\long\def\diag{\mathrm{diag}}%
\global\long\def\vectorized{\mathrm{vec}}%
\global\long\def\bone{\mathbf{1}}%
\global\long\def\ba{\boldsymbol{a}}%
\global\long\def\bbb{\boldsymbol{b}}%
\global\long\def\bF{\mathbf{F}}%
\global\long\def\bv{\mathbf{v}}%
\global\long\def\bD{\mathbf{D}}%
\global\long\def\bc{\boldsymbol{c}}%
\global\long\def\bA{\mathbf{A}}%
\global\long\def\bB{\mathbf{B}}%
\global\long\def\bk{\boldsymbol{k}}%
\global\long\def\bK{\mathbf{K}}%
\global\long\def\reals{\mathds{R}}%
\global\long\def\cD{\mathcal{D}}%
\global\long\def\bmu{\boldsymbol{\mu}}%
\global\long\def\bC{\mathbf{C}}%
\global\long\def\jax{j_{\ast}^{\mu}}%
\global\long\def\mua{\mu_{\ast}}%
\global\long\def\muta{\tilde{\mu}_{\ast}}%
\global\long\def\asymeq{\overset{.}{=}}%
\global\long\def\bh{\boldsymbol{h}}%
\global\long\def\iidsim{\overset{\text{i.i.d.}}{\sim}}%
\global\long\def\cA{\mathcal{A}}%
\global\long\def\cum#1{\langle\langle#1\rangle\rangle}%
\global\long\def\cZ{\mathcal{Z}}%

\title{The Distribution of Unstable Fixed Points in Chaotic Neural Networks\\
(Supplemental Material)}
\author{Jakob Stubenrauch}
\affiliation{Institute of Neuroscience and Medicine (INM-6) and Institute for Advanced
Simulation (IAS-6) and JARA-Institute Brain Structure-Function Relationships
(INM-10), J\"ulich Research Centre, J\"ulich, Germany}
\affiliation{Department of Physics, Faculty 1, RWTH Aachen University, Aachen,
Germany}
\author{Christian Keup}
\affiliation{Institute of Neuroscience and Medicine (INM-6) and Institute for Advanced
Simulation (IAS-6) and JARA-Institute Brain Structure-Function Relationships
(INM-10), J\"ulich Research Centre, J\"ulich, Germany}
\affiliation{Department of Physics, Faculty 1, RWTH Aachen University, Aachen,
Germany}
\author{Anno C.~Kurth}
\affiliation{Institute of Neuroscience and Medicine (INM-6) and Institute for Advanced
Simulation (IAS-6) and JARA-Institute Brain Structure-Function Relationships
(INM-10), J\"ulich Research Centre, J\"ulich, Germany}
\affiliation{Department of Physics, Faculty 1, RWTH Aachen University, Aachen,
Germany}
\author{Moritz Helias}
\affiliation{Institute of Neuroscience and Medicine (INM-6) and Institute for Advanced
Simulation (IAS-6) and JARA-Institute Brain Structure-Function Relationships
(INM-10), J\"ulich Research Centre, J\"ulich, Germany}
\affiliation{Department of Physics, Faculty 1, RWTH Aachen University, Aachen,
Germany}
\author{Alexander van Meegen}
\affiliation{Institute of Neuroscience and Medicine (INM-6) and Institute for Advanced
Simulation (IAS-6) and JARA-Institute Brain Structure-Function Relationships
(INM-10), J\"ulich Research Centre, J\"ulich, Germany}
\affiliation{Institute of Zoology, University of Cologne, 50674 Cologne, Germany}
\date{\today}

\maketitle
\tableofcontents{}

\subsection{Distribution of fixed points}

\subsubsection{Kac-Rice formula\label{subsec:Kac-Rice-formula}}

\paragraph{Counting roots}

To find the distribution of fixed points we first need to establish
how to count roots of a field $\by(\bx)$. The number of roots in
a volume $V$ is given by the Kac-Rice formula \citep{Kac43_314,Rice45,Azais09}
\begin{equation}
N_{\text{fp}}(V)=\int_{V}d\bx\,\delta\left[\by(\bx)\right]\left|\det\by^{\prime}(\bx)\right|\label{eq:root_counting}
\end{equation}
where $\left[\by^{\prime}(\bx)\right]_{ij}=\partial_{x_{j}}y_{i}(\bx)$
is the Jacobian of the field $\by(\bx)$. The Dirac delta $\delta\left[\by(\bx)\right]$
hits at every root of $\by(\bx)$ and the Jacobian determinant assures
that every root contributes $1$ to the total number of roots in $N_{0}$.
The latter is necessary due to the scaling property of the Dirac delta
\begin{equation}
\int_{V}d\bx\,\delta\left[\by(\bx)\right]\,h(\bx)=\sum_{\bx^{\ast}\in V\,\text{s.t.}\,\by(\bx^{\ast})=0}\frac{h(\bx^{\ast})}{\left|\det\by^{\prime}(\bx^{\ast})\right|}.
\end{equation}

\paragraph{Fixed point distribution}

Since the counting formula Eq. \eqref{eq:root_counting} holds for
every Borel set $V\subset\reals^{N}$ \citep{Azais09}, the integrand
in Eq. \eqref{eq:root_counting} can be interpreted as a distribution
of roots. On average over realizations $\bJ$ and $\bEta$, this distribution
is
\begin{equation}
\rho(\bx)=\left\langle \delta\left[\by(\bx)\right]\left|\det\by^{\prime}(\bx)\right|\right\rangle _{\bJ,\bEta}.\label{eq:fpd_def}
\end{equation}
Its norm $N_{\text{fp}}\equiv\left\langle N_{\text{fp}}(\reals^{N})\right\rangle =\int_{\mathds{R}^{N}}d\bx\rho(\bx)$
is the expected total number of fixed points. 

\paragraph{Joint distribution of velocity and Jacobian}

The expectation value in Eq. \eqref{eq:fpd_def} can be simplified
by transforming from the realization parameters $\bJ$ and $\bEta$
to the velocity
\begin{equation}
\by(\bx)\equiv\dot{\bx}=-\bx+\bJ\bphi+\bEta\label{eq:model}
\end{equation}
and Jacobian
\begin{equation}
\by^{\prime}(\bx)=-\id+\bJ\,\diag\left[\phi^{\prime}(\bx)\right]\label{eq:modeljac}
\end{equation}
as random fields. Here, $J_{ij}\overset{\text{i.i.d.}}{\sim}\cN(0,g^{2}/N)$
and $\eta_{i}\overset{\text{i.i.d.}}{\sim}\cN(0,D)$ where we use
the notation $\chi_{i}\overset{\text{i.i.d.}}{\sim}\cN(a,b)$ to denote
that the random variables $\chi_{i}$ are independently and identically
distributed (i.i.d.) by the Gaussian distribution with mean $a$ and
variance $b$. The joint probability distribution of $\by(\bx)$ and
$\by^{\prime}(\bx)$ follows by computing their first moments
\begin{equation}
\left\langle y_{i}(\bx)\right\rangle =-x_{i}\equiv\mu_{i}(\bx),\quad\quad\left\langle \left[\by^{\prime}(\bx)\right]_{ik}\right\rangle =-\delta_{ik}\equiv[\bmu_{i}(\bx)]_{k},\label{eq:first_moment}
\end{equation}
and second cumulants
\begin{equation}
\begin{aligned}\left\langle \left\langle y_{i}(\bx)y_{j}(\bx)\right\rangle \right\rangle  & =\delta_{ij}\left[\frac{g^{2}}{N}\sum_{k}\phi(x_{k})^{2}+D\right]\equiv\delta_{ij}[\kappa(\bx)+D]\\
\left\langle \left\langle y_{i}(\bx)\left[\by^{\prime}(\bx)\right]_{jk}\right\rangle \right\rangle  & =\frac{g^{2}}{N}\delta_{ij}\phi(x_{k})\phi^{\prime}(x_{k})\equiv\delta_{ij}[\bk(\bx)]_{k},\\
\left\langle \left\langle \left[\by^{\prime}(\bx)\right]_{ik}\left[\by^{\prime}(\bx)\right]_{jl}\right\rangle \right\rangle  & =\delta_{ij}\delta_{kl}\frac{g^{2}}{N}\phi^{\prime}(x_{k})^{2}\equiv\delta_{ij}[\bK(\bx)]_{kl}.
\end{aligned}
\label{eq:second_cumulant}
\end{equation}
Since all higher cumulants vanish, the joint distribution of $\by$
and $\by^{\prime}$ is Gaussian. Furthermore, it factorizes into velocity
components $y_{i}$ and respective gradients (rows of the Jacobian)
$\nabla y_{i}=\by_{i}^{\prime}$. Hence, we can write the probability
distribution of $\by$ and $\by^{\prime}$ at each point $\bx$ as
$p_{\bx}(\by,\by^{\prime})=\prod_{i=1}^{N}p_{\bx}(y_{i},\by_{i}^{\prime})$
with
\begin{equation}
\begin{pmatrix}y_{i}\\
\by_{i}^{\prime}
\end{pmatrix}\overset{\mathrm{i.i.d.}}{\sim}\mathcal{N}\left[\begin{pmatrix}\mu_{i}(\bx)\\
\bmu_{i}(\bx)
\end{pmatrix},\begin{pmatrix}\left[\kappa(\bx)+D\right] & \bk(\bx)^{\T}\\
\bk(\bx) & \bK(\bx)
\end{pmatrix}\right].\label{eq:joint_probability}
\end{equation}
A different formulation of the Kac-Rice formula \citep{Azais09,Stratonovich67}
gives the distribution of crossings of a Gaussian random process $(\by,\by^{\prime})$
through $\by=0$ in terms of the joint probability as
\begin{equation}
\rho(\bx)=\int d\by^{\prime}\,p_{\bx}(\by=0,\by^{\prime})\,\left|\det\by^{\prime}\right|.\label{eq:fpd_joint_pdf}
\end{equation}
Since the formulation in terms of the Dirac delta, Eq. \eqref{eq:fpd_def},
is self-explaining and Eq. \eqref{eq:fpd_joint_pdf} is harder to
motivate from scratch, we show their equivalence by explicit substitution
from $\bJ$ and $\bEta$ to $\by$ and $\by^{\prime}$. Solving Eqs.
\eqref{eq:model} and \eqref{eq:modeljac} for $\bJ$ and $\bEta$
yields
\begin{equation}
\begin{bmatrix}\bJ\\
\bEta
\end{bmatrix}(\by,\by^{\prime})=\begin{bmatrix}\left(\id+\by^{\prime}\right)\,\diag\left[1/\bphi^{\prime}\right]\\
\by+\bx-\left(\id+\by^{\prime}\right)\,\diag\left[1/\bphi^{\prime}\right]\bphi
\end{bmatrix}.\label{eq:solve_for_J_eta}
\end{equation}
The Jacobian of the substitution is
\begin{equation}
D_{\begin{bmatrix}\by^{\prime}\\
\by
\end{bmatrix}}\begin{bmatrix}\bJ\\
\bEta
\end{bmatrix}=\begin{bmatrix}D_{\by^{\prime}}\bJ & D_{\by}\bJ\\
D_{\by^{\prime}}\bEta & D_{\by}\bEta
\end{bmatrix}=\begin{bmatrix}\diag\left[1/\bphi^{\prime}\right]\\
 & \ddots &  & \boldsymbol{0}\\
 &  & \diag\left[1/\bphi^{\prime}\right]\\
 & D_{\by^{\prime}}\bEta &  & \id
\end{bmatrix}
\end{equation}
where the columns of $\bJ$ are stacked to a vector to get the usual
rule for substitution. Due to the upper right block being zero, the
lower left block does not contribute to the determinant. The integration
in Eq. \eqref{eq:fpd_def} can hence be written
\begin{align}
\rho(\bx) & =\left\langle \delta\left[\boldsymbol{y}(\boldsymbol{x})\right]\left|\det\boldsymbol{y}^{\prime}(\boldsymbol{x})\right|\right\rangle _{\bJ,\bEta}=\int d\bJ\,d\bEta\,p_{g}(\bJ)\,p_{D}(\bEta)\,\delta\left[\boldsymbol{y}(\boldsymbol{x})\right]\left|\det\boldsymbol{y}^{\prime}(\boldsymbol{x})\right|\\
 & =\int d\by^{\prime}\,d\by\,\left|\prod_{i=1}^{N}\frac{1}{\phi^{\prime}(x_{i})}\right|^{N}p_{g}\left[\bJ(\by^{\prime})\right]\,p_{D}\left[\bEta(\by,\by^{\prime})\right]\,\delta\left[\boldsymbol{y}(\bx)\right]\left|\det\boldsymbol{y}^{\prime}\right|\label{eq:fpd_substi_expand}
\end{align}
where $p_{g}(\bX)=\prod_{ij}\cN(J_{ij}|0,g^{2}/N)$ and $p_{D}(\bx)=\prod_{i}\cN(x_{i}|0,D)$
and $\cN(a|b,c)$ denotes the Gaussian distribution function with
random variable $a$, mean value $b$ and variance $c$.

By plugging Eq. \eqref{eq:solve_for_J_eta} into the first part of
the integrand in Eq. \eqref{eq:fpd_substi_expand}, one can see that
\begin{equation}
\left|\prod_{i=1}^{N}\frac{1}{\phi^{\prime}(x_{i})}\right|^{N}p_{g}\left[\bJ(\by^{\prime})\right]=\prod_{i}\cN\left[\left.\by_{i}^{\prime}\right|\bmu_{i}(\bx),\bK(\bx)\right]\label{eq:marginal_jac}
\end{equation}
with $\bmu$, $\bK$ as in Eqs. \eqref{eq:first_moment} and \eqref{eq:second_cumulant}.
Equation \eqref{eq:marginal_jac} is the marginal distribution of
the Jacobian $\by^{\prime}$, since marginalizing a Gaussian distribution
does not alter diagonal cumulants. A similar procedure, and using
the formula for the conditional probability of a Gaussian \citep[Appendix A2]{WilliamsRasmussen06}
\begin{equation}
p_{\bx}(y_{i}|\by_{i}^{\prime})=\mathcal{N}\left[\left.y_{i}\right|\mu_{i}+\bk^{\T}\bK^{-1}\left(\by_{i}^{\prime}-\bmu_{i}\right),\kappa+D-\bk^{\T}\bK^{-1}\bk\right],
\end{equation}
allows to identify, using that $\kappa=\bk^{\T}\bK^{-1}\bk$,
\begin{equation}
\int d\by\,p_{D}\left[\bEta(\by,\by^{\prime})\right]\,\delta\left[\boldsymbol{y}(\bx)\right]=p_{\boldsymbol{x}}(y_{i}=0|\boldsymbol{y}_{i}^{\prime})
\end{equation}
Putting all together, we get
\begin{equation}
\rho(\bx)=\int d\by^{\prime}p_{\bx}(\by=0|\by^{\prime})p_{\bx}(\by^{\prime})\left|\det\by^{\prime}\right|,
\end{equation}
which, using Bayes' law
\begin{equation}
p_{\bx}(\by,\by^{\prime})=p_{\bx}(\by^{\prime})\,p_{\bx}(\by|\by^{\prime})=p_{\bx}(\by)\,p_{\bx}(\by^{\prime}|\by),\label{eq:bayes_law}
\end{equation}
gives Eq. \eqref{eq:fpd_joint_pdf}.

\paragraph{Condition on the level}

Using the rightmost dissection in Eq. \eqref{eq:bayes_law}, i.e.
condition on the level $\by$, is advantageous since the \emph{level
probability} 
\begin{equation}
p_{\L}(\bx)\equiv p_{\bx}(\by=0)=\frac{1}{\sqrt{2\pi\left[\kappa(\bx)+D\right]}^{N}}e^{-\frac{\bx^{\T}\bx}{2\left[\kappa(\bx)+D\right]}},
\end{equation}
where $\kappa(\bx)=(g^{2}/N)\bphi^{\T}\bphi$, can be taken out of
the integral, i.e.
\begin{equation}
\rho(\bx)=p_{\L}(\bx)\,\left\langle \left|\det\by^{\prime}\right|\right\rangle _{\by^{\prime}\sim p_{\bx}(\by^{\prime}|\by=0)}.\label{eq:pL_JacDet}
\end{equation}
The level probability denotes the marginal probability of the field
$\by(\bx)$ to be zero at a point $\bx$. This alone is not the fixed
point distribution since the distribution of the Jacobian (think of
it as the slope) controls how often the field $\by(\bx)$ can cross
the zero level when it is close to zero.

\paragraph{Jacobian distribution}

To compute the expected determinant in Eq. \eqref{eq:pL_JacDet},
we have to characterize the conditional distribution $p_{\bx}(\by^{\prime}|\by=0)$
first. The conditional distribution of a Gaussian is again a Gaussian.
Its moments are given by \citep[Appendix A2]{WilliamsRasmussen06}
\begin{equation}
p_{\bx}(\by^{\prime}|\by=0)=\prod_{i}\cN\left[\by_{i}^{\prime}|\bM_{i}(\bx),\bC(\bx)\right]
\end{equation}
where the mean conditioned Jacobian is
\begin{equation}
M_{ij}(\bx)=-\delta_{ij}-\frac{g^{2}}{N}\frac{x_{i}\phi(x_{j})\phi^{\prime}(x_{j})}{\kappa(\bx)+D}\label{eq:mean_cond_jac}
\end{equation}
and each row $\by_{i}^{\prime}$ has the same covariance matrix
\begin{equation}
C_{nm}=\delta_{nm}\frac{g^{2}}{N}\phi^{\prime}(x_{n})^{2}-\frac{g^{4}}{N^{2}}\frac{\phi(x_{n})\phi^{\prime}(x_{n})\phi(x_{m})\phi^{\prime}(x_{m})}{\kappa(\boldsymbol{x})+D}
\end{equation}
It will prove useful to write the conditioned Jacobian in the compact
form
\begin{equation}
\boldsymbol{y}^{\prime}(\boldsymbol{y}=0)=\bM+\bX\bSigma
\end{equation}
with $X_{ij}\overset{\text{i.i.d.}}{\sim}\mathcal{N}(0,1/N)$ and
$\bSigma^{\T}\bSigma\equiv N\bC$. To see this, lets verify that it
has the correct moments (here we suppress the dependency $(\boldsymbol{y}=0)$):
\begin{equation}
\begin{aligned}\left\langle y_{ij}^{\prime}\right\rangle  & =M_{ij}\\
\left\langle y_{ij}^{\prime}y_{kl}^{\prime}\right\rangle -\left\langle y_{ij}^{\prime}\right\rangle \left\langle y_{kl}^{\prime}\right\rangle  & =\left\langle \sum_{n,m=1}^{N}X_{in}\Sigma_{nj}X_{km}\Sigma_{ml}\right\rangle =\sum_{n,m=1}^{N}\frac{1}{N}\delta_{ik}\delta_{nm}\Sigma_{nj}\Sigma_{ml}=\delta_{ik}\frac{1}{N}\sum_{n}\Sigma_{jn}^{\T}\Sigma_{nl}=\delta_{ik}C_{jl}
\end{aligned}
\end{equation}
Thus, as needed, the different rows are uncorrelated ($\delta_{ik}$)
and within a row the correlation structure is given by $\bC$. The
scaled square root $\bSigma$ of the covariance matrix
\begin{equation}
\begin{aligned}\bSigma^{\T}\bSigma= & N\boldsymbol{C}=\boldsymbol{\Lambda}(\mathds{1}-\boldsymbol{vv}^{\T})\boldsymbol{\Lambda},\\
\boldsymbol{\Lambda} & =g\,\text{diag}[\phi^{\prime}(\bx)],\\
v_{i} & =\frac{g}{\sqrt{N(\kappa+D)}}\phi(x_{i})
\end{aligned}
\end{equation}
can be computed by using 
\begin{equation}
\sqrt{\mathds{1}-\boldsymbol{vv}^{\T}}=\mathds{1}-\frac{\boldsymbol{vv}^{\T}}{1+\sqrt{1-\boldsymbol{v}^{\T}\boldsymbol{v}}},
\end{equation}
which can be checked for general vectors $\boldsymbol{v}$. We find
\begin{equation}
\begin{aligned}\bSigma(\bx) & =\left[\id-\frac{g^{2}}{N}\frac{\phi(\bx)\phi(\bx)^{\T}}{\kappa(\bx)+D\left[1+\sqrt{1+\kappa(\bx)/D}\right]}\right]\,\boldsymbol{\Lambda}(\bx)\end{aligned}
.
\end{equation}
Concluding, the fixed point distribution can be expressed as
\begin{equation}
\rho(\bx)=p_{\L}(\bx)\,\left\langle \left|\det\left[\bM(\bx)+\bX\,\bSigma(\bx)\right]\right|\right\rangle _{X_{ij}\overset{\mathrm{i.i.d.}}{\sim}\cN(0,1/N)}.\label{eq:fpd_exact}
\end{equation}
Note that both $\bM(\bx)$ and $\bSigma(\bx)$ consist of a sum of
a diagonal and a rank one matrix.

\subsubsection{Determinant of a non-hermitian random matrix}

We want to compute the expectation of the modulus determinant in Eq.
\eqref{eq:fpd_exact}, i.e. of the matrix
\begin{equation}
\by^{\prime}(\bx)=\bM(\bx)+\bX\,\bSigma(\bx),\quad X_{ij}\overset{\mathrm{i.i.d.}}{\sim}\cN(0,1/N).\label{eq:random_matrix}
\end{equation}
The determinant of a matrix is the product of its eigenvalues. We
use that asymptotically
\begin{equation}
\zeta(\bx)=\frac{1}{N}\ln\left\langle \left|\det\by^{\prime}(\bx)\right|\right\rangle =-\frac{1}{N}\ln\left\langle \left|\det\by^{\prime}(\bx)\right|^{-1}\right\rangle ,\label{eq:logdet}
\end{equation}
i.e., the determinant is strongly self-averaging. We first calculate
the determinant based on this assumption and justify the assumption
afterwards by a replica calculation. Briefly, the replica calculation
shows that the replicas decouple under a replica-symmetric Ansatz.

\paragraph{Determinant as Gaussian integral}

To compute $\zeta$, we build on the approach by \citep{Sommers88}.
First, we use $\det\matr A^{\T}=\det\matr A$ and $\det(\matr A\matr B)=\det\matr A\det\matr B$
to rewrite Eq. \eqref{eq:logdet} as 
\begin{equation}
\zeta=-\frac{1}{N}\ln\left\langle \frac{1}{\sqrt{\det(\by^{\prime\T}\by^{\prime}+\varepsilon\id)}}\right\rangle \label{eq:logdet_prereplica}
\end{equation}
where we also added a diagonal matrix with (infinitesimal) strength
$\varepsilon$ to avoid divergences. We note that $\by^{\prime\T}\by^{\prime}$
is positive semi-definite, and thus $\by^{\prime\T}\by^{\prime}+\varepsilon\id$
is positive definite, so we can use the Gaussian integral $\int d\bz\,e^{-\frac{1}{2}\bz^{\T}\matr A\bz}=\sqrt{\frac{(2\pi)^{N}}{\det\matr A}}$
to get
\begin{equation}
\zeta=-\frac{1}{N}\ln\left\langle \int d\bz\,e^{-\frac{1}{2}\bz^{\T}(\by^{\prime\T}\by^{\prime}+\varepsilon\id)\bz}\right\rangle +\frac{1}{2}\ln(2\pi).\label{eq:zeta_replica-1}
\end{equation}
The exponent in the expectation value in Eq. \eqref{eq:zeta_replica-1}
is at most quadratic in the random variables $X_{ij}.$ Hence, it
can be computed as a Gaussian integral and we arrive at
\begin{equation}
\zeta=-\frac{1}{N}\ln\int d\bz\,e^{-\frac{1}{2}\varepsilon\bz^{\T}\bz-\frac{\bz^{\T}\bM^{\T}\bM\bz}{2(1+\frac{1}{N}\bz^{\T}\bSigma^{\T}\bSigma\bz)}-\frac{N}{2}\ln(1+\frac{1}{N}\bz^{\T}\bSigma^{\T}\bSigma\bz)}+\frac{1}{2}\ln(2\pi).\label{eq:zeta_expc_solved}
\end{equation}

\paragraph{Collective variables}

We evaluate the integral in saddle point approximation. To this end,
we introduce the collective variables
\begin{equation}
m=\frac{1}{N}\bz^{\T}\bM^{\T}\bM\bz,\qquad\sigma=\frac{1}{N}\bz^{\T}\bSigma^{\T}\bSigma\bz\label{eq:collective_vars}
\end{equation}
Denoting all collective variables as $\bm{\theta}$, we arrive at
\[
\zeta=-\frac{1}{N}\ln\int d\bm{\theta}\,e^{-N[\frac{m}{2(1+\sigma)}+\frac{1}{2}\ln(1+\sigma)]}\int d\bz\,e^{-\frac{1}{2}\varepsilon\bz^{\T}\bz}\delta(\bm{\theta}-\bm{\theta}(\bz))+\frac{1}{2}\ln(2\pi).
\]
We can interpret
\[
p(\bm{\theta})=\Big(\frac{\varepsilon}{2\pi}\Big)^{N/2}\int d\bz\,e^{-\frac{1}{2}\varepsilon\bz^{\T}\bz}\delta(\bm{\theta}-\bm{\theta}(\bz))
\]
as the density of the transformed random variable $\bm{\theta}=\bm{\theta}(\bz)$
where $z_{i}\overset{i.i.d.}{\sim}\mathcal{N}(0,1/\varepsilon)$.
The corresponding characteristic function is
\[
\Phi(\tilde{\bm{\theta}})=\Big(\frac{\varepsilon}{2\pi}\Big)^{N/2}\int d\bz\,e^{-\frac{1}{2}\varepsilon\bz^{\T}\bz+i\tilde{\bm{\theta}}^{\T}\bm{\theta}(\bz)}=\sqrt{\frac{\varepsilon^{N}}{\det(\varepsilon\bm{1}-2i\tilde{m}\tfrac{1}{N}\bM^{\T}\bM-2i\tilde{\sigma}\tfrac{1}{N}\bSigma^{\T}\bSigma)}}
\]
such that we can write
\[
p(\bm{\theta})=\frac{N^{2}}{(2\pi)^{2}}\int d\tilde{\bm{\theta}}\,e^{-N[i\tilde{\bm{\theta}}^{\T}\bm{\theta}-\Omega(\tilde{\bm{\theta}})]}
\]
with a scaled cumulant generating function $\Omega(\tilde{\bm{\theta}})=\frac{1}{N}\ln\Phi(N\tilde{\bm{\theta}})$
which is given by
\[
\Omega(\tilde{\bm{\theta}})=-\frac{1}{2N}\ln\det(\varepsilon\id-2i\tilde{m}\bM^{\T}\bM-2i\tilde{\sigma}\bSigma^{\T}\bSigma)+\frac{1}{2}\ln\varepsilon.
\]
Inserting $p(\bm{\theta})$, we arrive at
\begin{equation}
\zeta=-\frac{1}{N}\ln\int d\bm{\theta}\int d\tilde{\bm{\theta}}\,e^{-NH(\bm{\theta},\tilde{\bm{\theta}})}+O(N^{-1}\ln N),
\end{equation}
\begin{equation}
H(\bm{\theta},\tilde{\bm{\theta}})=\frac{m}{2(1+\sigma)}+\frac{1}{2}\ln(1+\sigma)+i\tilde{\bm{\theta}}^{\T}\bm{\theta}+\frac{1}{2N}\ln\det(\varepsilon\bm{1}-2i\tilde{m}\bM^{\T}\bM-2i\tilde{\sigma}\bSigma^{\T}\bSigma).\label{eq:def_H}
\end{equation}
Note that the terms containing $\varepsilon$ or $2\pi$ cancel. Now,
we would like to perform a saddle-point approximation to get $\zeta=H(\bm{\theta}_{*},\tilde{\bm{\theta}}_{*})+\cO(N^{-1}\ln N)$
where $\bm{\theta}_{*}$ and $\tilde{\bm{\theta}}_{*}$ minimize $H(\bm{\theta},\tilde{\bm{\theta}})$.

\paragraph{Saddle-point approximation}

The corresponding saddle-point equations are
\begin{equation}
-2i\tilde{m}=\frac{1}{1+\sigma},\qquad-2i\tilde{\sigma}=-\frac{m}{(1+\sigma)^{2}}+\frac{1}{1+\sigma}=\frac{1+\sigma-m}{(1+\sigma)^{2}},\label{eq:SPE_tilde}
\end{equation}
\[
-2im=\frac{1}{N}\partial_{\tilde{m}}\ln\det(\varepsilon\id-2i\tilde{m}\bM^{\T}\bM-2i\tilde{\sigma}\bSigma^{\T}\bSigma),\qquad-2i\sigma=\frac{1}{N}\partial_{\tilde{\sigma}}\ln\det(\varepsilon\id-2i\tilde{m}\bM^{\T}\bM-2i\tilde{\sigma}\bSigma^{\T}\bSigma).
\]
To evaluate the derivatives, we use $\partial_{x}\ln\det(x\bm{X}+y\bm{Y})=\mathrm{tr}((x\bm{X}+y\bm{Y})^{-1}\bm{X})$
to obtain
\begin{equation}
m=\frac{1}{N}\mathrm{tr}\Big[(\varepsilon\id-2i\tilde{m}\bM^{\T}\bM-2i\tilde{\sigma}\bSigma^{\T}\bSigma)^{-1}\bM^{\T}\bM\Big],\qquad\sigma=\frac{1}{N}\mathrm{tr}\Big[(\varepsilon\id-2i\tilde{m}\bM^{\T}\bM-2i\tilde{\sigma}\bSigma^{\T}\bSigma)^{-1}\bSigma^{\T}\bSigma\Big].\label{eq:SPE_trace}
\end{equation}
We can combine Eqs. \eqref{eq:SPE_tilde} and \eqref{eq:SPE_trace}
to
\begin{eqnarray}
m & = & \frac{1}{N}\mathrm{tr}\Big[\big(\varepsilon\id+(1+\sigma)^{-1}\bM^{\T}\bM+(1+\sigma-m)(1+\sigma)^{-2}\bSigma^{\T}\bSigma\big)^{-1}\bM^{\T}\bM\Big],\label{eq:SPE_m}\\
\sigma & = & \frac{1}{N}\mathrm{tr}\Big[\big(\varepsilon\id+(1+\sigma)^{-1}\bM^{\T}\bM+(1+\sigma-m)(1+\sigma)^{-2}\bSigma^{\T}\bSigma\big)^{-1}\bSigma^{\T}\bSigma\Big].\label{eq:SPE_sig}
\end{eqnarray}
Putting it all together, we arrive at
\begin{equation}
\zeta=-\frac{\sigma_{*}(1+\sigma_{*}-m_{*})}{2(1+\sigma_{*})^{2}}+\frac{1}{2N}\ln\det[\varepsilon(1+\sigma_{*})\id+\bM^{\T}\bM+(1+\sigma_{*}-m_{*})(1+\sigma_{*})^{-1}\bSigma^{\T}\bSigma]+\cO(N^{-1}\ln N)\label{eq:zeta_full}
\end{equation}
where we used $i\tilde{m}_{*}m_{*}=-\frac{m_{*}}{2(1+\sigma_{*})}$
and $i\tilde{\sigma}_{*}\sigma_{*}=-\frac{\sigma_{*}(1+\sigma_{*}-m_{*})}{2(1+\sigma_{*})^{2}}$
due to Eq. \eqref{eq:SPE_tilde}.

\paragraph{Diagonal case}

We first consider the case where $\bM=-\id$ and $\bSigma=\bLambda\equiv g\,\diag[\phi^{\prime}(\bx)]$,
i.e., we neglect the rank 1 parts. As we will see later, this already
yields the leading order contribution to the determinant.

The resulting saddle-point equations are
\[
m=\frac{1}{N}\sum_{i=1}^{N}\frac{1}{\varepsilon+\frac{1}{1+\sigma}+\frac{1+\sigma-m}{(1+\sigma)^{2}}\lambda_{i}^{2}},\qquad\sigma=\frac{1}{N}\sum_{i=1}^{N}\frac{\lambda_{i}^{2}}{\varepsilon+\frac{1}{1+\sigma}+\frac{1+\sigma-m}{(1+\sigma)^{2}}\lambda_{i}^{2}},
\]
and the determinant is
\[
\zeta=-\frac{\sigma_{*}(1+\sigma_{*}-m_{*})}{2(1+\sigma_{*})^{2}}+\frac{1}{2N}\sum_{i=1}^{N}\ln[\varepsilon(1+\sigma_{*})+1+(1+\sigma_{*}-m_{*})(1+\sigma_{*})^{-1}\lambda_{i}^{2}]+\cO(N^{-1}\ln N).
\]
For $\varepsilon\to0$, $m,\sigma\to\infty$ is a possible solution
to the saddle-point equations. 

Assuming the diverging solution is the relevant one, the determinant
simplifies to
\[
\zeta=-\frac{1}{2}\Big(1-\frac{m_{*}}{\sigma_{*}}\Big)+\frac{1}{2N}\sum_{i=1}^{N}\ln\bigg[\varepsilon\sigma_{*}+1+\Big(1-\frac{m_{*}}{\sigma_{*}}\Big)\lambda_{i}^{2}\bigg]+\cO(N^{-1}\ln N).
\]
The saddle-point equations determine $\frac{m}{\sigma}$ and $\varepsilon\sigma$
via
\[
\frac{m}{\sigma}=\frac{1}{N}\sum_{i=1}^{N}\frac{1}{\varepsilon\sigma+1+(1-\frac{m}{\sigma})\lambda_{i}^{2}},\qquad1=\frac{1}{N}\sum_{i=1}^{N}\frac{\lambda_{i}^{2}}{\varepsilon\sigma+1+(1-\frac{m}{\sigma})\lambda_{i}^{2}}
\]
which can be combined to $(\varepsilon\sigma+1)\frac{m}{\sigma}+(1-\frac{m}{\sigma})=1$.
The latter equation leads to $\varepsilon\sigma=0$, thus the saddle-point
equations reduce further to
\[
\frac{m}{\sigma}=\frac{1}{N}\sum_{i=1}^{N}\frac{1}{1+(1-\frac{m}{\sigma})\lambda_{i}^{2}},\qquad1=\frac{1}{N}\sum_{i=1}^{N}\frac{\lambda_{i}^{2}}{1+(1-\frac{m}{\sigma})\lambda_{i}^{2}}.
\]
These two equations are equivalent; for convenience we choose the
second and introduce $z=1-\frac{m}{\sigma}$ which is determined by
\begin{equation}
1=\frac{1}{N}\sum_{i=1}^{N}\frac{g^{2}\phi^{\prime}(x_{i})^{2}}{1+zg^{2}\phi^{\prime}(x_{i})^{2}}\label{eq:def_of_z_diagonal}
\end{equation}
where we plugged in the explicit Jacobian eigenvalues in the diagonal
approximation $\lambda_{i}=g\phi^{\prime}(x_{i})$. This equation
needs to be solved numerically. In terms of the solution $z_{*}$,
the determinant follows from
\begin{equation}
\zeta=-\frac{1}{2}z_{*}+\frac{1}{2N}\sum_{i=1}^{N}\ln(1+z_{*}g^{2}\phi^{\prime}(x_{i})^{2})+\cO(N^{-1}\ln N).\label{eq:zeta_diagonal}
\end{equation}
The result can numerically be shown to be equivalent to results based
on the spectral density from Ahmadian et al. \citep{Ahmadian15_012820}.

\paragraph{Exact determinant}

Here, we compute the determinant for general matrices $\bM$, $\bSigma$
to cover the case beyond the diagonal approximation, including the
rank 1 terms.

If we assume that $m,\sigma\to\infty$ is also in the general case
the relevant solution, the determinant simplifies to 
\[
\zeta=-\frac{1}{2}\Big(1-\frac{m_{*}}{\sigma_{*}}\Big)+\frac{1}{2N}\ln\det\bigg[\varepsilon\sigma_{*}\id+\bM^{\T}\bM+\Big(1-\frac{m_{*}}{\sigma_{*}}\Big)\bSigma^{\T}\bSigma\bigg]+\cO(N^{-1}\ln N).
\]
The saddle-point equations for the relevant quantities are
\begin{eqnarray*}
\frac{m}{\sigma} & = & \frac{1}{N}\mathrm{tr}\bigg[\Big(\varepsilon\sigma\id+\bM^{\T}\bM+\big(1-\frac{m}{\sigma}\big)\bSigma^{\T}\bSigma\Big)^{-1}\bM^{\T}\bM\bigg],\\
1 & = & \frac{1}{N}\mathrm{tr}\bigg[\Big(\varepsilon\sigma\id+\bM^{\T}\bM+\big(1-\frac{m}{\sigma}\big)\bSigma^{\T}\bSigma\Big)^{-1}\bSigma^{\T}\bSigma\bigg].
\end{eqnarray*}
We can combine both equations to 
\[
\frac{m}{\sigma}+(1-\frac{m}{\sigma})=\frac{1}{N}\mathrm{tr}\bigg[\Big(\varepsilon\sigma\id+\bM^{\T}\bM+\big(1-\frac{m}{\sigma}\big)\bSigma^{\T}\bSigma\Big)^{-1}\Big(\bM^{\T}\bM+\big(1-\frac{m}{\sigma}\big)\bSigma^{\T}\bSigma\Big)\bigg]
\]
which is fulfilled for $\varepsilon\sigma=0$. For $\varepsilon\sigma=0$,
the remaining saddle-point equations are equivalent; hence, we opt
again for the second one and introduce $z=1-\frac{m}{\sigma}$ which
obeys
\begin{equation}
1=\frac{1}{N}\mathrm{tr}\bigg[\Big(\bM^{\T}\bM+z\bSigma^{\T}\bSigma\Big)^{-1}\bSigma^{\T}\bSigma\bigg].\label{eq:z_final}
\end{equation}
In terms of the solution $z_{*}$, the determinant is given by
\begin{equation}
\zeta=-\frac{1}{2}z_{*}+\frac{1}{2N}\ln\det(\bM^{\T}\bM+z_{*}\bSigma^{\T}\bSigma)+\cO(N^{-1}\ln N).\label{eq:zeta_final}
\end{equation}

\paragraph{Leading order determinant}

Equations \eqref{eq:z_final} and \eqref{eq:zeta_final} are exact.
However, they require computing the trace and the determinant of a
matrix. To capture the $N\gg1$ limit, we express the equations in
terms of network sums $\sum_{i=1}^{N}f(x_{i})$. Here, we will also
see that to leading order they correspond to the diagonal case above.

To this end, we first need an expression in terms of network sums
of
\begin{equation}
\ln\det\left(\bM^{\T}\bM+z_{\ast}\bSigma^{\T}\bSigma\right).
\end{equation}
We recall \prettyref{eq:mean_cond_jac}
\begin{align}
M_{ij}(\bx) & =-\delta_{ij}-\frac{g^{2}}{N}\frac{x_{i}\phi(x_{j})\phi^{\prime}(x_{j})}{\kappa(\bx)+D}=-\left(\id+\frac{1}{\kappa+D}\bx\bk^{\T}\right)
\end{align}
where $[\bk(\bx)]_{k}=\frac{g^{2}}{N}\phi(x_{k})\phi^{\prime}(x_{k})$,
and
\begin{equation}
\begin{aligned}\bSigma(\bx) & =\left[\id-\frac{g^{2}}{N}\frac{\phi(\bx)\phi(\bx)^{\T}}{\kappa(\bx)+D\left(1+\sqrt{1+\kappa(\bx)/D}\right)}\right]\,\boldsymbol{\Lambda}(\bx)\\
\bLambda(\bx) & =g\,\text{diag}[\phi^{\prime}(\bx)].
\end{aligned}
\end{equation}
Hence, the squares are
\begin{equation}
\bM^{\T}\bM=\id+\frac{1}{\kappa+D}\left(\bk\bx^{\T}+\bx\bk^{\T}+\frac{\bx^{\T}\bx}{\kappa+D}\bk\bk^{\T}\right),
\end{equation}
and
\begin{equation}
\bSigma^{\T}\bSigma=N\left[\bK-\frac{1}{\kappa+D}\bk\bk^{\T}\right]
\end{equation}
where $[\bK(\bx)]_{kl}=\delta_{kl}\frac{g^{2}}{N}\phi^{\prime}(x_{k})^{2}$.

Thus, we need the determinant of a matrix of the form 
\begin{equation}
\bA=\bD+\ba\bx^{\T}+\bx\ba^{\T}+\bbb\bbb^{\T}
\end{equation}
where $\bD=\id+z_{\ast}N\bK$ is diagonal, $\ba=\frac{1}{\kappa+D}\bk$,
$\bbb=\sqrt{\frac{\bx^{\T}\bx}{\left(\kappa+D\right)^{2}}-\frac{Nz_{\ast}}{\kappa+D}}\bk=\sqrt{\bx^{\T}\bx-\left(\kappa+D\right)Nz_{\ast}}\ba\equiv\sqrt{\vartheta(\bx)}\ba$.
Threefold application of the matrix determinant lemma yields
\begin{align}
\det\bA & =\left(1+\bbb^{\T}\left(\bD+\ba\bx^{\T}+\bx\ba^{\T}\right)^{-1}\bbb\right)\left(1+\bx^{\T}\left(\bD+\bx\ba^{\T}\right)^{-1}\ba\right)\left(1+\ba^{\T}\bD^{-1}\bx\right)\det\bD
\end{align}
The inverse matrices can be computed with the Sherman Morrison formula
\begin{equation}
\left(\boldsymbol{\mathcal{A}}+\boldsymbol{u}\boldsymbol{v}^{\T}\right)^{-1}=\boldsymbol{\mathcal{A}}^{-1}-\frac{\boldsymbol{\mathcal{A}}^{-1}\boldsymbol{u}\boldsymbol{v}^{\T}\boldsymbol{\mathcal{A}}^{-1}}{1+\boldsymbol{v}^{\T}\boldsymbol{\mathcal{A}}^{-1}\boldsymbol{u}}
\end{equation}
where $\boldsymbol{\mathcal{A}}$ is an invertible square matrix and
$\boldsymbol{u}$ and $\boldsymbol{v}$ are column vectors. Up to
threefold application of the Sherman Morrison formula and some simplifications
lead to
\begin{equation}
\zeta=-\frac{1}{2}z_{\ast}+\frac{1}{N}\sum_{i=1}^{N}\ln\left[1+z_{\ast}g^{2}\phi^{\prime}(x_{i})^{2}\right]+\frac{1}{N}\ln\left[(1+\beta)^{2}+\alpha(\vartheta-\gamma)\right]\label{eq:zeta_exact_nw_averg}
\end{equation}
where 
\begin{align}
\alpha(\bx) & =\frac{g^{4}}{\left[\kappa(\bx)+D\right]^{2}}\frac{1}{N^{2}}\sum_{i=1}^{N}\frac{\phi(x_{i})^{2}\phi^{\prime}(x_{i})^{2}}{1+z_{\ast}g^{2}\phi^{\prime}(x_{i})^{2}}\\
\beta(\bx) & =\frac{g^{2}}{\kappa(\bx)+D}\frac{1}{N}\sum_{i=1}^{N}\frac{x_{i}\phi(x_{i})\phi^{\prime}(x_{i})}{1+z_{\ast}g^{2}\phi^{\prime}(x_{i})^{2}}\\
\gamma(\bx) & =\sum_{i=1}^{N}\frac{x_{i}^{2}}{1+z_{\ast}g^{2}\phi^{\prime}(x_{i})^{2}}
\end{align}
Note that $\alpha=\cO(N^{-1})$, $\beta=\cO(1)$, and $\gamma,\vartheta=\cO(N)$.
Hence, to leading order, Eq. \eqref{eq:zeta_exact_nw_averg} equals
the diagonal approximation Eq. \eqref{eq:zeta_diagonal}. This correspondence,
however, only holds if the same is true for the defining Equation
of $z_{\ast}$, which we show next.

The exact definition of $z_{\ast}$ is the solution of Eq. \eqref{eq:z_final}.
Using the Sherman Morrison formula, we find
\begin{equation}
\Big(\bM^{\T}\bM+z\bSigma^{\T}\bSigma\Big)^{-1}=\bB^{-1}-\frac{\bB^{-1}\ba\ba^{\T}\bB^{-1}}{\vartheta^{-1}+\ba^{\T}\bB^{-1}\ba}
\end{equation}
where $\bB=\bD+\ba\bx^{\T}+\bx\ba^{\T}$. Hence, the right hand side
of Eq. \eqref{eq:z_final} is
\begin{align}
 & \frac{1}{N}\tr\left[\Big(\bB^{-1}-\frac{\bB^{-1}\ba\ba^{\T}\bB^{-1}}{\vartheta^{-1}+\ba^{\T}\bB^{-1}\ba}\Big)\bSigma^{\T}\bSigma\right]\\
= & \frac{1}{N}\text{tr}\left[\bB^{-1}N\left(\bK-\frac{1}{\kappa+D}\bk\bk^{\T}\right)\right]-\frac{1}{N}\frac{1}{\vartheta^{-1}+\ba^{\T}\bB^{-1}\ba}\left(\bSigma^{\T}\bSigma\bB^{-\T}\ba\right)^{\T}\bB^{-1}\ba\label{eq:sherMorrExpanded}
\end{align}
where for the second line, we used the properties of the trace. Applying
twice the Sherman-Morrison formula and identifying the fields from
above, the inverse of $\bB$ is
\begin{equation}
\bB^{-1}=\bD^{-1}-\frac{\bD^{-1}\bx\ba^{\T}\bD^{-1}}{1+\beta}-\frac{\bD^{-1}\left(\ba\bx^{\T}-\frac{\gamma}{1+\beta}\ba\ba^{\T}-\frac{\alpha}{1+\beta}\bx\bx^{\T}+\frac{\alpha\gamma}{\left(1+\beta\right)^{2}}\bx\ba^{\T}\right)\bD^{-1}}{1+\beta-\frac{\gamma\alpha}{1+\beta}}.
\end{equation}
We have a lot of contributions. The only full rank part however is
\begin{equation}
\frac{1}{N}\text{tr}\left(\bD^{-1}N\bK\right)=\frac{1}{N}\sum_{i=1}^{N}\frac{g^{2}\phi^{\prime}(x_{i})^{2}}{1+zg^{2}\phi^{\prime}(x_{i})^{2}}=\cO(1).\label{eq:def_of_z_leading_order}
\end{equation}
The other parts (rank 1 parts in the trace and scalar products to
the right of \prettyref{eq:sherMorrExpanded}) yield scalar products
that are all $\cO(N^{-1})$. To see this, we count each inner product
that arises as a factor $N$, track the explicit appearances of $N$
and use the orders known for $\alpha$, $\beta$, and $\gamma$.
Hence, the equation defining $z_{\ast}$, \prettyref{eq:z_final},
is to leading order equivalent to \prettyref{eq:def_of_z_diagonal}.

Lastly, we have to check that a $\cO(N^{-1})$ correction to Eq. \eqref{eq:def_of_z_diagonal}
only yields a $\cO(N^{-1})$ correction to $z_{\ast}$, and furthermore
to $\zeta$ in Eq. \eqref{eq:zeta_diagonal}. To this end, assume
that $z_{\ast}^{0}$ solves the approximate Eq. \eqref{eq:def_of_z_diagonal}.
Adding a correction term of $\cO(N^{-1})$ to the approximate defining
\prettyref{eq:def_of_z_diagonal} leads to a shift $z_{\ast}^{0}\rightarrow z_{\ast}^{0}+\delta$,
where $\delta$ is the solution of
\begin{align}
1 & =\frac{1}{N}\sum_{i=1}^{N}\frac{\lambda_{i}^{2}}{1+\left(z_{\ast}^{0}+\delta\right)\lambda_{i}^{2}}+\cO(N^{-1})\\
\Leftrightarrow\frac{1}{N}\sum_{i=1}^{N}\frac{\lambda_{i}^{4}}{\left(1+z_{\ast}^{0}\lambda_{i}^{2}\right)^{2}}\delta & =\cO(N^{-1})+\cO(\delta^{2})
\end{align}
where we Taylor expanded around $\delta=0$. Since the prefactor in
front of $\delta$ is $\cO(1)$, we find that $\delta=\cO(N^{-1})$
is self-consistent. The change to $\zeta$ is then
\begin{align}
\zeta & =-\frac{1}{2}\left(z_{\ast}^{0}+\delta\right)+\frac{1}{2N}\sum_{i=1}^{N}\ln\left[1+\left(z_{\ast}^{0}+\delta\right)\lambda_{i}^{2}\right]\\
 & =-\frac{1}{2}z_{\ast}^{0}+\frac{1}{2N}\sum_{i=1}^{N}\ln(1+z_{*}\lambda_{i}^{2})+\cO(\delta).
\end{align}
The leading order contribution to the scaled log determinant is hence
indeed Eq. \eqref{eq:zeta_diagonal} with $z_{\ast}$ being the solution
of Eq. \eqref{eq:def_of_z_diagonal}.

\paragraph{Replica calculation}

We use the replica formalism for random matrix theory \citep{Edwards76_1595,Livan2018}
to justify Eq.~\eqref{eq:logdet}. The starting point is
\begin{equation}
\langle|\det(\bM+\bX\bSigma)|\rangle=\lim_{n\to-1}\Big\langle\frac{1}{|\det(\bM+\bX\bSigma)|^{n}}\Big\rangle
\end{equation}
Again we use a Gaussian integral with a complex $N\times n$ matrix
$\bZ$ to rewrite the absolute value of the determinant
\begin{equation}
\Big\langle\frac{1}{|\det(\bM+\bX\bSigma)|^{n}}\Big\rangle=\frac{1}{(2\pi)^{nN/2}}\int d\bZ\,\Big\langle\exp\Big[-\frac{\varepsilon}{2}\tr\bZ^{\T}\bZ-\frac{1}{2}\tr\bZ^{\T}(\bM+\bX\bSigma)^{\T}(\bM+\bX\bSigma)\bZ\Big]\Big\rangle
\end{equation}
where $\varepsilon\to0+$ avoids divergences. Solving the Gaussian
expectation over $\bX$ leads to
\begin{align}
\Big\langle\frac{1}{|\det(\bM+\bX\bSigma)|^{n}}\Big\rangle=\frac{1}{(2\pi)^{nN/2}}\int d\bZ\,\exp\Big[ & -\frac{\varepsilon}{2}\tr\bZ^{\T}\bZ-\frac{1}{2}\tr(\id+\frac{1}{N}\bZ^{\T}\bSigma^{\T}\bSigma\bZ)^{-1}\bZ^{\T}\bM^{\T}\bM\bZ\nonumber \\
 & -\frac{N}{2}\ln\det(\id+\frac{1}{N}\bZ^{\T}\bSigma^{\T}\bSigma\bZ)\Big]
\end{align}
akin to \eqref{eq:zeta_expc_solved}. We introduce the $n\times n$
dimensional order parameters
\begin{equation}
\bem=\frac{1}{N}\bZ^{\T}\bM^{\T}\bM\bZ,\qquad\bsigma=\frac{1}{N}\bZ^{\T}\bSigma^{\T}\bSigma\bZ,
\end{equation}
leading to
\begin{align}
\Big\langle\frac{1}{|\det(\bM+\bX\bSigma)|^{n}}\Big\rangle= & \int d\bem d\bsigma d\tilde{\bem}d\tilde{\bsigma}\,\exp\Big[-\frac{N}{2}\tr(\id+\bsigma)^{-1}\bem-\frac{N}{2}\ln\det(\id+\bsigma)-iN\tr[\tilde{\bem}^{\T}\bem+\tilde{\bsigma}^{\T}\bsigma]\Big]\nonumber \\
 & \times\frac{1}{(2\pi)^{nN/2}}\int d\bZ\,\exp\Big[-\frac{\varepsilon}{2}\tr\bZ^{\T}\bZ+i\tr\tilde{\bem}\bZ^{\T}\bM^{\T}\bM\bZ+i\tr\tilde{\bsigma}\bZ^{\T}\bSigma^{\T}\bSigma\bZ\Big].
\end{align}
We rewrite the exponent of the $\bZ$ integral in terms of Kronecker
products (see, e.g., \citep{Gupta1999}),
\begin{equation}
-\frac{\varepsilon}{2}\tr\bZ^{\T}\bZ+i\tr\tilde{\bem}\bZ^{\T}\bM^{\T}\bM\bZ+i\tr\tilde{\bsigma}\bZ^{\T}\bSigma^{\T}\bSigma\bZ=-\frac{1}{2}\vectorized(\bZ)^{\T}[\varepsilon(\id\otimes\id)-2i(\tilde{\bem}\otimes\bM^{\T}\bM)-2i(\tilde{\bsigma}\otimes\bSigma^{\T}\bSigma)]\vectorized(\bZ),
\end{equation}
where $\vectorized(\bZ)$ denotes the vectorization of the matrix
$\bZ$, such that the integral is a $nN$-dimensional Gaussian:
\begin{equation}
\frac{1}{(2\pi)^{nN/2}}\int d\bZ\,e^{-\frac{1}{2}\vectorized(\bZ)^{\T}[\varepsilon(\id\otimes\id)-2i(\tilde{\bem}\otimes\bM^{\T}\bM)-2i(\tilde{\bsigma}\otimes\bSigma^{\T}\bSigma)]\vectorized(\bZ)}=\frac{1}{\sqrt{\det[\varepsilon(\id\otimes\id)-2i(\tilde{\bem}\otimes\bM^{\T}\bM)-2i(\tilde{\bsigma}\otimes\bSigma^{\T}\bSigma)]}}
\end{equation}
In summary, we arrive at
\begin{align}
\Big\langle\frac{1}{|\det(\bM+\bX\bSigma)|^{n}}\Big\rangle=\int d\bem d\bsigma d\tilde{\bem}d\tilde{\bsigma}\,\exp\Big[ & -\frac{N}{2}\tr(\id+\bsigma)^{-1}\bem-\frac{N}{2}\ln\det(\id+\bsigma)-iN\tr[\tilde{\bem}^{\T}\bem+\tilde{\bsigma}^{\T}\bsigma]\nonumber \\
 & -\frac{1}{2}\ln\det[\varepsilon(\id\otimes\id)-2i(\tilde{\bem}\otimes\bM^{\T}\bM)-2i(\tilde{\bsigma}\otimes\bSigma^{\T}\bSigma)]\Big].
\end{align}
We proceed with a replica-symmetric (RS) Ansatz for the order parameters.

The RS Ansatz is
\begin{equation}
\bem=(m_{0}-m)\id+m\bone\bone^{\T},\qquad\tilde{\bem}=(\tilde{m}_{0}-\tilde{m})\id+\tilde{m}\bone\bone^{\T},\qquad\bsigma=(\sigma_{0}-\sigma)\id+\sigma\bone\bone^{\T},\qquad\tilde{\bsigma}=(\tilde{\sigma}_{0}-\tilde{\sigma})\id+\tilde{\sigma}\bone\bone^{\T},
\end{equation}
where $\bone$ denotes the vector containing only ones. The RS Ansatz
simplifies the terms in the action:
\begin{align}
\tr(\id+\bsigma)^{-1}\bem & =\frac{1}{1+\sigma_{0}-\sigma}\Big(nm_{0}-\frac{m_{0}+(n-1)m}{1+\sigma_{0}+(n-1)\sigma}\sigma n\Big),
\end{align}
\begin{equation}
\ln\det(\id+\bsigma)=\ln(1+\frac{\sigma}{1+\sigma_{0}-\sigma}n)+n\ln(1+\sigma_{0}-\sigma),
\end{equation}
\begin{equation}
\tr\tilde{\bem}^{\T}\bem=\tilde{m}_{0}m_{0}n+\tilde{m}m(n-1)n,\qquad\tr\tilde{\bsigma}^{\T}\bsigma=\tilde{\sigma}_{0}\sigma_{0}n+\tilde{\sigma}\sigma(n-1)n,
\end{equation}
\begin{equation}
\varepsilon(\id\otimes\id)-2i(\tilde{\bem}\otimes\bM^{\T}\bM)-2i(\tilde{\bsigma}\otimes\bSigma^{\T}\bSigma)=\id\otimes[\varepsilon\id-2i(\tilde{m}_{0}-\tilde{m})\bM^{\T}\bM-2i(\tilde{\sigma}_{0}-\tilde{\sigma})\bSigma^{\T}\bSigma]-2i\bone\bone^{\T}\otimes[\tilde{m}\bM^{\T}\bM+\tilde{\sigma}\bSigma^{\T}\bSigma].
\end{equation}
The corresponding saddle-point equations are
\begin{equation}
2i\tilde{m}_{0}=-\frac{1}{1+\sigma_{0}-\sigma}\frac{1+\sigma_{0}+(n-2)\sigma}{1+\sigma_{0}+(n-1)\sigma},\label{eq:replicaSPE_m0}
\end{equation}
\begin{equation}
2i\tilde{m}=\frac{1}{1+\sigma_{0}-\sigma}\frac{\sigma}{1+\sigma_{0}+(n-1)\sigma},\label{eq:replicaSPE_m}
\end{equation}
\begin{equation}
2i\tilde{\sigma}_{0}(1+\sigma_{0}-\sigma)=\frac{1}{1+\sigma_{0}-\sigma}\Big(m_{0}-\frac{m_{0}+(n-1)m}{1+\sigma_{0}+(n-1)\sigma}\sigma\Big)-\frac{m_{0}+(n-1)m}{(1+\sigma_{0}+(n-1)\sigma)^{2}}\sigma+\frac{\sigma}{1+\sigma_{0}+(n-1)\sigma}-1,\label{eq:replicaSPE_sig0}
\end{equation}
\begin{align}
2i\tilde{\sigma}(n-1)(1+\sigma_{0}-\sigma)= & -\frac{1}{1+\sigma_{0}-\sigma}\Big(m_{0}-\frac{m_{0}+(n-1)m}{1+\sigma_{0}+(n-1)\sigma}\sigma\Big)+\frac{m_{0}+(n-1)m}{1+\sigma_{0}+(n-1)\sigma}-\frac{m_{0}+(n-1)m}{(1+\sigma_{0}+(n-1)\sigma)^{2}}\sigma(n-1)\nonumber \\
 & -\frac{1+\sigma_{0}-\sigma}{1+\sigma_{0}+(n-1)\sigma}-\frac{\sigma}{1+\sigma_{0}+(n-1)\sigma}+1,\label{eq:replicaSPE_sig}
\end{align}
\begin{equation}
m_{0}=\frac{1}{Nn}\tr\Big[\big(\id\otimes[\varepsilon\id-2i(\tilde{m}_{0}-\tilde{m})\bM^{\T}\bM-2i(\tilde{\sigma}_{0}-\tilde{\sigma})\bSigma^{\T}\bSigma]-2i\bone\bone^{\T}\otimes[\tilde{m}\bM^{\T}\bM+\tilde{\sigma}\bSigma^{\T}\bSigma]\big)^{-1}\big(\id\otimes\bM^{\T}\bM\big)\Big],\label{eq:replicaSPE_tm0}
\end{equation}
\begin{equation}
m=-\frac{1}{Nn(n-1)}\mathrm{tr}\Big[\big(\id\otimes[\varepsilon\id-2i(\tilde{m}_{0}-\tilde{m})\bM^{\T}\bM-2i(\tilde{\sigma}_{0}-\tilde{\sigma})\bSigma^{\T}\bSigma]-2i\bone\bone^{\T}\otimes[\tilde{m}\bM^{\T}\bM+\tilde{\sigma}\bSigma^{\T}\bSigma]\big)^{-1}\big([\id-\bone\bone^{\T}]\otimes\bM^{\T}\bM\big)\Big],\label{eq:replicaSPE_tm}
\end{equation}
\begin{equation}
\sigma_{0}=\frac{1}{Nn}\tr\Big[\big(\id\otimes[\varepsilon\id-2i(\tilde{m}_{0}-\tilde{m})\bM^{\T}\bM-2i(\tilde{\sigma}_{0}-\tilde{\sigma})\bSigma^{\T}\bSigma]-2i\bone\bone^{\T}\otimes[\tilde{m}\bM^{\T}\bM+\tilde{\sigma}\bSigma^{\T}\bSigma]\big)^{-1}\big(\id\otimes\bSigma^{\T}\bSigma\big)\Big],\label{eq:replicaSPE_tsig0}
\end{equation}
\begin{equation}
\sigma=-\frac{1}{N(n-1)n}\tr\Big[\big(\id\otimes[\varepsilon\id-2i(\tilde{m}_{0}-\tilde{m})\bM^{\T}\bM-2i(\tilde{\sigma}_{0}-\tilde{\sigma})\bSigma^{\T}\bSigma]-2i\bone\bone^{\T}\otimes[\tilde{m}\bM^{\T}\bM+\tilde{\sigma}\bSigma^{\T}\bSigma]\big)^{-1}\big([\id-\bone\bone^{\T}]\otimes\bSigma^{\T}\bSigma\big)\Big].\label{eq:replicaSPE_tsig}
\end{equation}
The saddle-point equations admit a solution with decoupled replicas,
$m=\tilde{m}=\sigma=\tilde{\sigma}=0$. In this case Eqs.~\eqref{eq:replicaSPE_m}
and \eqref{eq:replicaSPE_sig} are trivially fulfilled; for Eqs.~\eqref{eq:replicaSPE_tm}
and \eqref{eq:replicaSPE_tsig} we obtain
\begin{equation}
\frac{1}{nN}\mathrm{tr}\Big[\big(\id\otimes[\varepsilon\id-2i\tilde{m}_{0}\bM^{\T}\bM-2i\tilde{\sigma}_{0}\bSigma^{\T}\bSigma]^{-1}\big)\big([\id-\bone\bone^{\T}]\otimes\bM^{\T}\bM\big)\Big]=\frac{1}{n}\mathrm{tr}[\id-\bone\bone^{\T}]m_{0}=0,
\end{equation}
\begin{equation}
\frac{1}{nN}\tr\Big[\big(\id\otimes[\varepsilon\id-2i\tilde{m}_{0}\bM^{\T}\bM-2i\tilde{\sigma}_{0}\bSigma^{\T}\bSigma]^{-1}\big)\big([\id-\bone\bone^{\T}]\otimes\bSigma^{\T}\bSigma\big)\Big]=\frac{1}{n}\mathrm{tr}[\id-\bone\bone^{\T}]\sigma_{0}=0,
\end{equation}
i.e., the decoupled solution $m=\tilde{m}=\sigma=\tilde{\sigma}=0$
is indeed a self-consistent solution. The remaining saddle-point equations
reduce to the familiar saddle-point equations \eqref{eq:SPE_tilde}
and \eqref{eq:SPE_trace},
\begin{equation}
2i\tilde{m}_{0}=-\frac{1}{1+\sigma_{0}},\qquad2i\tilde{\sigma}_{0}=\frac{m_{0}}{(1+\sigma_{0})^{2}}-\frac{1}{1+\sigma_{0}},
\end{equation}
\begin{equation}
m_{0}=\frac{1}{N}\tr\Big[[\varepsilon\id-2i\tilde{m}_{0}\bM^{\T}\bM-2i\tilde{\sigma}_{0}\bSigma^{\T}\bSigma]^{-1}\bM^{\T}\bM\Big],\qquad\sigma_{0}=\frac{1}{N}\tr\Big[[\varepsilon\id-2i(\tilde{m}_{0}-\tilde{m})\bM^{\T}\bM-2i(\tilde{\sigma}_{0}-\tilde{\sigma})\bSigma^{\T}\bSigma]^{-1}\bSigma^{\T}\bSigma\Big].
\end{equation}
Furthermore the terms in the action simplify to
\begin{align}
\tr(\id+\bsigma)^{-1}\bem & =\frac{nm_{0}}{1+\sigma_{0}},\qquad\ln\det(\id+\bsigma)=n\ln(1+\sigma_{0})
\end{align}
\begin{equation}
\tr\tilde{\bem}^{\T}\bem=\tilde{m}_{0}m_{0}n,\qquad\tr\tilde{\bsigma}^{\T}\bsigma=\tilde{\sigma}_{0}\sigma_{0}n
\end{equation}
\begin{equation}
\ln\det[\varepsilon(\id\otimes\id)-2i(\tilde{\bem}\otimes\bM^{\T}\bM)-2i(\tilde{\bsigma}\otimes\bSigma^{\T}\bSigma)]=n\ln\det[\varepsilon\id-2i\tilde{m}_{0}\bM^{\T}\bM-2i\tilde{\sigma}_{0}\bSigma^{\T}\bSigma]
\end{equation}
In summary, we obtain
\begin{equation}
\Big\langle\frac{1}{|\det(\bM+\bX\bSigma)|^{n}}\Big\rangle=\exp\Big[-\frac{nN}{2}\frac{m_{0}}{1+\sigma_{0}}-\frac{nN}{2}\ln(1+\sigma_{0})-inN[\tilde{m}_{0}m_{0}+\tilde{\sigma}_{0}\sigma_{0}]-\frac{n}{2}\ln\det[\varepsilon\id-2i\tilde{m}_{0}\bM^{\T}\bM-2i\tilde{\sigma}_{0}\bSigma^{\T}\bSigma]\Big];
\end{equation}
taking the limit $n\to-1$ we arrive at
\begin{equation}
\zeta=\frac{1}{2}\frac{m_{0}}{1+\sigma_{0}}+\frac{1}{2}\ln(1+\sigma_{0})+i[\tilde{m}_{0}m_{0}+\tilde{\sigma}_{0}\sigma_{0}]+\frac{1}{2N}\ln\det[\varepsilon\id-2i\tilde{m}_{0}\bM^{\T}\bM-2i\tilde{\sigma}_{0}\bSigma^{\T}\bSigma]
\end{equation}
which recovers Eq.~\eqref{eq:def_H} and thus Eq.~\eqref{eq:zeta_full}.

To summarize the progress so far, we found an expression for the fixed
point density following \eqref{eq:fpd_exact},
\begin{equation}
\rho(\bx)=\frac{1}{\sqrt{2\pi\left[\kappa(\bx)+D\right]}^{N}}e^{-\frac{q(\bx)}{2\left[\kappa(\bx)+D\right]}+N\zeta(\bx)},\label{eq:fpd_asinmain}
\end{equation}
where $\kappa(\bx)=\frac{g^{2}}{N}\sum_{i=1}^{N}\phi(x_{i})^{2}$,
$q(\bx)=\frac{1}{N}\sum_{i=1}^{N}x_{i}^{2}$, and $\zeta(\bx)$ is
to leading order given by Eqs. \eqref{eq:zeta_diagonal} and \eqref{eq:def_of_z_diagonal}.

\subsection{Empirical measure}

So far we derived the distribution of fixed points $\rho(\bx)$. In
this Section we discuss the \emph{empirical measure} 
\begin{equation}
\mu_{\bx}(y)=\frac{1}{N}\sum_{i=1}^{N}\delta(y-x_{i}).
\end{equation}
It is a function of $y$ that is parameterized by the position $\bx$.
It contains all vector elements of $\bx$ but forgets about their
order. We want to understand the distribution of empirical measures
when $\bx\sim\rho(\bx)/\int d\bz\rho(\bz)$. We will see that this
distribution (of distributions) is strongly peaked at the expected
empirical measure
\begin{equation}
\mu_{\ast}(y)=\left\langle \mu_{\bx}(y)\right\rangle _{\bx\sim\rho(\bx)}
\end{equation}
where $\left\langle \circ\right\rangle _{\bx\sim\rho(\bx)}$ denotes
an average with respect to $\rho(\bx)/\left(\int d\bz\,\rho(\bz)\right)$.
From the distribution of the empirical measure, the distribution of
certain network-averaged quantities can be nicely characterized. Particularly,
the expectation of a network average is given by the expected empirical
measure
\begin{equation}
\left\langle \frac{1}{N}\sum_{i=1}^{N}f(x_{i})\right\rangle _{\bx\sim\rho(\bx)}=\int dy\,\mu_{\ast}(y)f(y),
\end{equation}
which can be seen by plugging in the definition. Similarly, we can
also express $\kappa(\bx)$ as a functional of $\mu_{\bx}$
\begin{equation}
\kappa(\bx)=\frac{g^{2}}{N}\bphi^{\T}\bphi=g^{2}\int dy\,\mu_{\bx}(y)\phi(y)^{2}=\kappa[\mu_{\bx}].
\end{equation}
Furthermore, we can express the log determinant $\zeta(\bx)$ by $\mu_{\bx}$
\begin{align}
\zeta(\bx) & =-\frac{1}{2}z_{*}(\bx)+\frac{1}{2N}\sum_{i=1}^{N}\ln(1+z_{*}(\bx)g^{2}\phi^{\prime}(x_{i})^{2})\\
 & =-\frac{1}{2}z_{*}[\mu_{\bx}]+\frac{1}{2}\int dy\,\mu_{\bx}(y)\ln(1+z_{*}[\mu_{\bx}]g^{2}\phi^{\prime}(y)^{2})\label{eq:zeta_mu}\\
 & =\zeta[\mu_{\bx}]
\end{align}
where $z_{\ast}[\mu_{\bx}]$ is the solution of
\begin{equation}
1=\frac{1}{N}\sum_{i=1}^{N}\frac{g^{2}\phi^{\prime}(x_{i})^{2}}{1+z_{\ast}\,g^{2}\phi^{\prime}(x_{i})^{2}}=\int dy\,\mu_{\bx}(y)\frac{g^{2}\phi^{\prime}(y)^{2}}{1+z_{\ast}\,g^{2}\phi^{\prime}(y)^{2}}.\label{eq:def_z_ast_mu}
\end{equation}
Summing up, the complete fixed point distribution can be expressed
in terms of the empirical measure $\mu_{\bx}$
\begin{equation}
\rho(\bx)=\rho[\mu_{\bx}].
\end{equation}

\subsubsection{Expected empirical measure\label{subsec:Most-likely-empirical}}

In this Section, we compute the expected empirical measure $\mu_{\ast}(y)=\left\langle \mu_{\bx}(y)\right\rangle _{\bx\sim\rho(\bx)}$.
To this end, we follow the method presented in \citep{vanMeegen21_158302}
and \citep{Touchette09}. We consider the characteristic functional
\begin{equation}
Z[j]=\left\langle e^{ij^{\T}\mu_{\bx}}\right\rangle _{\bx\sim\rho(\bx)}=\left\langle e^{i\frac{1}{N}\sum_{i=1}^{N}j(x_{i})}\right\rangle _{\bx\sim\rho(\bx)}
\end{equation}
where $j(y)$ is an auxiliary external source field and $j^{\T}\mu_{\bx}=\int dy\,j(y)\mu_{\bx}(y)$
denotes a functional scalar product (this notation will be implicit
in the following). By $\left\langle \circ\right\rangle _{\bx\sim\rho(\bx)}$
we mean average with respect to $\rho(\bx)/\left(\int d\bz\,\rho(\bz)\right)$.
We define the scaled cumulant generating functional as $W_{N}[j]=\frac{1}{N}\ln Z[Nj]$.
The expected empirical measure is the first Taylor coefficient of
$W_{N}$
\begin{equation}
\mu_{\ast}(y)=\left.\frac{\delta}{\delta j(y)}W_{N}[j]\right|_{j(y)=0}.\label{eq:em_first_moment}
\end{equation}
Plugging in the explicit result for the fixed point distribution,
we have
\begin{align}
W_{N}[j] & =\frac{1}{N}\ln\int d\bx\,\frac{1}{\left[2\pi\left(\kappa[\mu_{\bx}]+D\right)\right]^{N/2}}e^{S[j;\bx]}-c,\label{eq:scgf}\\
S[j;\bx] & =-\frac{\bx^{\T}\bx}{2\left(\kappa[\mu_{\bx}]+D\right)}+N\zeta[\mu_{\bx}]+i\sum_{i}j(x_{i}),\\
c & =\frac{1}{N}\ln\int d\bx\rho(\bx)=\frac{1}{N}\ln N_{\text{fp}},\label{eq:def_topcomp}
\end{align}
where the expected number of fixed points $N_{\text{fp}}$ is the
norm of the fixed point distribution $\rho(\bx)$. The rate $c$ is
known as the \emph{topological complexity} \citep{wainrib13_118101}.

We want to evaluate the integral in Eq. \eqref{eq:scgf}. To this
end, we introduce an auxiliary field $\mu(y)$ that we use to replace
the $\bx$-dependent field $\mu_{\bx}(y)$. To ensure that this is
still correct, we multiply the integrand by the functional Dirac constraint
$\delta[\mu-\mu_{\bx}]\equiv\lim_{M\rightarrow\infty}\prod_{i=1}^{M}\delta\left[\mu(y_{i})-\mu_{\bx}(y_{i})\right]$
where $\left\{ y_{1},...,y_{M}\right\} \overset{M\rightarrow\infty}{\rightarrow}\reals$
is a discretization of the real line. Then, we have to integrate over
$\mu(y)$ in a functional sense $\int\cD\mu\equiv\lim_{M\rightarrow\infty}\int_{-\infty}^{\infty}\prod_{i=1}^{M}d\mu(y_{i})$
for the replacement to be correct at every point in the $\bx$ integration.
Lastly, we replace the functional Dirac constraint by its Fourier
integral representation

\begin{equation}
\begin{aligned}\delta\left[\mu-\mu_{\bx}\right] & =\int\cD\tilde{\mu}\,e^{-iN\tilde{\mu}^{\T}\left(\mu-\mu_{\bx}\right)}\\
 & =\int\cD\tilde{\mu}\,e^{-iN\tilde{\mu}^{\T}\mu+i\sum_{i=1}^{N}\tilde{\mu}(x_{i})},
\end{aligned}
\end{equation}
where 
\begin{equation}
\int\cD\tilde{\mu}\equiv N\lim_{M\rightarrow\infty}\int_{-\infty}^{\infty}\prod_{i=1}^{M}\frac{d\tilde{\mu}(y_{i})}{2\pi}.\label{eq:def_Dmut}
\end{equation}
 With the auxiliary fields $\mu$ and $\tilde{\mu}$, the $\bx$-integral
in Eq. \eqref{eq:scgf} formally factorizes
\begin{align}
W_{N}[j] & =\frac{1}{N}\ln\int\cD\mu\cD\tilde{\mu}\,e^{-iN\tilde{\mu}^{\T}\mu+N\ln\Omega[\mu,\tilde{\mu},j]}-c\label{eq:scgf_pre_spa}\\
\Omega[\mu,\tilde{\mu},j] & =\int\frac{dx}{\sqrt{2\pi\left(\kappa[\mu]+D\right)}}e^{-\frac{x^{2}}{2\left(\kappa[\mu]+D\right)}+\zeta[\mu]+ij(x)+i\tilde{\mu}(x)}.
\end{align}
Note that this factorization into identical integrals $\int dx$ is
only formal: The integrals are still coupled through their common
dependence on the fields $\mu$ and $\tilde{\mu}$.

\paragraph{Saddle point approximation}

The exponent of the integrand in Eq. \eqref{eq:scgf_pre_spa} is proportional
to $N$. We are interested in large networks where $N\gg1$. In this
regime we can perform a saddle point approximation which yields to
leading order in $N^{-1}$
\begin{equation}
W_{N}[j]=-i\tilde{\mu}_{\ast}[j]^{\T}\mu_{\ast}[j]+\ln\Omega\left\{ \mu_{\ast}[j],\tilde{\mu}_{\ast}[j],j\right\} -c\label{eq:scgf_spa}
\end{equation}
where

\begin{equation}
\begin{aligned}\mu_{\ast}[j](y) & =\left.\frac{\delta\ln\Omega}{\delta\,i\tilde{\mu}(y)}\right|_{\mu_{\ast}[j],\tilde{\mu}_{\ast}[j]},\quad i\tilde{\mu}_{\ast}[j](y)=\left.\frac{\delta\ln\Omega}{\delta\,\mu(y)}\right|_{\mu_{\ast}[j],\tilde{\mu}_{\ast}[j]}\end{aligned}
\label{eq:spes_abstract}
\end{equation}
are the maxima of the exponent of the integrand in Eq. \eqref{eq:scgf_pre_spa}.
With this, we can compute the expectation value of $\mu_{\bx}$ in
saddle point approximation: Evaluating Eq. \eqref{eq:em_first_moment}
gives 
\begin{equation}
\begin{aligned}\mu_{\ast}(y) & =\left.\frac{\delta\ln\Omega\left\{ \mu_{\ast}[0],\tilde{\mu}_{\ast}[0],j\right\} }{\delta\,j(y)}\right|_{j=0}\\
 & =\mu_{\ast}[0](y)
\end{aligned}
\label{eq:expect_of_em}
\end{equation}
where we used Eqs. \eqref{eq:spes_abstract} in the first step to
eliminate the chain-rule derivatives and in the second step to identify
the result with $\mu_{\ast}[0]$. Therefore, the derivative in Eq.
\eqref{eq:expect_of_em} only acts on the explicit dependency of $\Omega$
on $j$.

\paragraph{Saddle point equations}

Next, to get the expected empirical measure, we discuss the solution
of the saddle point Equations \eqref{eq:spes_abstract} for $j=0$.

The first saddle point Equation follows from straight forward differentiation
\begin{equation}
\mu_{\ast}(y)=\frac{e^{-\frac{y^{2}}{2\left(\kappa\left[\mu_{\ast}\right]+D\right)}+\zeta\left[\mu_{\ast}\right]+i\tilde{\mu}_{\ast}(y)}}{\sqrt{2\pi\left(\kappa\left[\mu_{\ast}\right]+D\right)}\Omega\left[\mu_{\ast},\tilde{\mu}_{\ast}\right]}.\label{eq:spe_mu}
\end{equation}
The second saddle point Equation
\begin{equation}
i\tilde{\mu}_{\ast}(y)=\left.\frac{1}{\Omega}\left(\frac{\partial\Omega}{\partial\kappa}\,\frac{\delta\kappa[\mu]}{\delta\mu(y)}+\frac{\partial\Omega}{\partial\zeta}\,\frac{\delta\zeta[\mu]}{\delta\mu(y)}\right)\right|_{\mu_{\ast},\tilde{\mu}_{\ast}}
\end{equation}
involves some chain rule derivatives. We need the derivative of $\kappa$,
\begin{equation}
\frac{\delta\kappa[\mu]}{\delta\mu(y)}=g^{2}\phi(y)^{2},
\end{equation}
and we need the derivative of $\zeta[\mu]$ as in Eq. \eqref{eq:zeta_mu}.
We find
\begin{equation}
\frac{\delta\zeta[\mu]}{\delta\mu(y)}=-\frac{1}{2}\frac{\delta z_{\ast}[\mu]}{\delta\mu(y)}\left(1-\int dx\,\mu(x)\frac{g^{2}\phi^{\prime}(x)^{2}}{1+z_{*}[\mu]g^{2}\phi^{\prime}(x)^{2}}\right)+\frac{1}{2}\ln(1+z_{*}[\mu]g^{2}\phi^{\prime}(y)^{2})
\end{equation}
where the first part vanishes due to the definition of $z_{\ast}$,
see Eq. \eqref{eq:def_z_ast_mu}.

\begin{figure}
\includegraphics{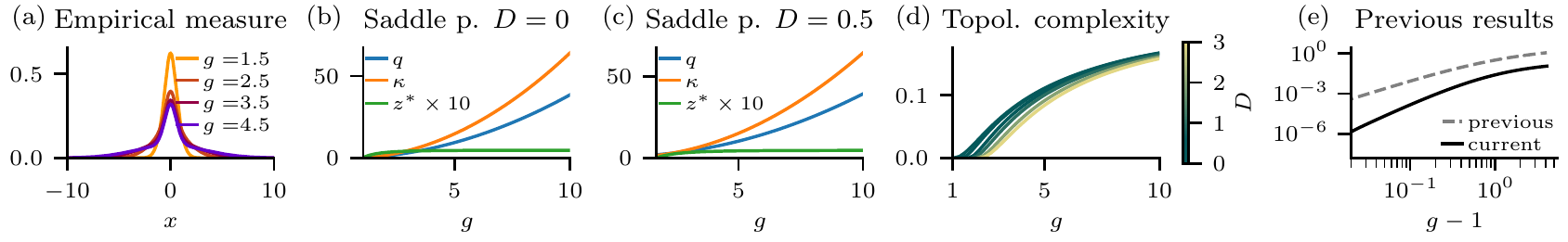}

\caption{\label{fig:Sol-SPA}Solution of the saddle point approximation. (a)
Expected empirical measure \prettyref{eq:EM_compact} for $D=10^{-3}$.
(b) Saddle point fields \prettyref{eq:sp_eq_kappa} for $D=0$. (c)
Same as (a), but $D=0.5$. (d) Topological complexity \prettyref{eq:topcomp_spa}
for several values of $g$ and $D$. The transition to a positive
complexity shifts for increasing noise strength. (e) Topological complexity
without noise $D=0$ on a $\log$-$\log$-scale following the saddle
point equations presented here (black solid line) and the previous
result presented in Ref. \citep{wainrib13_118101}, and numerical
results for the topological complexity (blue dots).}

\end{figure}

Concluding, the saddle point equation for $\tilde{\mu}$ is
\begin{equation}
i\tilde{\mu}_{\ast}(y)=\frac{g^{2}\phi(y)^{2}}{2(\kappa[\mu_{\ast}]+D)}\left(\frac{\left\langle x^{2}\right\rangle _{\mu_{\ast}}}{\kappa[\mu_{\ast}]+D}-1\right)+\frac{1}{2}\ln(1+z_{*}[\mu_{\ast}]g^{2}\phi^{\prime}(y)^{2}).\label{eq:spe_mut}
\end{equation}
The expected empirical measure is determined by the simultaneous solution
of Eqs. \eqref{eq:spe_mu} and \eqref{eq:spe_mut}. Combining them,
we find
\begin{equation}
\mu_{\ast}(y)=\mathcal{\mathcal{Z}}^{-1}\sqrt{1+\alpha\phi^{\prime}(y)^{2}}e^{-\frac{y^{2}+\gamma\phi(y)^{2}}{2\beta}}\label{eq:EM_compact}
\end{equation}
where 
\begin{align}
\alpha & =z_{*}[\mu_{\ast}]g^{2},\quad\beta=\kappa[\mu_{\ast}]+D,\quad\gamma=g^{2}\left(1-\frac{q[\mu_{\ast}]}{\beta}\right)\label{eq:SP_eqs_compact}
\end{align}
and $\mathcal{Z}$ normalizes the empirical measure. Here, $q[\mu_{\ast}]\equiv\left\langle y^{2}\right\rangle _{y\sim\mu_{\ast}}$.
The parametrized form of the empirical measure in Eq. \eqref{eq:EM_compact}
allows to formulate the saddle point approximation in terms of the
scalars $z_{\ast}$, $\kappa$ and $q$. The respective set of equations
can be formulated in terms of standard Gauss integrals 
\begin{align}
q & =\sqrt{2\pi\beta(\kappa)}\mathcal{Z}^{-1}\left\langle \beta x^{2}\sqrt{1+\alpha(z_{\ast})\phi^{\prime}(\sqrt{\beta(\kappa)}x)^{2}}e^{-\frac{\gamma(\kappa,q)}{2\beta(\kappa)}\phi(\sqrt{\beta(\kappa)}x)^{2}}\right\rangle _{x\sim\cN(0,1)}\label{eq:sp_eq_q}\\
\kappa & =g^{2}\sqrt{2\pi\beta(\kappa)}\mathcal{Z}^{-1}\left\langle \phi(\sqrt{\beta(\kappa)}x)^{2}\sqrt{1+\alpha(z_{\ast})\phi^{\prime}(\sqrt{\beta(\kappa)}x)^{2}}e^{-\frac{\gamma(\kappa,q)}{2\beta(\kappa)}\phi(\sqrt{\beta(\kappa)}x)^{2}}\right\rangle _{x\sim\cN(0,1)}\label{eq:sp_eq_kappa}\\
1 & =g^{2}\sqrt{2\pi\beta(\kappa)}\mathcal{Z}^{-1}\left\langle \frac{\phi^{\prime}(\sqrt{\beta(\kappa)}x)^{2}}{\sqrt{1+\alpha(z_{\ast})\phi^{\prime}(\sqrt{\beta(\kappa)}x)^{2}}}e^{-\frac{\gamma(\kappa,q)}{2\beta(\kappa)}\phi(\sqrt{\beta(\kappa)}x)^{2}}\right\rangle _{x\sim\cN(0,1)}\label{eq:sp_eq_zast}
\end{align}
where the norm $\mathcal{Z}$ in Eq. \eqref{eq:EM_compact} can be
written as
\begin{align}
\mathcal{Z} & =\int dy\,\sqrt{1+\alpha\phi^{\prime}(y)^{2}}e^{-\frac{y^{2}+\gamma\phi(y)^{2}}{2\beta}}\\
 & =\sqrt{2\pi\beta(\kappa)}\left\langle \sqrt{1+\alpha(z_{\ast})\phi^{\prime}(\sqrt{\beta(\kappa)}x)^{2}}e^{-\frac{\gamma(\kappa,q)}{2\beta(\kappa)}\phi(\sqrt{\beta(\kappa)}x)^{2}}\right\rangle _{x\sim\cN(0,1)}.
\end{align}
Equations \eqref{eq:sp_eq_q}, \eqref{eq:sp_eq_kappa}, and \eqref{eq:sp_eq_zast}
can be solved efficiently by a damped iteration and using Gauss-Hermite
quadrature for the integrals due to the formulation in terms of standard
Gaussians. They are equivalent to the compact equations in the main
text. The resulting saddle point fields and the expected empirical
measure are shown in \prettyref{fig:Sol-SPA}.

\begin{figure}
\includegraphics{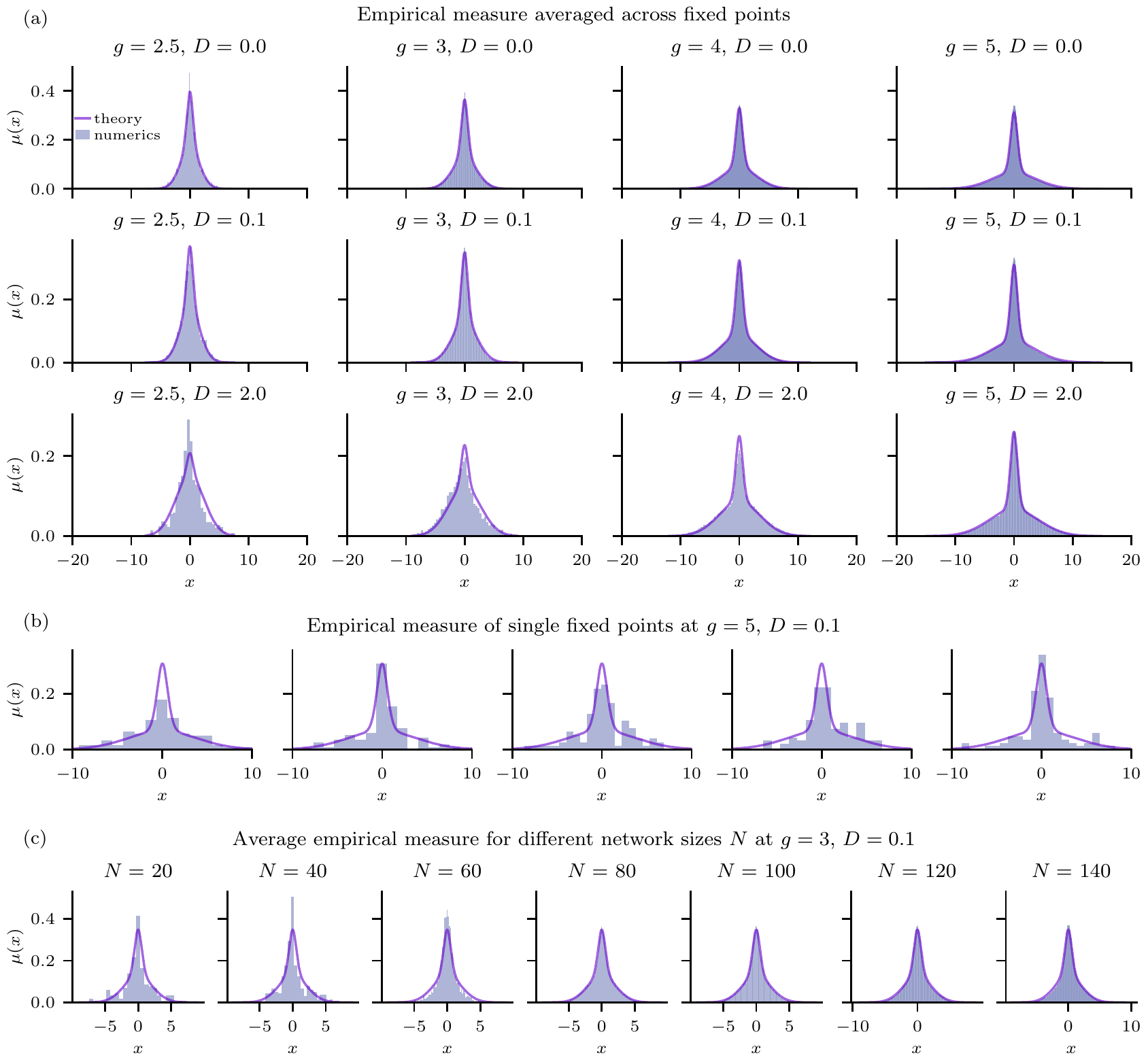}

\caption{\label{fig:EM-and-Num}Empirical measure for different parameters.
(a) Numerical verification of \prettyref{eq:EM_compact} by averaging
across all fixed point found at saturation for a fixed realization
(see \prettyref{subsec:Finding-fixed-points}). (b) Same as (a), but
using single fixed points instead of averaging across all. We note
that despite the low resolution due to numerical limitations, even
single fixed points resemble the expected empirical measure. This
is discussed in \prettyref{subsec:Rate-function-for}. (c) Finite
size analysis of the empirical measure.}

\end{figure}

Furthermore, the expected empirical measure Eq. \eqref{eq:EM_compact}
is compared to numerical results (see Sec. \ref{subsec:Finding-fixed-points})
in \prettyref{fig:EM-and-Num}.

\paragraph{Asymptotic solutions of the Saddle Point Equations \label{par:Solutions-large_g}}

Here we present closed form solutions of Equations \eqref{eq:sp_eq_q},
\eqref{eq:sp_eq_kappa}, and \eqref{eq:sp_eq_zast} in the limit $g\rightarrow\infty$
and in the limit $g\rightarrow1+$. We restrict the analysis to $D=0$.
For both limits, it is useful to rewrite the expected empirical measure.
 The normalized empirical measure is
\begin{equation}
\mu_{\ast}(y)=\frac{1}{\mathcal{Z}}\sqrt{1+\alpha\phi^{\prime}(y)^{2}}e^{-\frac{y^{2}+\gamma\phi(y)^{2}}{2\beta}},\qquad\mathcal{Z}=\int dy\,\sqrt{1+\alpha\phi^{\prime}(y)^{2}}e^{-\frac{y^{2}+\gamma\phi(y)^{2}}{2\beta}}\label{eq:def_of_norm_of_EM}
\end{equation}
for which the parameters $\alpha$, $\beta$, and $\gamma$ are determined
by 
\begin{equation}
1=g^{2}\left\langle \frac{\phi^{\prime}(y)^{2}}{1+\alpha\phi^{\prime}(y)^{2}}\right\rangle _{\mu_{\ast}},\qquad\beta=g^{2}\big\langle\phi(y)^{2}\big\rangle_{\mu_{\ast}},\qquad\gamma=g^{2}\left(1-\beta^{-1}\big\langle y^{2}\big\rangle_{\mu_{\ast}}\right).\label{eq:speqs_compact}
\end{equation}
We can express expectations as derivatives of $\mathcal{Z}$:
\begin{equation}
2\partial_{\alpha}\ln\mathcal{Z}=\langle\frac{\phi^{\prime}(y)^{2}}{1+\alpha\phi^{\prime}(y)^{2}}\rangle_{\mu}=\frac{1}{g^{2}}
\end{equation}
\begin{equation}
2\beta\partial_{\beta}\ln\mathcal{Z}=\frac{1}{\beta}\langle y^{2}+\gamma\phi(y)^{2}\rangle_{\mu}=1
\end{equation}
\begin{equation}
2\partial_{\gamma}\ln\mathcal{Z}=-\frac{1}{\beta}\langle\phi(y)^{2}\rangle_{\mu}=-\frac{1}{g^{2}}
\end{equation}
where we used Eqs. \eqref{eq:speqs_compact} in the last step. From
these identities follows
\begin{equation}
\ln\mathcal{Z}=\frac{\alpha-\gamma}{2g^{2}}+\frac{1}{2}\ln2\pi\beta+c,\qquad\mathcal{Z}=\sqrt{2\pi\beta}e^{\frac{\alpha-\gamma}{2g^{2}}+c}\label{eq:norm_of_mu}
\end{equation}
where $c$ does not depend on $\alpha,\beta,\gamma$ (but still on
$g$).

\textbf{Large $g$ asymptotics}

After rescaling $\alpha/g^{2}\to\alpha,\beta/g^{2}\to\beta,\gamma/g^{2}\to\gamma$,
and $y\to gy$ we notice that we can split the integrand in Eq. \eqref{eq:def_of_norm_of_EM}
into a slim peak of width $\cO(1/g)$ and a broad base of width $\cO(1)$,
thus for $g\gg1$,
\begin{equation}
\mathcal{Z}=\int dy\,\left[\sqrt{\alpha}g\phi^{\prime}(gy)e^{-\frac{\gamma\phi(gy)^{2}}{2\beta}}+e^{-\frac{\gamma}{2\beta}}e^{-\frac{y^{2}}{2\beta}}\right]
\end{equation}
 which can be evaluated
\begin{equation}
\mathcal{Z}=\sqrt{2\pi\beta}\left[\sqrt{\frac{\alpha}{\gamma}}\,\mathrm{erf}(\sqrt{\gamma/2\beta})+e^{-\frac{\gamma}{2\beta}}\right].\label{eq:norm_of_mu_large_g}
\end{equation}
Next, we derive saddle point equations that are consistent with the
approximation of $\mathcal{Z}$ by exploiting Eq. \eqref{eq:norm_of_mu}
\begin{equation}
2\partial_{\alpha}\ln\mathcal{Z}=\frac{\sqrt{\frac{1}{\alpha\gamma}}\,\mathrm{erf}(\sqrt{\gamma/2\beta})}{\sqrt{\frac{\alpha}{\gamma}}\,\mathrm{erf}(\sqrt{\gamma/2\beta})+e^{-\frac{\gamma}{2\beta}}}=1
\end{equation}
\begin{equation}
2\beta\partial_{\beta}\ln\mathcal{Z}=1+\frac{-\sqrt{\frac{2\alpha}{\pi\beta}}e^{-\frac{\gamma}{2\beta}}+\frac{\gamma}{\beta}e^{-\frac{\gamma}{2\beta}}}{\sqrt{\frac{\alpha}{\gamma}}\,\mathrm{erf}(\sqrt{\gamma/2\beta})+e^{-\frac{\gamma}{2\beta}}}=1
\end{equation}
\begin{equation}
2\partial_{\gamma}\ln\mathcal{Z}=\frac{-\sqrt{\frac{\alpha}{\gamma^{3}}}\,\mathrm{erf}(\sqrt{\gamma/2\beta})+\sqrt{\frac{2\alpha}{\pi\beta\gamma^{2}}}e^{-\frac{\gamma}{2\beta}}-\frac{1}{\beta}e^{-\frac{\gamma}{2\beta}}}{\sqrt{\frac{\alpha}{\gamma}}\,\mathrm{erf}(\sqrt{\gamma/2\beta})+e^{-\frac{\gamma}{2\beta}}}=-1
\end{equation}
The equations can be solved; back in the original scaling the solution
is
\begin{equation}
\beta=\frac{\pi}{2}\gamma,\qquad\gamma=\alpha,\qquad\gamma=g^{2}\frac{x}{1+x},\qquad x=\mathrm{erf}(\sqrt{1/\pi})e^{\frac{1}{\pi}}\label{eq:saddle_point_large_g}
\end{equation}
which is shown numerically in Fig.~\ref{fig:Asymptotic}.

\textbf{Small $g$ asymptotics}

We write $g=1+\epsilon$ and assume that $\alpha,\beta,\gamma=\cO(\epsilon)$.
Due to the Gaussian part, $y$ only varies on $\cO(\sqrt{\epsilon})$.
We include terms of $\cO(\epsilon^{3})$. In the measure Eq. \eqref{eq:def_of_norm_of_EM}
we write $\phi(y)^{2}=y^{2}+[\phi(y)^{2}-y^{2}]$ and expand 
\begin{equation}
e^{-\frac{\gamma}{2\beta}[\phi(y)^{2}-y^{2}]}=1-\frac{\gamma}{2\beta}[\phi(y)^{2}-y^{2}]+\cO(\epsilon^{4})=1+\frac{1}{3}\frac{\gamma}{\beta}y^{4}-\frac{17}{90}\frac{\gamma}{\beta}y^{6}+\cO(\epsilon^{4})
\end{equation}
Also expanding 
\begin{equation}
\sqrt{1+\alpha\phi^{\prime}(y)^{2}}=\sqrt{1+\alpha}[1-\frac{\alpha}{1+\alpha}y^{2}+\frac{\alpha(4\alpha+7)}{6(1+\alpha)^{2}}y^{4}+\cO(\epsilon^{4})]=\sqrt{1+\alpha}[1-\alpha(1-\alpha)y^{2}+\frac{7}{6}\alpha y^{4}+\cO(\epsilon^{4})]
\end{equation}
we get
\begin{align}
\mua(y) & =\frac{1}{\mathcal{Z}}\sqrt{2\pi\beta\frac{1+\alpha}{1+\gamma}}\mathcal{N}(y\,|\,0,\beta/(1+\gamma))\Big(1-\alpha(1-\alpha)y^{2}+\big(\frac{7}{6}\alpha+\frac{1}{3}\frac{\gamma}{\beta}\big)y^{4}-\frac{17}{90}\frac{\gamma}{\beta}y^{6}+\cO(\epsilon^{4})\Big)
\end{align}
\begin{align}
\mathcal{Z} & =\sqrt{2\pi\beta\frac{1+\alpha}{1+\gamma}}\Big(1+(\gamma-\alpha)\beta+(\alpha^{2}+\frac{7}{2}\alpha\beta+\alpha\gamma-\frac{17}{6}\beta\gamma-2\gamma^{2})\beta+\cO(\epsilon^{4})\Big)\label{eq:norm_of_mu_low_g}
\end{align}
Thus, the saddle point equations can again be derived using Eq. \eqref{eq:norm_of_mu}
\begin{equation}
2\partial_{\alpha}\ln\mathcal{Z}=\frac{1}{1+\alpha}+2\frac{(-1+2\alpha+\frac{7}{2}\beta+\gamma)\beta}{1+(\gamma-\alpha)\beta+(\alpha^{2}+\frac{7}{2}\alpha\beta+\alpha\gamma-\frac{17}{6}\beta\gamma-2\gamma^{2})\beta}=\frac{1}{g^{2}}
\end{equation}
\begin{equation}
2\beta\partial_{\beta}\ln\mathcal{Z}=1+2\beta\frac{\gamma-\alpha+\alpha^{2}+\frac{7}{2}\alpha\beta+\alpha\gamma-\frac{17}{6}\beta\gamma-2\gamma^{2}+\frac{7}{2}\alpha\beta-\frac{17}{6}\gamma\beta}{1+(\gamma-\alpha)\beta+(\alpha^{2}+\frac{7}{2}\alpha\beta+\alpha\gamma-\frac{17}{6}\beta\gamma-2\gamma^{2})\beta}=1
\end{equation}
\begin{equation}
2\partial_{\gamma}\ln\mathcal{Z}=-\frac{1}{1+\gamma}+2\frac{(1+\alpha-\frac{17}{6}\beta-4\gamma)\beta}{1+(\gamma-\alpha)\beta+(\alpha^{2}+\frac{7}{2}\alpha\beta+\alpha\gamma-\frac{17}{6}\beta\gamma-2\gamma^{2})\beta}=-\frac{1}{g^{2}}
\end{equation}
Dropping higher order terms, the equations evaluate to
\begin{equation}
\alpha=\gamma+\cO(\epsilon^{2}),\qquad\gamma=\beta+\cO(\epsilon^{2}),\qquad\alpha=\frac{2}{3}\epsilon+\cO(\epsilon^{2})\label{eq:saddle_point_low_g}
\end{equation}
which is shown in Fig.~\ref{fig:Asymptotic}.

\subsubsection{Fluctuations of the empirical measure}

Here, we discuss some properties of the expected empirical measure.

\paragraph{Expected empirical measure versus marginal distribution}

In the context of self-averaging variables one often wants to know
if certain distributions can be swapped when computing the expected
value of the respective variable. In this spirit, we here show that
the expected empirical measure is not only the expected distribution
of vector components at a fixed point, but also the expected marginal
distribution of one vector component across all fixed points. To see
this, consider the definition of the expected empirical measure

\begin{equation}
\begin{aligned}\mu_{\ast}(y) & =\frac{1}{\int d\bx^{\prime}\rho(\bx^{\prime})}\int d\bx\,\rho(\bx)\,\mu_{\bx}(y)=\frac{1}{\int d\bx^{\prime}\rho(\bx^{\prime})}\frac{1}{N}\sum_{i=1}^{N}\int d\bx\,\rho(\bx)\,\delta(x_{i}-y).\end{aligned}
\label{eq:ml_em_marginal}
\end{equation}
Recall that due to the realization average, the fixed point distribution
is symmetric under permutations of the units. Hence, every part of
the sum in Eq. \eqref{eq:ml_em_marginal} is equal, and we can write
\begin{equation}
\mu_{\ast}(y)=\frac{1}{\int d\bx^{\prime}\rho(\bx^{\prime})}\int d\bx\,\rho(\bx)\,\delta(x_{1}-y)
\end{equation}
where instead of $x_{1}$ any other unit could be chosen as well.
Carrying out the integration over $x_{1}$, we find the usual expression
for marginal distributions
\begin{equation}
\mu_{\ast}(y)=\frac{1}{\int d\bx^{\prime}\rho(\bx^{\prime})}\int\left(\prod_{i=2}^{N}dx_{i}\right)\rho(y,x_{2},...,x_{N}).
\end{equation}

\paragraph{Origin of the variability in $\rho(\protect\bx)$ \label{subsec:Origin-of-the}}

So far we computed the expected empirical measure. In the next Section,
we estimate fluctuations of the empirical measure around the expected
empirical measure. In order to interpret them, we here study the origin
of these fluctuations. When considering the empirical measure of single
fixed points in a fixed realization, the fluctuations could be across
fixed points \emph{within} the realization. When considering the average
empirical measure over all fixed points in one realization, the fluctuations
of this average could be \emph{across} realizations.

We find that the fixed point density accounts for both, within- and
across-realization fluctuations. To see this we derive the \emph{law
of total variance}. The variance of the empirical measure is 
\begin{equation}
\begin{aligned}T(y) & =\left.\frac{\delta^{2}}{\delta j(y)^{2}}\ln\left\langle e^{ij^{\T}\mu_{\bx}}\right\rangle _{\bx\sim\rho(\bx)}\right|_{j=0}=\left\langle \left(\mu_{\bx}(y)-\left\langle \mu_{\bz}(y)\right\rangle _{\rho(\bz)}\right)^{2}\right\rangle _{\rho(\bx)}\end{aligned}
.
\end{equation}
 Recall that the fixed point distribution is the realization average
$\rho(\bx)=\left\langle \rho_{\bJ,\bEta}(\bx)\right\rangle _{\bJ,\bEta}$.
In that spirit, we can dissect the total variance $T$ (by adding
a zero term) into
\begin{equation}
T(y)=A(y)+\left\langle W_{\bJ,\bEta}(y)\right\rangle _{\bJ,\bEta}\label{eq:law_total_variance}
\end{equation}
where
\begin{equation}
A(y)=\left\langle \left(\left\langle \mu_{\bx}(y)\right\rangle _{\rho_{\bJ,\bEta}(\bx)}-\left\langle \left\langle \mu_{\bz}(y)\right\rangle _{\rho_{\bJ^{\prime},\bEta^{\prime}}(\bz)}\right\rangle _{\bJ^{\prime},\bEta^{\prime}}\right)^{2}\right\rangle _{\bJ,\bEta}
\end{equation}
 is the \emph{across realizations variance} of the expected empirical
measure and 
\begin{equation}
W_{\bJ,\bEta}(y)=\left\langle \left(\mu_{\bx}(y)-\left\langle \mu_{\bz}(y)\right\rangle _{\rho_{\bJ,\bEta}(\bz)}\right)^{2}\right\rangle _{\rho_{\bJ,\bEta}(\bx)}
\end{equation}
is the \emph{within realizations variance} of the empirical measure.
Equation \eqref{eq:law_total_variance} reflects the law of total
variance.

We conclude that the variability we get from the fixed point distribution
accounts for both -- across- and within-realizations -- variances.
This means that the expected empirical measure is also the expected
distribution of vector components of \emph{every} fixed point. We
expect that for our case the within realizations part $\left\langle W_{\bJ,\bEta}\right\rangle _{\bJ,\bEta}$
strongly dominates for both the empirical measure and the norm due
to a self-averaging property in the sense that $\left\langle G(\bx)\right\rangle _{\rho_{\bJ,\bEta}(\bx)}\approx\left\langle G(\bx)\right\rangle _{\rho(\bx)};$
proving this analytically would require to compute the joint density
$\left\langle \rho_{\bJ,\bEta}(\bx)\rho_{\bJ,\bEta}(\bz)\right\rangle _{\bJ,\bEta}$
akin to the computation of $\left\langle \rho_{\bJ,\bEta}(\bx)\right\rangle _{\bJ,\bEta}$
in Sec.~\ref{subsec:Kac-Rice-formula} and show that it is very close
to $\rho(\bx)\rho(\bz)$.

Computing across-realization fluctuations numerically is resource-intensive
because one has to find a representative set of fixed points for several
realizations to track the variability of the within-realization mean
but even finding a representative set for one realization is difficult
for resource reasons, see Sec. \ref{subsec:Finding-fixed-points}.

However, for the case of norms we can compare the rate function we
compute in Sec. \ref{subsec:Norm-distribution} with the distribution
of numerical fixed point norms in one realization. Here we find that
most of the variability seems to be already within the realization,
see \prettyref{fig:Distance-distribution}, underlying our claim that
the across-variances part $A$ is subdominant.

\paragraph{Rate function for the empirical measure\label{subsec:Rate-function-for}}

In \prettyref{fig:EM-and-Num} we observe that the expected empirical
measure is in excellent agreement with the empirical measure averaged
across fixed point of a fixed realization. We further observe that
even single fixed point's empirical measures resemble the expected
empirical measure. To understand this, we here investigate the likeliness
of deviations of the empirical measure from the expected empirical
measure. In deriving the saddle point Equations \eqref{eq:spe_mu}
and \eqref{eq:spe_mut} we showed the differentiability of the scaled
cumulant generating functional Eq. \eqref{eq:scgf_spa}.  Due to
the existence and the differentiability of the scaled cumulant generating
functional, the G\"artner-Ellis theorem holds \citep{Touchette09},
\citep[Sec. 5]{Ellis95_97}, stating that the empirical measure fulfills
a \emph{large deviation principle}: The family of measures $\mu$
converges to the expected measure $\mu_{\ast}$ as $N\rightarrow\infty$
in the sense of distributions.

According to the G\"artner-Ellis theorem, the probability distribution
functional of all empirical measures at fixed points is of the form
$P[\mu]\asymeq\exp(-NH[\mu])$, and the \emph{rate functional} $H[\mu]$
is the Legendre transform of the scaled cumulant generating functional.
Here, as in the main text, we use the notation $a\overset{.}{=}e^{Nb}$
to denote $\lim_{N\to\infty}\frac{1}{N}\ln a=b$. The rate functional
quantifies the probability of deviations of $\mu$ from the minimum
$\mu_{\ast}$ of the rate functional. These are rare for large $N$
since they are exponentially suppressed in probability. 

In this Section, we compute the rate functional. Since the scaled
cumulant generating functional is differentiable, the Legendre transform
is
\begin{equation}
H[\mu]=i\mu^{\T}\jax-W_{N}\left[\jax\right]\label{eq:legendre_transform}
\end{equation}
where $\jax$ is determined by the stationarity condition
\begin{equation}
\mu(y)=\left.\frac{\delta}{\delta\,ij(y)}W_{N}[j]\right|_{\jax}.
\end{equation}
Analogously to Eq. \eqref{eq:expect_of_em}, we get
\begin{equation}
\begin{aligned}\mu(y) & =\frac{e^{-\frac{y^{2}}{2\left(\kappa\left\{ \mua[\jax]\right\} +D\right)}+\zeta\left\{ \mua[\jax]\right\} +i\jax(y)+i\muta[\jax](y)}}{\Omega_{\ast}\sqrt{2\pi\left(\kappa\left[\mua[\jax]\right]+D\right)}}\end{aligned}
\label{eq:stationarity_expanded}
\end{equation}
where we introduced the short hand notation $\Omega_{\ast}=\Omega\left\{ \mu_{\ast}[\jax],\muta[\jax],\jax\right\} $.
Comparing with Eq. \eqref{eq:expect_of_em}, we also see the identity
\begin{equation}
\mu(y)=\mu_{\ast}[\jax](y)
\end{equation}
reflecting that $\mu$ is canonically conjugate to $j$. Formally
solving Eq. \eqref{eq:stationarity_expanded} for the appearance of
$\jax$ in the exponent, and plugging into Eq. \eqref{eq:legendre_transform},
yields
\begin{equation}
H[\mu]=D_{\text{KL}}(\mu\|\nu)-W_{N}\left[\jax\right]\label{eq:rate_function}
\end{equation}
where $D_{\text{KL}}(\rho_{1}\|\rho_{2})=\left\langle \ln\left[\rho_{1}(x)/\rho_{2}(x)\right]\right\rangle _{\rho_{1}(x)}$
is the Kullback-Leibler divergence between the probability distribution
functions $\rho_{1}$ and $\rho_{2}$. As a functional of $\rho_{1}$
it is convex and has a single minimum at $\rho_{1}=\rho_{2}$. The
reference function here is 
\begin{equation}
\nu(y)=\frac{e^{-\frac{y^{2}}{2\left(\kappa\left[\mu\right]+D\right)}+\zeta\left[\mu\right]+i\muta[\jax](y)}}{\Omega_{\ast}\sqrt{2\pi\left(\kappa\left[\mu\right]+D\right)}}.
\end{equation}

A necessary condition for the expected measure $\mua$ is that it
minimizes the rate functional 
\begin{equation}
\left.\frac{\delta}{\delta\mu}H[\mu]\right|_{\mu=\mua}=ij_{\ast}^{\mua}\overset{!}{=}0.
\end{equation}
Indeed, since $j=0$ maximizes $W_{N}[j]$, namely $W_{N}[0]=0$,
the expected empirical measure determined through the self consistent
requirement of the Kullback-Leibler divergence is
\begin{equation}
\mua(y)=\frac{e^{-\frac{y^{2}}{2\left(\kappa\left[\mua\right]+D\right)}+\zeta\left[\mu\right]+i\muta[0](y)}}{\Omega_{\ast}\sqrt{2\pi\left(\kappa\left[\mua\right]+D\right)}},
\end{equation}
in line with the saddle point Equation \eqref{eq:spe_mu}.

Equation \eqref{eq:rate_function} expresses that points $\mu\neq\mua$
are exponentially suppressed for two reasons: First, simply because
of the difference in terms of the Kullback-Leibler divergence. Second,
because of a mismatch of the extremum of the scaled cumulant generating
functional.

\subsubsection{Permutation symmetry: approximate independence and convergence to
$\protect\mua$}

To interpret the permutation symmetry of the fixed point density \prettyref{eq:fpd_asinmain},
we here recall connections between exchangeability and statistical
independence.

\paragraph{Prelude: Exchangeability}

Consider a random vector $X=(X_{1},...,X_{n})\in\mathbb{R}^{n}$.
$X$ is called $l$-exchangeable \citep{Aldous85_1} where $l\leq n$
if 
\begin{equation}
\mathbb{P}[X_{1}\in A_{1},...,X_{l}\in A_{l}]=\mathbb{P}[X_{\sigma(1)}\in A_{1},...,X_{\sigma(l)}\in A_{l}]
\end{equation}
for all permutations $\sigma\in S(\{1,...,l\})$ (i.e., the symmetric
group of the set $\{1,...,l\}$) and $A_{1},...,A_{l}$ measurable.
Note that equivalently, we could have required 
\begin{equation}
\mathbb{P}[X_{1}\in A_{1},...,X_{l}\in A_{l}]=\mathbb{P}[X_{1}\in A_{\sigma(1)},...,X_{l}\in A_{\sigma(l)}].
\end{equation}
We remark that if $l^{\prime}\leq l$, $l$-exchangeability implies
$l^{\prime}$-exchangeability. We call $X$ exchangeable if $X$ is
$n$-exchangeable. An infinite sequence of random variables $(X_{1},X_{2},...)$
is called exchangeable if it is $l$-exchangeable for all $l\geq0$.

If $(X_{1},...,X_{n})$ or $(X_{1},X_{2},...)$ is a finite or infinite
sequence of i.i.d.~random variables it is exchangeable; exchangeability
is a natural generalization of the i.i.d.~condition since it already
implies an identical distribution of the coordinates.

Let $p(\bx)=p(x_{1},...,x_{n})$ be the density of a random vector
$X=(X_{1},...,X_{n})$; $X$ is exchangeable if and only if $p$ is
symmetric, meaning 
\begin{equation}
p(x_{1},...,x_{n})=p(x_{\sigma(1)},...,x_{\sigma(n)})
\end{equation}
for all permutations $\sigma\in S(n)$. This is due to the fact that
the matrix representation $\mathbf{M}_{\sigma}$ of the permutation
$\sigma$ is an orthogonal matrix, i.e.
\begin{align}
\mathbb{P}[X_{1}\in A_{1},...,X_{n}\in A_{n}] & =\int_{A_{1}}...\int_{A_{n}}p(\bx)\,d\bx=\int_{A_{1}}...\int_{A_{n}}p(\mathbf{M}_{\sigma}\bx)|\mathrm{det}(\mathbf{M}_{\sigma})|\,d\bx\\
 & =\int_{A_{\sigma(1)}}...\int_{A_{\sigma(n)}}p(\bx)\,d\bx=\mathbb{P}[X_{1}\in A_{\sigma(1)},...,X_{n}\in A_{\sigma(n)}].
\end{align}
Conversely, 
\begin{align}
p(x_{1},...,x_{n})\epsilon^{n} & =\mathbb{P}[X_{1}\in[x_{1},x_{1}+\epsilon),...,X_{n}\in[x_{n},x_{n}+\epsilon)]\\
 & =\mathbb{P}[X_{1}\in[x_{\sigma(1)},x_{\sigma(1)}+\epsilon),...,X_{n}\in[x_{\sigma(1)},\sigma_{n}+\epsilon)]=p(x_{\sigma(1)},...,x_{\sigma(n)})\epsilon^{n}.
\end{align}
Alternatively, we could have noted that the cumulative distribution
function 
\begin{equation}
F(x_{1},...,x_{n})=\mathbb{P}[X_{1}\in(-\infty,x_{1}],...,X_{n}\in(-\infty,x_{n}]]
\end{equation}
is symmetric in the above sense and $p(x_{1},...,x_{n})=\partial_{1}...\partial_{n}F(x_{1},...,x_{n})$.

\paragraph{Exchangeability and statistical independence}

The strongest connection between exchangeability and statistical independence
is given for \emph{infinite} sequences by the \emph{Theorem of de
Finetti} \citep{Aldous85_1}: For any exchangeable random sequence
$(X_{1},X_{2},...)$ with values in a state space $\Sigma$, there
exists a probability measure $P$ on the space of probability measures
on $\Sigma$ such that 
\begin{equation}
\mathbb{P}[X_{1}\in A_{1},...,X_{l}\in A_{l}]=\int\mu(A_{1})\cdot...\cdot\mu(A_{l})P[d\mu]
\end{equation}
for any $l\in\mathds{N}$. Moreover, $P$ is the distribution function
of the empirical measure. Phrasing differently, conditioned on $P=\mu$,
the coordinates are independent and indentically distributed: 
\begin{equation}
\mathbb{P}[X_{1}\in A_{1},...,X_{l}\in A_{l}|P=\mu]=\mu(A_{1})\cdot...\cdot\mu(A_{l})
\end{equation}
A consequence of this theorem is that an exchangeable infinite random
sequence $(X_{1},X_{2},...)$, the elements of which are pairwise
independent, is already i.i.d.. The Theorem of de Finetti is also
often interpreted in the following way: `A sequence of exchangeable
random variables is a mixture of i.i.d.'s'.

The theorem, however, does not hold true in this nice form for \emph{finite}
exchangeable sequences like the random positions of fixed points which
are distributed by \prettyref{eq:fpd_asinmain}. However, there is
a slightly weaker but still useful result by\emph{ Diaconis and Freedman}
\citep{Diaconis80_745}: Given an exchangeable finite random vector
$X=(X_{1},...,X_{n})$ with state space $\Sigma$ and let $k\leq n$.
We denote the probability measure projected onto the first $k$-coordinates
by $\mathbb{P}_{k}$. Then there exists a probability measure $P$
on the space of probability measures on $\Sigma$ with 
\begin{equation}
\bigg\vert\mathbb{P}_{k}[X_{1}\in A_{1},...,X_{k}\in A_{k}]-\int\mu(A_{1})\cdot...\cdot\mu(A_{k})P[\mathrm{d}\mu]\bigg\vert\leq\frac{k(k-1)}{n}
\end{equation}
for all $A_{1},...,A_{n}$ measurable and $P$ is the law of the empirical
measures of $X$.

\paragraph{Distribution of Coordinates for a Single Fixed-Point}

Recall that the $N$-dimensional distribution of fixed-points is given
by $\rho(\bx)=\exp(-NS(\bx))$ where 
\begin{equation}
S(\bx)=\frac{q(\bx)}{2[\kappa(\bx)+D]}+\ln\{2\pi[\kappa(\bx)+D]\}-\zeta(\bx).
\end{equation}
Here 
\begin{equation}
q(\bx)=\frac{1}{N}\sum_{i=1}^{N}x_{i}^{2},\qquad\kappa(\bx)=\frac{g^{2}}{N}\sum_{i=1}^{N}\phi(x_{i})^{2},\qquad\zeta(\bx)=-\frac{1}{2}z_{*}+\frac{1}{2N}\sum_{i=1}^{N}\ln[1+z_{*}g^{2}\phi^{\prime}(x_{i})^{2}],
\end{equation}
where $z_{*}$ is the solution of 
\begin{equation}
1=\frac{1}{N}\sum_{i=1}^{N}\frac{g^{2}\phi^{\prime}(x_{i})^{2}}{1+z_{*}g^{2}\phi^{\prime}(x_{i})^{2}}.
\end{equation}
Note that $q(\bx)$ and $\kappa(\bx)$ as well as $z_{*}$ and $\zeta(\bx)$
are invariant under a relabeling of the coordinates, i.e. the action
of the symmetric group, and thus is $S(\bx)$ and $\rho(\bx)$. Hence,
the $N$-dimensional random vector $X=(X_{1},...,X_{N})$ with the
density $\rho(\bx)$ describing the distribution of the fixed-points
is exchangeable.

Since the probability distribution function of the empirical measure
of $X$ takes the form $P[\mu]\asymeq\exp(-NH[\mu])$, it is concentrated
in the vicinity of the minimum of $H[\mu]$, which we denote by $\mu_{*}$.
This together with the above result on exchangeable random vectors
implies for a sufficiently small $k$ 
\begin{align}
\mathbb{P}_{k}[X_{1}\in A_{1},...,X_{k}\in A_{k}] & \approx\int\mu(A_{1})...\mu(A_{k})P[\mathrm{d}\mu]\\
 & \approx\mu_{*}(A_{1})...\mu_{*}(A_{k}).
\end{align}
For $N\gg1$ we note (1) individual coordinates have the distribution
$\mu_{*}$, (2) $k$ arbitrary coordinates are i.i.d., and (3) in
particular the coordinates are pairwise independent. Moreover, in
the limit $N\to\infty$ the coordinates are globally i.i.d..

Next we turn to the distribution of the coordinates of a given fixed-point.
Denoting by $F_{*}(x)=\mu_{*}((-\infty,x])$ the cumulative distribution
function of $\mu_{*}$, we first note that the above conclusions imply
$\mathbb{P}[X_{i}\leq x]=F_{*}(x)$ as well as $\mathbb{P}[X_{i}\leq x,X_{j}\leq x]=F_{*}(x)^{2}$.
The squared mismatch of the empirical cumulative distribution and
its i.i.d. limit result is thus
\begin{align}
\mathbb{E}\bigg[\big(\frac{1}{N}\sum_{i=1}^{N}1_{X_{i}\leq x}-F_{*}(x)\big)^{2}\bigg]= & \mathbb{E}\bigg[\frac{1}{N^{2}}\sum_{i,j=1}^{N}1_{X_{i}\leq x}1_{X_{j}\leq x}-\frac{2F_{*}(x)}{N}\sum_{i=1}^{N}1_{X_{i}\leq x}+F_{*}(x)^{2}\bigg]=\frac{1}{N}(F_{*}(x)-F_{*}(x)^{2})
\end{align}
which implies that the empirical measure converges to the expected
empirical measure uniformly in $x$. This in turn implies convergence
in probability due to Markov's inequality: For all $\epsilon>0$ 
\begin{equation}
\mathbb{P}\bigg[\big\vert\frac{1}{N}\sum_{i}^{N}1_{X_{i}\leq x}-F_{*}(x)\big\vert>\epsilon\bigg]\leq\frac{1}{\epsilon^{2}}\frac{F_{*}(x)-F_{*}(x)^{2}}{N}\leq\frac{1}{4\epsilon^{2}}\frac{1}{N}\overset{N\rightarrow\infty}{\rightarrow}0
\end{equation}
where we used that for a function $f$ with range in $[0,1]$ we have
$\max\big(f(x)-f(x)^{2}\big)\leq1/4$. From this we conclude that
for $N\gg1$, the empirical distribution function and thus the empirical
measure of the coordinates of a single fixed point are approximately
distributed as $F_{*}$ resp. $\mu_{*}$. Moreover, the coordinates
of any sufficiently small subsample of the fixed point are i.i.d.
with distribution $\mu_{*}$.

This also explains why sometimes the distributions do not match so
nicely: The convergence we get is weaker and the bound still allows
for larger fluctuations for sufficiently many realizations, especially
in comparison with the situation studied in the setting of the Glivenko-Cantelli
Theorem, where i.i.d. sequences $(X_{1},X_{2},...)$ are studied and
the Dvoretzky--Kiefer--Wolfowitz inequality gives a sharp bound
\begin{equation}
\mathbb{P}\bigg[\sup_{x}\big\vert\frac{1}{N}\sum_{i}^{N}1_{X_{i}\leq x}-F_{*}(x)\big\vert>\epsilon\bigg]\leq2e^{-2N\epsilon^{2}}.
\end{equation}
The i.i.d. condition here is crucial in the proof.

\subsection{Further observables}

\subsubsection{Distance distribution, separation of shells\label{subsec:Norm-distribution}}

Here, we ask about the distance $d_{\bx}=\sqrt{\bx^{\T}\bx}$ of fixed
points to the origin. The expected distance is
\begin{equation}
d_{\ast}=\sqrt{N\int dy\,\mua(y)y^{2}}.\label{eq:mean_distance_to_orig}
\end{equation}
According to the \emph{contraction principle }\citep{Touchette09},
the distance inherits the large deviation principle from the empirical
measure and thus, Eq. \eqref{eq:mean_distance_to_orig} is dominant
for large $N$. Hence, the fixed points live on a thin shell of an
$N$-dimensional sphere of radius $d_{\ast}$. The thickness of the
spherical shell decays exponentially with $N$.

\begin{figure}
\includegraphics{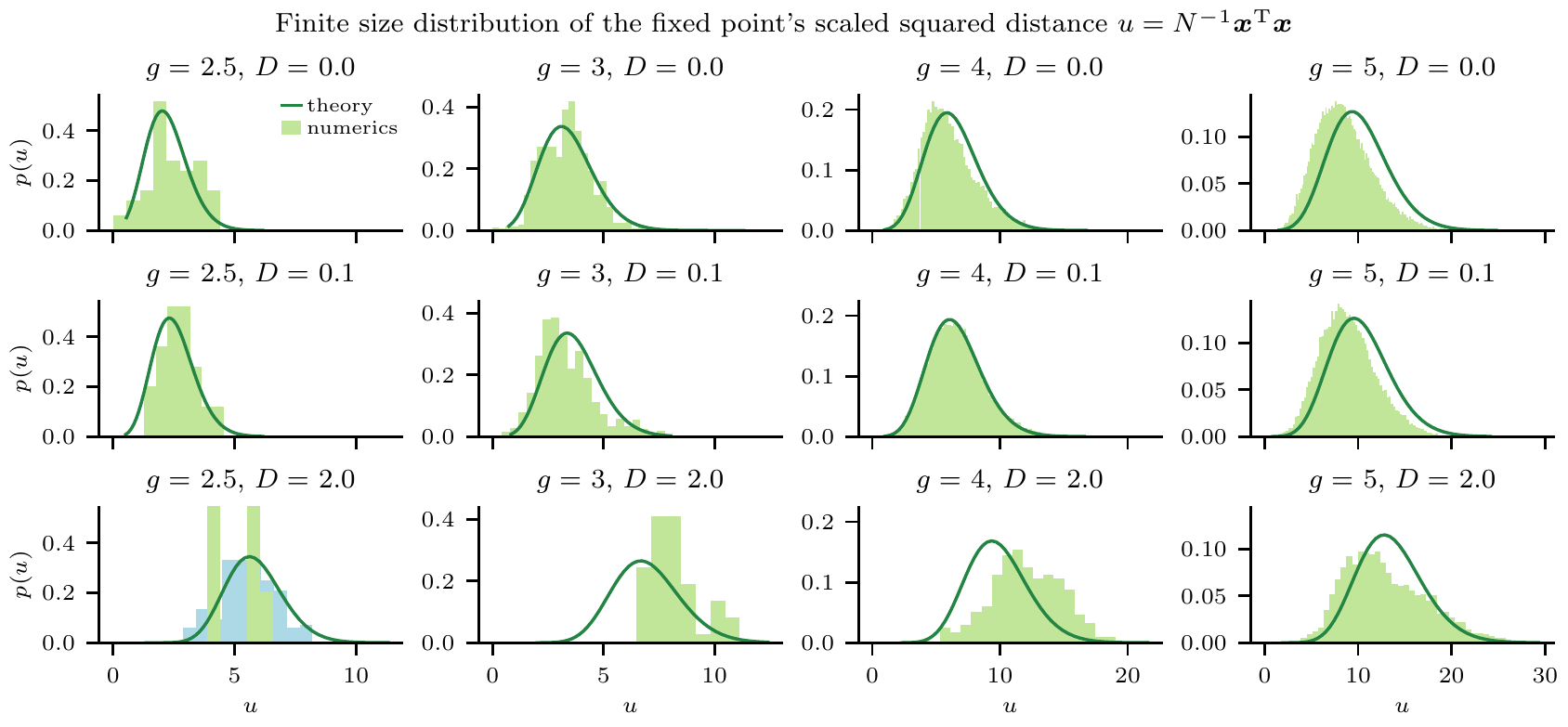}

\caption{\label{fig:Distance-distribution}Distance distribution. Using \prettyref{eq:rate_function_for_u}
and \prettyref{eq:optimize_for_rf_for_u}, we compute the distribution
of the scaled squared distance for several parameters (solid line)
and compare it to the numerical findings (green histograms) for fixed
realizations of $\protect\bJ$ and $\protect\bEta$. For high noise
$D$ and low $g$, there is only a small number of fixed points for
the moderate $N=100$ used for the numerical results. For parameters
$g=2.5$ and $D=2.0$ (lower left) we exemplify that the analytical
results are matched better when the distribution is averaged over
multiple realizations of $\protect\bJ$ and $\protect\bEta$ (blue
histogram).}
\end{figure}

We want to compute the finite size fluctuations of the distance. To
this end, we compute the rate function of the squared scaled distance
$u=d^{2}/N=\int dy\,\mu(y)y^{2}$. The contraction principle states
that the rate function for $u$ can be derived from the rate functional
for $\mu$
\begin{equation}
I(u)=\inf_{\mu:\,v^{\T}\mu=u}H[\mu]
\end{equation}
where $v(y)=y^{2}$. The minimization is carried out using a Lagrange
multiplier, hence we have to optimize
\begin{equation}
\mathcal{L}[\mu;\lambda]=H[\mu]-\lambda\left(u-v^{\T}\mu\right)
\end{equation}
for both, $\lambda$ and $\mu$. Recalling the Legendre transform
for $H[\mu]=\inf_{j}\left(i\mu^{\T}j-W[j]\right)$ in Eq. \eqref{eq:legendre_transform}
and swapping the optimizations, we have
\begin{equation}
\frac{\delta}{\delta\mu(y)}\left[i\mu^{\T}j-\lambda\left(u-v^{\T}\mu\right)\right]\overset{!}{=}0
\end{equation}
which constraints the source term $ij\overset{!}{=}\lambda v$. Since
$v(y)=y^{2}$ is a fixed function, $\inf_{j}\rightarrow\inf_{\lambda}$.
We have

\begin{equation}
\begin{aligned}I(u) & =\left.\inf_{\lambda}\left(\lambda\mu^{T}v-W[-i\lambda v]\right)\right|_{v^{\T}\mu\overset{!}{=}u}=\inf_{\lambda}\left(\lambda u-W_{N}[-i\lambda v]\right)\end{aligned}
\label{eq:rate_function_for_u}
\end{equation}
 where we were able to explicitly plug in the condition $v^{\T}\mu=u$.
Thus, we have to solve 
\begin{equation}
\begin{aligned}u & =\frac{\partial}{\partial\lambda}W_{N}[-i\lambda v]=v^{\T}\mua[-i\lambda v]\end{aligned}
.\label{eq:optimize_for_rf_for_u}
\end{equation}
In practice, we solve this by computing $\mua[-i\lambda v]$ in the
same way as we computed $\mua[0]$ in Sec. \ref{subsec:Most-likely-empirical}
for a sequence of values $\lambda$. Then, we check which $u$ they
correspond to using Eq. \eqref{eq:optimize_for_rf_for_u}. Lastly,
we plug into Eq. \eqref{eq:rate_function_for_u}. In the main text,
we show the respective variances for $D=0.1$ and several values of
$g$. Here, in \prettyref{fig:Distance-distribution} we also show
the whole distribution for several parameters and compare it to the
numerical results.

\paragraph{Deviations from numerical results}

Here, we discuss the deviations of the distance distribution derived
as above from the numerical findings using brute force fixed point
finding, which can be observed in \prettyref{fig:Distance-distribution}.

For low values of $g$, only few fixed points can be found due to
only $N=100$ and $c(g\approx1)\approx0$ and $N_{\text{fp}}\asymeq e^{cN}$.
Hence, the distance distribution relies on a low number of samples.
In contrast, for the empirical measure, we have $N$ times more samples,
since every vector component is a sample.

For high values of $g$, the system has many fixed points. So many
that only a tiny subset of the complete number can be found within
reasonable time and we therefore have to stop the fixed point finding
routine before saturation. Hence, the resulting numerical distribution
is strongly dependent on possible unknown biases of the Levenberg-Marquart
fixed point finder (possibly due to choice of initial values) to reach
fixed points with certain properties more often than others.

For intermediate values of $g$ we should expect the best agreement,
but also both effects from above might become relevant.

Furthermore, the bottom left panel in Fig. \ref{fig:Distance-distribution}
illustrates that when averaging over several realizations of $\bJ$
and $\bEta$, deviations can decrease. For parameters that exhibit
more fixed points, like the top right panel in Fig. \ref{fig:Distance-distribution},
we spared this check for resource reasons.

\subsubsection{Distribution of the spectral radius\label{subsec:Distribution-of-spectral-radius}}

The procedure that we used above for the distribution of the scaled
squared distance can be used equivalently for the spectral radius,
since it is also a contraction of the empirical measure $R^{2}=g^{2}\int dy\,\mua(y)\phi^{\prime}(y)^{2}$.
The rate function of the spectral radius follows as
\begin{equation}
J(R^{2})=\inf_{\mu:\,w^{\T}\mu=r}H[\mu]\label{eq:rate_function_spectral_radius}
\end{equation}
where $w(y)=g^{2}\phi^{\prime}(y)^{2}$. An analogous argument to
Sec. \ref{subsec:Norm-distribution} above shows that we need to compute
$\mua[-i\lambda w]$ for a sequence of values $\lambda$, then solve
$R^{2}=w^{\T}\mua[-i\lambda w]$ for $\lambda(R^{2})$ and get the
rate function as $J(R^{2})=\lambda(R^{2})R^{2}-W_{N}[-i\lambda(R^{2})w]$.
This procedure is used for Fig. \ref{fig:Finite-size-effects}(a).

\subsubsection{Topological complexity}

In this Section, we discuss the \emph{topological complexity}. It
was introduced by Wainrib and Touboul \citep{wainrib13_118101} to
describe the rate by which the expected number of fixed points grows
with the number of units $N_{\text{fp}}\asymeq e^{c(g,D)\,N}$. Wainrib
and Touboul proposed the topological complexity as a structural indicator
of chaoticity, complementary to the maximum Lyapunov exponent which
they refer to as \emph{dynamical complexity}.

By construction in Eq. \eqref{eq:fpd_def}, the expected number of
fixed points is the norm of the fixed point distribution. We therefore
encountered the topological complexity in the scaled cumulant generating
functional $W_{N}$ in Eq. \eqref{eq:def_topcomp}. We get the topological
complexity $c$ in saddle point approximation by the normalization
of the scaled cumulant generating functional Eq. \eqref{eq:scgf_spa}
$W_{N}[0]\overset{!}{=}0$ as
\begin{equation}
c=-i\tilde{\mu}_{\ast}^{\T}\mu_{\ast}+\ln\Omega\left[\mu_{\ast},\tilde{\mu}_{\ast}\right].\label{eq:topcomp_spa}
\end{equation}
Plugging in the expressions for $\tilde{\mu}_{\ast}$ in Eq. \eqref{eq:spe_mut}
and $\Omega$ in Eq. \eqref{eq:scgf_pre_spa} with $j=0$, we get
\begin{equation}
c=\frac{\gamma-\alpha}{2g^{2}}-D\frac{\gamma}{2\beta g^{2}}+\ln\mathcal{Z}-\frac{1}{2}\ln(2\pi\beta)\label{eq:topcomp_explicit}
\end{equation}
where
\begin{equation}
\mathcal{Z}=\int dx\sqrt{1+\alpha\phi^{\prime}(x)^{2}}e^{-\frac{x^{2}+\gamma\phi(x)^{2}}{2\beta}}.
\end{equation}
Based on the results of Sec. \ref{subsec:Most-likely-empirical},
the topological complexity is shown in \prettyref{fig:Sol-SPA}(d).
This result deviates from the result in \citep{wainrib13_118101},
even in the asymptotic limit $g=1+\epsilon$, $0<\epsilon\ll1$, where
the authors expect validity. For the two limits $\epsilon\ll1$ and
$g\gg1$ we derive closed form expressions below.

\paragraph{Asymptotic complexity for $g\rightarrow\infty$}

\begin{figure}
\includegraphics{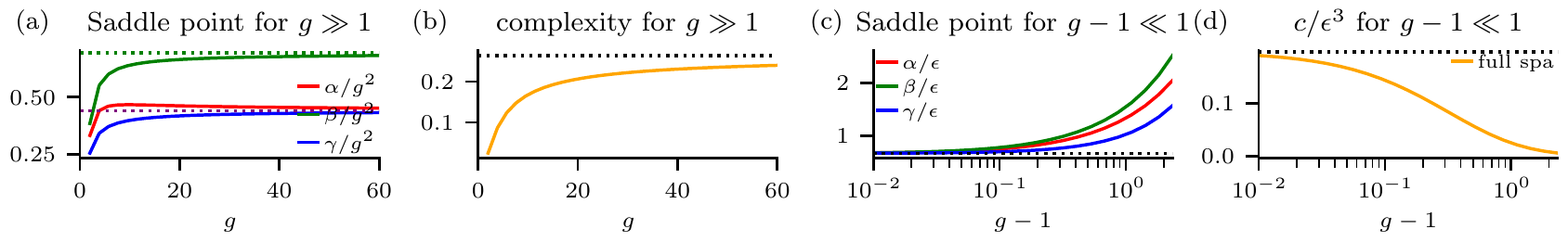}

\caption{\label{fig:Asymptotic}Asymptotic results for large $g$ and low $g$.
(a) Numerical solutions of the saddle point equations \prettyref{eq:SP_eqs_compact}
(solid lines) and asymptotic $g\gg1$ result Eq. \eqref{eq:saddle_point_large_g}
(horizontal dotted lines). (b) Saddle point result of the topological
complexity \prettyref{eq:topcomp_explicit} (solid line) and asymptotic
$g\gg1$ result Eq. \eqref{eq:topcomp_large_g} (horizontal dotted
line). (c) Numerical solutions of the saddle point equations \prettyref{eq:SP_eqs_compact}
(solid lines) and asymptotic $g-1\ll1$ result Eq. \eqref{eq:saddle_point_low_g}
(horizontal dotted line). (d) Saddle point result of the topological
complexity \prettyref{eq:topcomp_explicit} (solid line) and asymptotic
$g-1\ll1$ result Eq. \eqref{eq:topcomp_low_g}. Here, $D=0$.}
\end{figure}

For the strong chaos limit, the closed form expressions from Sec.
\ref{par:Solutions-large_g} can be leveraged, in fact comparing Eqs.
\eqref{eq:topcomp_explicit} and \eqref{eq:norm_of_mu} we note that
in Eq. \eqref{eq:norm_of_mu} $c$ is the complexity. Plugging the
asymptotic fields \eqref{eq:saddle_point_large_g} into \eqref{eq:norm_of_mu_large_g}
we get (for $D=0$)
\begin{equation}
c=\frac{\gamma-\alpha}{2g^{2}}+\ln\left[\sqrt{\frac{\alpha}{\gamma}}\,\mathrm{erf}(\sqrt{\gamma/2\beta})+e^{-\frac{\gamma}{2\beta}}\right]=-\frac{1}{\pi}+\ln\left[1+\mathrm{erf}(\sqrt{1/\pi})e^{1/\pi}\right]\label{eq:topcomp_large_g}
\end{equation}
which is presented in Fig. \ref{fig:Asymptotic}. Here, we used that
the factor in $\sqrt{2\pi\beta}$ in Eq. \eqref{eq:norm_of_mu_large_g}
cancels the $+\frac{1}{2}\ln2\pi\beta$ in \prettyref{eq:norm_of_mu}.

\paragraph{Asymptotic complexity for $g\rightarrow0$}

Using Eq. \eqref{eq:saddle_point_low_g}, i.e. to leading order $\alpha=\beta=\gamma$,
the low $g$ expression for $\mathcal{Z}$ (cf. \prettyref{eq:norm_of_mu_low_g})
is
\begin{align}
\mathcal{Z} & =\sqrt{2\pi\beta}\left[1+\frac{1}{2}(\alpha-\gamma)(1-\alpha)+O(\epsilon^{4})\right]\left[1+(\gamma-\alpha)\alpha+\left(\frac{2}{3}\right)^{4}\epsilon^{3}+O(\epsilon^{4})\right]\label{eq:topcomp_low_g}
\end{align}
such that the complexity is (again, for $D=0$)
\begin{equation}
c=\frac{\gamma-\alpha}{2g^{2}}+\ln\mathcal{Z}-\frac{1}{2}\ln(2\pi\beta)=\frac{1}{2}(\alpha-\gamma)(1-\alpha-\frac{1}{g^{2}}-2\alpha)+\big(\frac{2}{3}\big)^{4}\epsilon^{3}=\big(\frac{2}{3}\big)^{4}\epsilon^{3}+O(\epsilon^{4}),
\end{equation}
which we present in Fig. \ref{fig:Asymptotic}.

\paragraph{Critical $g$.--}

In the main text we discuss the transition from a system without fixed
points to a positive topological complexity $c>0$. The transition
point is computed as the root of \prettyref{eq:topcomp_spa}.

Next, in the main text, we show the transition line to chaos. The
result presented there is based on the dynamic mean-field computation
in \citep{Schuecker18_041029}. The case of \emph{quenched noise},
as \citep{Schuecker18_041029} puts it, which we consider here, is
slightly differs from the white noise case. This is for two reasons:
First, at infinity, the autocorrelation function does not drop to
zero but remains positive. Second, the stability of the quenched-noise
system is completely determined by the Jacobian of the dynamics. Hence,
the criterion for the transition to chaos is that of the spectral
radius of the Jacobian crossing unity
\begin{equation}
1=g^{2}\left\langle \phi^{\prime}(\sqrt{c_{0}}z)^{2}\right\rangle \label{eq:qn_trans_stability}
\end{equation}
where $z$ is standard normal. Here, $c_{0}$ is the equal-time autocorrelation,
which follows from ``energy conservation'' in the particle-in-a-potential
analogy \citep{Schuecker18_041029} as
\begin{equation}
-\frac{1}{2}c_{\infty}^{2}+g^{2}f_{\Phi}(c_{\infty},c_{0})+c_{\infty}D=-\frac{1}{2}c_{0}^{2}+g^{2}f_{\Phi}(c_{0},c_{0})+c_{0}D\label{eq:qn_trans_conservation}
\end{equation}
where $f_{u}(x,y)=\iint Dz_{1}Dz_{2}u\left(\sqrt{y-\frac{x^{2}}{y}}z_{1}+\frac{x}{\sqrt{y}}z_{2}\right)u\left(\sqrt{y}z_{2}\right)$
and $Dz=dz\frac{1}{\sqrt{2\pi}}e^{-z^{2}/2}$. Lastly, demanding a
constant autocorrelation $c_{\infty}$ at infinite timelag $\tau=\infty$
gives
\begin{equation}
c_{\infty}=g^{2}f_{\phi}(c_{\infty},c_{0})+D.\label{eq:qn_trans_inf_timelag}
\end{equation}
We solve Eqs. \eqref{eq:qn_trans_stability}, \eqref{eq:qn_trans_conservation},
and \eqref{eq:qn_trans_inf_timelag} with a damped fixed point iteration.
This yields the critical coupling strength $g_{c}$ at which the transition
to chaos occurs, and which is shown in the main text.

\subsubsection{Jacobian spectrum at fixed points}

Here, we want to characterize the spectrum of the Jacobian Eq. \eqref{eq:modeljac}
at fixed points. The Jacobian reads
\begin{equation}
\by^{\prime}(\bx)=-\id+\bJ\,\diag\left[\phi^{\prime}(\bx)\right]\label{eq:modeljac_reprint}
\end{equation}
where $J_{ij}\overset{\mathrm{i.i.d.}}{\sim}\cN(0,g^{2}/N)$. For
random matrices of this type, it is known that the spectrum of eigenvalues
is self-averaging \citep[Theorem 1.14]{Tao10_2023}. Ahmadian \emph{et
al.} \citep{Ahmadian15_012820} provide a general framework for computing
the spectrum of arbitrary random matrices of the form $\boldsymbol{\mathcal{M}}+\boldsymbol{\mathcal{L}}\bX\boldsymbol{\mathcal{R}}$
where $X_{ij}$ are i.i.d. with zero mean and variance $1/N$. Since
the identity in Eq. \eqref{eq:modeljac_reprint} only contributes
a shift of $-1$ of the eigenvalues of $\by^{\prime}(\bx)$, we can
use the method by Ahmadian \emph{et al.} for the simpler case $\boldsymbol{\mathcal{M}}=0$,
$\boldsymbol{\mathcal{L}}=\id$, and $\boldsymbol{\mathcal{R}}=g\,\text{diag}[\phi^{\prime}(\bx)]$,
where we pulled the factor $g$ out of $\bJ$ explicitly to match
the notation by Ahmadian \emph{et al.} The special case of random
matrices with zero mean is described by Ahmadian \emph{et al.} starting
from \citep[Eq. (2.13)]{Ahmadian15_012820} for the support $R(\bx)=\sqrt{\frac{1}{N}\sum_{i=1}^{N}\sigma_{i}^{2}}$
where $\sigma_{i}$ are the singular values of $\boldsymbol{\mathcal{LR}}$,
thus the eigenvalues of \textbf{$\bLambda$} since $\Lambda_{ij}=\delta_{ij}g\phi^{\prime}(x_{i})>0$.
In terms of the distribution of $\bX\bLambda(\bx)$, $\sigma_{i}=g\phi^{\prime}(x_{i})$
is the standard deviation of the matrix elements in the $i$th column.
Thus, we have the support
\begin{equation}
R(\bx)=g\sqrt{\frac{1}{N}\phi^{\prime}(\bx)^{\T}\phi^{\prime}(\bx)}.\label{eq:spectral_radius_x}
\end{equation}
Outside of this support, the spectral density vanishes. Within the
support, the spectral density is isotropic around the center at $-1+0i$,
which comes from the identity $-\id$ that we pulled out. The spectral
density 
\begin{equation}
\nu(r)=-\frac{1}{2\pi r}\partial_{r}n_{\bx}(r)
\end{equation}
can hence be derived from the \emph{radial tail distribution}, i.e.
the proportion of eigenvalues further than $r$ from the center, $n_{\bx}(r)$
\citep[Eq. (2.14)]{Ahmadian15_012820}. This is the fundamental theorem
of calculus in polar coordinates. The proportion $n_{\bx}(r)$ is
given by \citep[Eq. (2.15)]{Ahmadian15_012820}
\begin{equation}
\begin{aligned}1=\frac{1}{N}\sum_{i=1}^{N}\frac{1}{\frac{r^{2}}{\left[g\phi^{\prime}(x_{i})\right]^{2}}+n_{\bx}(r)} & \qquad\text{for }r<R\\
\text{and }n_{\bx}(r)=0 & \qquad\text{for }r\geq R
\end{aligned}
\label{eq:hardnut}
\end{equation}
Note that for the case of identically distributed columns ($x_{i}=x$
for all $i$), Eq. \eqref{eq:hardnut} can be solved $n(r)=1-r^{2}/\left[g\phi^{\prime}(x)\right]^{2}$,
hence $\nu(r)=1/\left[\pi R(\bx)\right]^{2}=\mathrm{const.}$ w.r.t.
$r$, and reflects the circular law which states that the spectrum
of i.i.d. random matrices is uniform \citep{Tao10_2023}.

We are, however, interested in the case where $x_{i}$ are different,
namely sampled from the expected empirical measure to give the expected
spectral distribution of the Jacobian at fixed points. Since both
the spectral radius and the defining equation of $n_{\bx}$ depend
on $\bx$ through network sums $\sum_{i=1}^{N}f(x_{i})$ only, both
can be expressed in terms of the empirical measure
\begin{equation}
R[\mu_{\bx}]=g\sqrt{\int dy\,\mu_{\bx}(y)\phi^{\prime}(y)^{2}},\qquad1=\int dy\,\mu_{\bx}(y)\frac{1}{\frac{r^{2}}{\left[g\phi^{\prime}(y)\right]^{2}}+n_{\mu_{\bx}}(r)}.
\end{equation}
Consequently, their expected values are given by $R[\mu_{\ast}]$
and $n_{\mu_{\ast}}(r)$ respectively. The latter is computed using
a bisection algorithm. The results are shown in the main text Fig.
2.

\subsection{Correlation despite separation}

In the main text we discuss the apparent contradiction that the dynamics'
velocity is strongly correlated with the prediction by the nearest
fixed point despite the separation of shells. Here, we give some more
technical details on this discussion.

\begin{figure}
\includegraphics{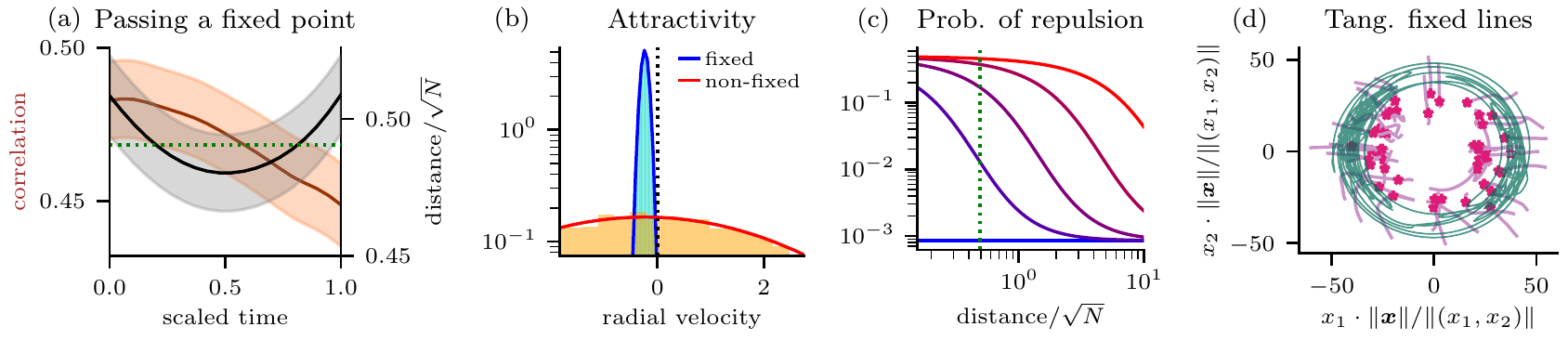}

\caption{\label{fig:Relation-of-fixed}Relation of fixed points to the dynamics.
(a) The trajectory passes a fixed point. While the specific fixed
point is closest, the Pearson correlation \prettyref{eq:pearson_corr}
(brown line, left axis) and the distance to the nearest fixed point
(black line, right axis) is shown. Shadings denote standard deviation
of the mean. (b) Distribution of the radial velocity away from a fixed
point \{theory based on \prettyref{eq:rad_vel_dist} (blue line),
and numerical result {[}velocity of the nonlinear model $\protect\bx^{\protect\T}\protect\by(\protect\bx^{\ast}+\protect\bx)${]}
(blue histogram)\} for sample points\textbf{ $\protect\bx$} of with
scaled squared norm $\gamma$, where $\gamma$ is chosen to correspond
to the distance at which the trajectory typically passes fixed points
{[}see green dotted line in (a), right axis, and green dotted line
in (c){]}. Respective result using non-fixed points {[}theory (red
line) and numerical result (orange histogram){]}. For the non-fixed
point we use the same points as for the blue histogram but redraw
the connectivity. (c) Probability of a Gaussian sample point of fixed
expected distance (x-axis) to be repelled from a fixed point (blue
line), from a non-fixed point (red line) with $\|\protect\by(\boldsymbol{p})\|=\left\langle \|\protect\by(\protect\bx^{\ast})\|\right\rangle _{x_{i}^{\ast}\sim\mu_{\ast}}\equiv y_{0}$
and from almost-fixed points with $\|\protect\by(\boldsymbol{p})\|\in\{y_{0}/1000,y_{0}/100,y_{0}/10\}$
(blue-purple to red-purple). The green dashed lines marks the distance
at which the trajectory typically passes fixed points. (d) Tangentially
fixed lines, i.e. lines along which the velocity is purely radial,
growing out of fixed points, connecting the phase space radially.
Parameters: $g=5$, $D=0$, $N=100$.}
\end{figure}

\subsubsection{Distance of fixed points to the dynamics\label{subsec:Distance-of-fixed}}

To measure the impact of fixed points we fix a realization of the
connectivity $\bJ$ and the static noise $\bEta$. Then, we use a
high performance cluster, as outlined in \prettyref{subsec:Finding-fixed-points},
to get a large fraction of the fixed points $\{\bx_{i}^{\ast}\}$
for that realization. Next, we integrate the model for the same realization
and ask at every time point along the trajectory $\bx(t)$ which one
of the many fixed points is the closest. The distance $d(t)=\min_{i}\|\bx(t)-\bx_{i}^{\ast}\|$,
where the minimization is carried out at every time step $t$, is
shown in \prettyref{fig:Relation-of-fixed} (a) and in the main text
(orange histograms in Fig. 3).

We now argue that the magnitude of $d(t)$ attests an attractive effect
of the fixed points. Numerically, we find that when replacing the
trajectory $\bx(t)$ by a set of random points $\bx_{j}$ whose statistics
are given by dynamic mean-field theory \citep{Sompolinsky88_259},
the respective distance $d_{j}=\min_{i}\|\bx_{j}-\bx_{i}^{\ast}\|$
is larger than that between fixed points and true dynamics. In the
main text Fig. 3, blue histograms, this is presented in terms of both,
radial and rotational separation.

We conclude an attractive effect of the fixed points. Indeed, due
to the majority of stable directions (the eigenvalue distribution
has most probability mass in the half plane of negative real parts)
and due to the absence of a constant field $\by(\bx^{\ast})$ (by
definition of fixed points), random points sampled around the fixed
point will almost always be drawn toward the fixed point -- before
being repelled along the unstable directions. To what degree is that
a sole feature of fixed points? We may interpolate between fixed points
and equivalent non-fixed points by adding a constant field $\bh$
with $h_{i}\iidsim\cN(0,\vartheta/N)$ and varying $\vartheta$. To
understand the fraction of points that are drawn toward a fixed point,
we study the distribution of the radial velocity $\cA=\bx^{\T}\bv_{\text{lin}}(\bx)=\bx^{\T}\left(\by^{\prime}(\bx^{\ast})\bx+\bh\right)$
at sample positions $\bx$ with $x_{i}\iidsim\cN(0,\gamma/N)$; for
the linear velocity field $\bv_{\text{lin}}$ to be accurate we require
small $\gamma$. We will see that for typical $\gamma$ {[}matched
to $d(t)$ above{]}, this approach accurately captures the attractivity
of the true nonlinear model. Performing the average w.r.t. $\bx$
\emph{before} averaging over the disorder $\bJ$ and $\bx^{\ast}$
(changing the order makes a difference here) we find
\begin{equation}
\langle\cA\rangle=-\gamma,\quad\cum{\cA}=\left[\gamma^{2}\left(R^{\ast2}+2\right)+\gamma\vartheta\right]N^{-1}+\cO(N^{-2})\label{eq:rad_vel_dist}
\end{equation}
where $R^{\ast}$ is the expected spectral radius at a fixed point;
all other cumulants vanish. The fraction of repulsive points is
\begin{equation}
\mathcal{R}=\int_{0}^{\infty}d\cA\,p(\cA)=\frac{1}{2}\left[1-\erf\sqrt{\frac{N}{2(R^{\ast2}+2+\vartheta/\gamma)}}\right].
\end{equation}
The distribution $p(\cA)$ is presented in \prettyref{fig:Relation-of-fixed}
(b) for a fixed point and a non-fixed point with $\|\bh\|=\sqrt{N\int dx\,x^{2}\mu^{\ast}(x)}$
and the fraction of repulsive samples is presented in \prettyref{fig:Relation-of-fixed}
(c) for fixed points, non-fixed points and almost-fixed points. Note
that the linear approximation leading to \prettyref{eq:rad_vel_dist}
accurately captures the nonlinear radial velocity at the relevant
distance {[}\prettyref{fig:Relation-of-fixed}(b){]}. For true fixed
points ($\vartheta=0$), $\mathcal{R}(\gamma)=\text{const.}$ is scale
free, i.e. at every distance $\sqrt{\gamma}$ to a fixed point the
probability to be drawn towards it (not considering the prehistory)
is equally large. This changes for non-fixed points $\vartheta>0$:
Here, the fraction of repulsive points grows monotonously when decreasing
the distance $\sqrt{\gamma}$ and saturates at $1/2$, which marks
the limit where the constant external field -- which repels the whole
hemisphere toward which it points -- dominates against the attractive
effect by the Jacobian. At the distance at which the trajectory typically
passes the fixed points {[}green dotted line in \prettyref{fig:Relation-of-fixed}(c){]},
even small levels of non-fixedness diminish the attractivity.

\subsubsection{Predictability at fixed points}

We found that the distance at which the trajectory passes the fixed
points is below odds. Here, we contrast this distance to the length-scale
of linear predictability: At any time $t$, we can estimate the predictability
of the trajectory $\bx(t)$ at a point $\bx_{1}$ by the Pearson correlation
\begin{equation}
c(t)=\frac{\by[\bx(t)]^{\T}\left[\by(\bx_{1})+\by^{\prime}(\bx_{1})(\bx-\bx_{1})\right]}{\|\by[\bx(t)]\|\|\by(\bx_{1})+\by^{\prime}(\bx_{1})(\bx-\bx_{1})\|}\label{eq:pearson_corr}
\end{equation}
of the actual velocity and the one predicted at $\bx_{1}$. For non-fixed
points $\bx_{1}$ that are radially or rotationally perturbed w.r.t.
$\bx(t)$, the resulting correlation is shown in the main text Fig.
3. In radial direction, prediction is possible across large gaps,
whereas a rotation quickly degrades the predictability, with almost
all predictability lost when $|\bx_{1}^{\T}\bx(t)|$ is small. For
the nearest fixed points $\bx_{1}=\bx^{\ast}$, as outlined in \prettyref{subsec:Distance-of-fixed},
and where the zeroth order in \prettyref{eq:pearson_corr} vanishes,
the Pearson correlation can be computed numerically and is presented
as the horizontal line in the main text Fig. 3; We make two observations:
(1) the magnitude of the Pearson correlation is substantial and hence
in terms of linear predictability the distance at which the trajectory
passes the fixed point can be regarded small, and (2) the magnitude
of the Pearson correlation is explained by the rotational rather than
the radial separation of nearest fixed points and dynamics.

The predictability during a `drive-by', i.e., while a specific fixed
point is closest, is shown in \prettyref{fig:Relation-of-fixed} (a).
Here, we observe that the fixed point predicts the dynamics better
during the attractive part of the drive-by than during the repulsive
part.

\subsubsection{Tangentially fixed points and lines}

Here, we propose a topological understanding of the strong radial
predictability. Based on the assumption that fixed points largely
determine the dynamics, one would assume (in light of the fixed points
concentration to a shell) that their impact is capable of bridging
radial gaps.

To bridge this gap, we consider \emph{tangentially} fixed points,
i.e., points $\bx$ where the velocity is purely radial $\by(\bx)\propto\bx$,
or, put differently,
\begin{equation}
0=\by(\bx)-\frac{\bx\bx^{\T}}{\|\bx\|^{2}}\by(\bx).
\end{equation}
The solutions can also be seen as conventional fixed points of an
equivalent model, wherein the distance is constrained $\|\bx(t)\|\overset{!}{=}r$.
The dynamics of the constraint model follows by subtracting the radial
velocity $\bx\bx^{\T}\by(\bx)/\|\bx\|^{2}$ from the original model,
yielding
\begin{equation}
\dot{\bx}=-\frac{\bx^{\T}\bJ\bphi}{\bx^{\T}\bx}\bx+\bJ\phi(\bx).
\end{equation}
Thus, tangentially fixed points are solutions of 
\begin{equation}
\bJ\phi(\bx)=\frac{\bx^{\T}\bJ\bphi}{\bx^{\T}\bx}\bx\quad\Leftrightarrow\quad\bx\bx^{\T}\bJ\phi(\bx)=r^{2}\bJ\phi(\bx).\label{eq:tangentially_fixed_points}
\end{equation}
Since \prettyref{eq:tangentially_fixed_points} is the eigenvalue
equation of a rank $1$ matrix, the solution is unique up to a scalar
factor, $\bJ\phi(\bx)=(1+s)\bx$.

The true fixed points are also tangentially fixed with vanishing radial
velocity. Do the true fixed points correspond to tangentially fixed
points on the dynamics shell? We start from true fixed points and
move outwards (inwards) into the unique direction where only the radial
velocity decreases (increases); all points along this path thus have
vanishing tangential velocity, so we term the paths \emph{tangentially
fixed lines}, see \prettyref{fig:Relation-of-fixed} (d).

To compute the tangentially fixed lines, we start at fixed points
$\bx$ and their exact distance to the origin $r=\|\bx\|$ (here,
$s=0$). Then we incrementally change this distance $r\rightarrow r+\Delta r$
and compute the solution of \prettyref{eq:tangentially_fixed_points}
using the Levenberg-Marquart rootfinder initialized at the last known
tangentially fixed point. Moving to larger (smaller) radii $r$ along
the fixed line corresponds to building up a radial velocity towards
(away from) the origin. Towards the inside, all tangentially fixed
lines stop; numerically this is detected by noting that no tangentially
fixed points can be found anymore in the vicinity and under the distance-constraint.
Note that in the linear regime of the phase space, tangentially fixed
points cannot exist in exponential abundance anymore, hence tangentially
fixed lines have to end. This is in line with the eventual depletion
of predictability at points near the origin {[}main text Fig. 3 (a){]}.
Towards the outside, most fixed lines did not end within the interval
of integration, some lines however did in the same manner as towards
the inside. An illustration of tangentially fixed lines is \prettyref{fig:Relation-of-fixed}(d).

\subsection{Numerical fixed point finding\label{subsec:Finding-fixed-points}}

To test the analytical results about the statistics of the fixed points,
we compare them to statistics of large numbers of fixed points obtained
for specific realizations of the random connectivity $\bJ$ and noise
$\bEta$ by brute force fixed point finding. To this end, we first
sample $\bJ$ and $\bEta$ randomly using the variances $g^{2}/N$
and $D$ respectively.

Next, we employ a Levenberg-Marquart minimizer on the velocity field
$\by(\bx)=-\bx+\bJ\phi(\bx)+\bEta$ starting from $1.5\times10^{9}$
to $10\times10^{9}$ initial values $\bx$ with $x_{i}\overset{\text{i.i.d.}}{\sim}\cN(0,200^{2})$.
The exact number of initial values is chosen such that the procedure,
if possible, saturates, see below.

Since Levenberg-Marquart is a minimizer, we next sort out slow points,
i.e., points at which the norm of the velocity is locally minimal
but not zero; for a point to be counted as fixed point, the norm of
the velocity is required to be $<10^{-6}$. We call the resulting
list of fixed points \emph{candidates}.

\begin{figure}
\includegraphics{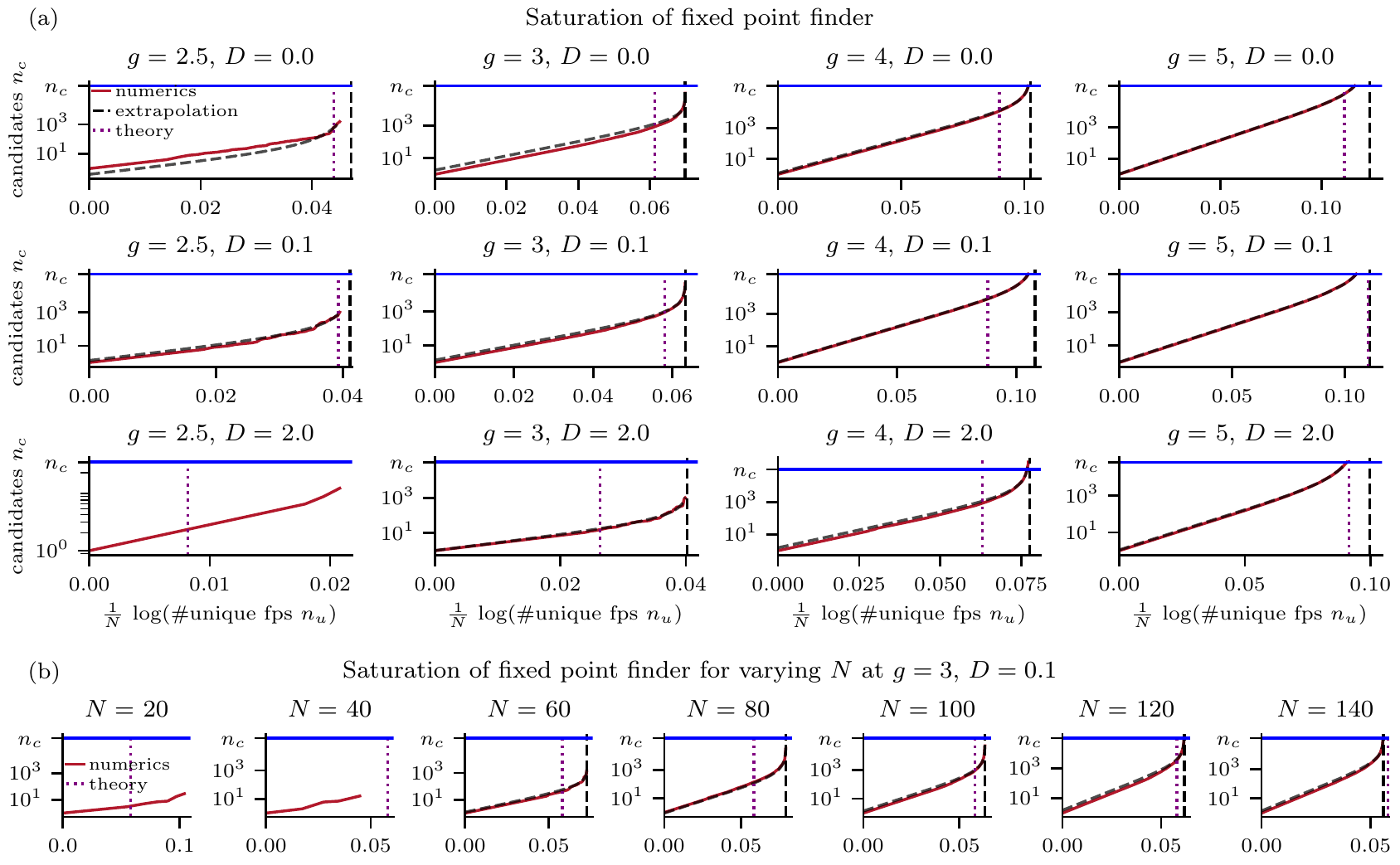}

\caption{\label{fig:Saturation}Saturation of the fixed point finder. For each
number of unique fixed points ($x$-axis) we show the number of candidates
needed ($y$-axis, red solid line). The black dashed curve shows the
fit to the function $n_{c}(n_{u})=p_{1}\left((n_{u}-p_{2})^{-p_{3}}-p_{2}^{-p_{3}}\right)$.
The dashed vertical line denotes the pole position $p_{2}$. It estimates
the expected number of unique fixed points. (a)$N=100$; $g$ and
$D$ are varied. (b) $N$ is varied; $g=3$, $D=0.1$.}
\end{figure}

\begin{figure}
\includegraphics{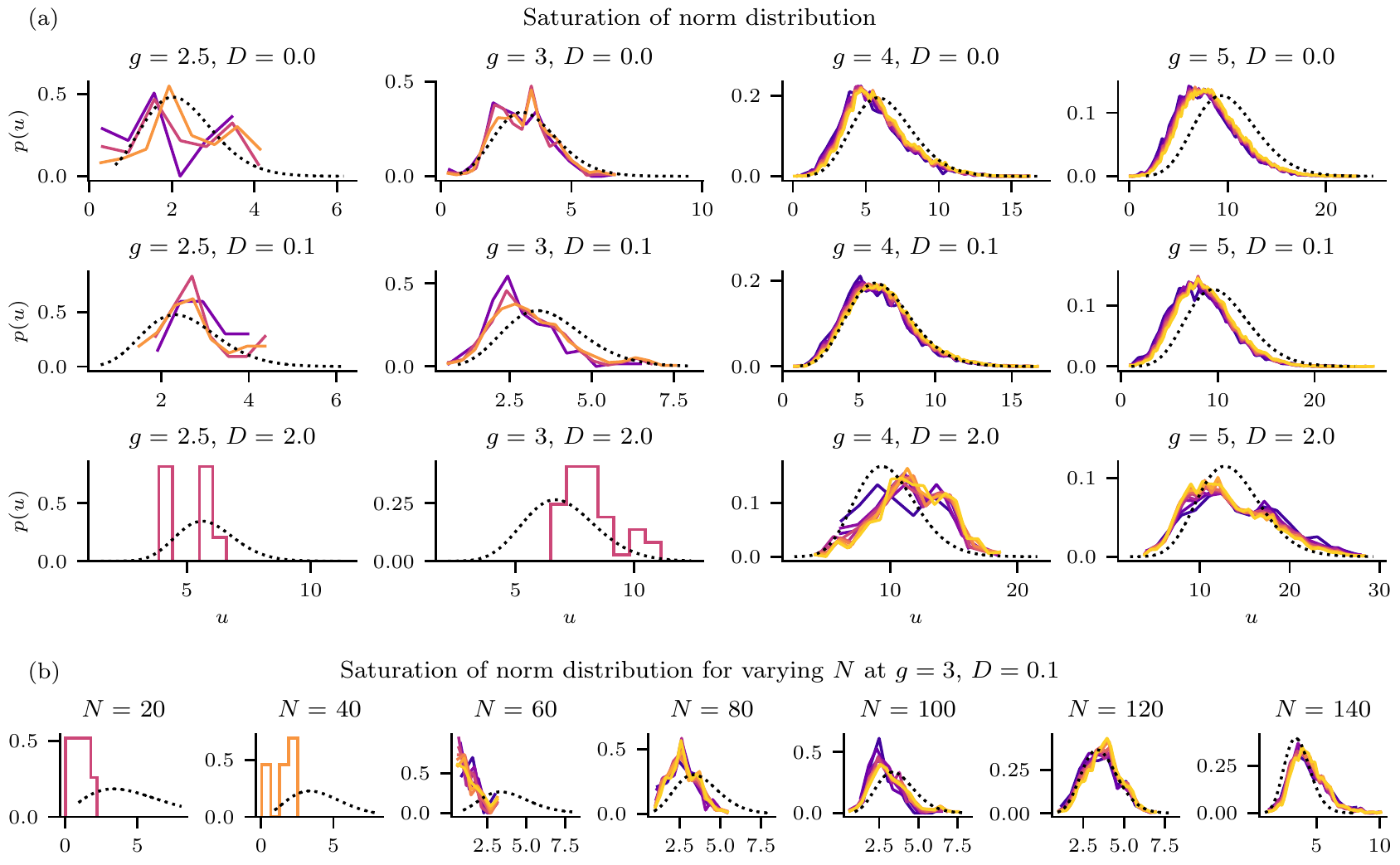}

\caption{\label{fig:Norm-saturation}Norm saturation. Distribution of the scaled
squared norm $u=\frac{1}{N}\protect\bx^{\ast\protect\T}\protect\bx^{\ast}$
of fixed points. The color denotes the number of fixed points taken
into account: The purple line denotes the distribution of a first
batch of fixed points. The yellow line uses all fixed points found.
The dotted lines show the theory based on Eqs. \eqref{eq:rate_function_for_u}
and \eqref{eq:optimize_for_rf_for_u}. (a) $N=100$; $g$ and $D$
are varied. All fixed points for a given realization are split into
$10$ batches (for most parameters; exceptions: $4$ batches for $g\protect\leq3$
and $2$ batches for $g\protect\leq3$ and $D=2$) of increasing appearance
during fixed point finding. (b) $N$ is varied; $g=3$, $D=0.1$.
Here, all fixed points are split into $10$ batches (for most parameters;
exceptions: $4$ batches for $N=40$, $2$ batches for $N=20$).}
\end{figure}

\subsubsection{Saturation}

Due to the large number of initial conditions, some fixed points are
found repeatedly in the list of candidates. Furthermore, fixed points
with certain properties (e.g. a small norm) seem to be found more
likely than others. To remove this bias, resulting from the choice
of the ensemble of initial conditions and from the mechanics of Levenberg-Marquart,
we use a high number of initial conditions and sort out duplicates.

To test whether the number of initial conditions is high enough, we
monitor the saturation: Iterating through the list of candidates,
for each new unique fixed point found (say, number $n_{u}$), we count
the number $n_{c}$ of candidates needed to find $n_{u}$ unique fixed
points. At the beginning of the iteration, $n_{c}\approx n_{u}$,
because almost every candidate has been unknown before. Along the
list of candidates, $n_{c}$ grows supra-linearly, as more and more
candidates are already contained in the list of unique fixed points
and hence more candidates are needed to find a new unique fixed point.
When $n_{c}\gg n_{d}$, the procedure is saturated in the sense that
a large fraction of the fixed points accessible to Levenberg-Marquart
seem to be found. The flow of the saturation $n_{c}$ is shown in
\prettyref{fig:Saturation}.

Assuming a finite number of fixed points, we expect a pole in the
function $n_{c}(n_{u})$ at the position of the (unknown) complete
number of unique fixed points $n_{u}\equiv p_{2}$. A generic function
for such a pole is $(n_{u}-p_{2})^{-p_{3}}$ where $p_{3}>0$. Setting
the $y$-axis intercept to $0$ requires subtracting $p_{2}^{-p_{3}}$.
To allow for an arbitrary overall scale, we further multiply the function
by $p_{1}$. Hence, we fit the saturation flow to the function $n_{c}(n_{u})=p_{1}\left[(n_{u}-p_{2})^{-p_{3}}-p_{2}^{-p_{3}}\right]$,
see \prettyref{fig:Saturation}. With this, $p_{2}$ is the extrapolated
estimate of the number of unique fixed points. This number is shown
in the main text in Fig. 1(e).

To understand the bias of the fixed point finder, we further show
in \prettyref{fig:Norm-saturation} how the norm distribution of the
fixed points changes after acquiring more and more fixed points. Specifically,
we sort the fixed points by order of unique appearance into equally
sized sets (batches) and compute the norm distribution of the first
batch, the first two batches, and so on. The size of the batches depends
on the number of fixed points found and is specified in the caption
of \prettyref{fig:Norm-saturation}. Here, we observe that especially
for poorly saturated searches {[}e.g. $g=5$, $D=0$, see upper right
panel in \prettyref{fig:Saturation}(a){]}, the procedure finds fixed
points with a small norm more likely. For very low numbers of fixed
points (e.g. for $g=5$, $D=2$), such a conclusion cannot be made.
Overall, the dynamic changes of the norm distribution during saturation
show that fixed points with certain features are found more often,
by identifying and sorting out these duplicates, the bias of the optimizer
can be reduced.

\subsubsection{Finite size effects}

\begin{figure}
\includegraphics{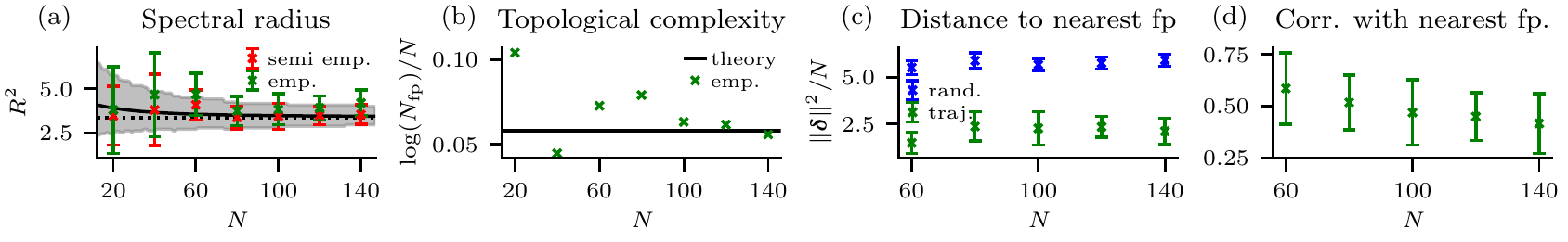}

\caption{\label{fig:Finite-size-effects}Finite size effects for the example
$g=3$, $D=0.1$. The fixed point finding was repeated for the network
sizes depicted on the $x$-axes. (a) Spectral radius. Large $N$ limit
{[}main Eq. (11), black dotted line{]} and finite $N$ result {[}\prettyref{eq:rate_function_spectral_radius},
mean value (black solid line), and asymmetric standard deviation i.e.
values $R_{\pm}^{2}$ s.t. $\int_{R_{-}^{2}}^{\text{mean}}p(R^{2})\,dR^{2}=\int_{\text{mean}}^{R_{+}^{2}}p(R^{2})\,dR^{2}=0.34$
(gray shading){]}. Empirical results are split into 1) the semi empirical
spectral radius based on numerically found fixed points and random
matrix theory $\left\langle R_{\text{semi emp.}}^{2}\right\rangle =g^{2}\left\langle \phi^{\prime}(\protect\bx)^{\protect\T}\phi^{\prime}(\protect\bx)\right\rangle _{\text{empirical fps.}\,\protect\bx}$
(red errorbars); and 2) the fully empirical spectral radius $\left\langle R_{\text{emp.}}\right\rangle =\left\langle \max_{\text{eigvals }\lambda}|\lambda+1|\right\rangle _{\text{empirical fps.}}$
(green errorbars). Both empirical error estimates are standard deviations
across fixed points. The finite size correction of the theory is only
expected to capture the semi empirical fluctuations. (b) Topological
complexity $N^{-1}\log(N_{\text{fp}}$). Large $N$ limit (theory,
black horizontal line), and empirical result based on extrapolation
for $N\protect\geq60$ and based on total number of fixed points found
for $N\protect\leq40$, see \prettyref{fig:Saturation}(b). (c) Distance
of the trajectory to the nearest fixed points (green errorbars) compared
to the distance of random points on the dynamics shell to the nearest
fixed points (blue errorbars). (d) Pearson correlation of the velocity
of the integrated model with the linear predictor by the nearest fixed
point.}
\end{figure}

Here, we explore some finite size effects numerically, which are summarized
in Fig. \ref{fig:Finite-size-effects}.

The empirical spectral radius (based on the maximum deviation of eigenvalues
from the center) is in line with the semi empirical spectral radius
(based on random matrix theory) within their respective fluctuations.
Yet, there seems to be a slight shift of the standard deviation of
the mean. The theory we present only captures the fluctuations of
the semi empirical spectral radius.

Beyond the summary in Fig. \ref{fig:Finite-size-effects}, there is an overrepresentation near $0$ in
the empirical measure \prettyref{fig:EM-and-Num}(c), especially for
low $N$, which remains up to $N=140$, above which we did not simulate.
We currently lack an explanation of this.

In conclusion, the analysis of the finite size effects underlines the robustness of our results.